# 博士學位論文

# 펨토셀 및 매크로셀 네트워크에서 멀티미디어 응용을 위한 적응적인 무선자원관리

# Adaptive Resource Management for Multimedia Applications in Femtocellular and Macrocellular Networks

**Graduate School, Kookmin University**
**Department of Electronics Engineering**
**Mostafa Zaman Chowdhury**
**2011**

# 博士學位論文

펨토셀 및 매크로셀 네트워크에서 멀티미디어 응용을 위한 적응적인 무선자원관리

Adaptive Resource Management for Multimedia Applications in Femtocellular and Macrocellular Networks

Supervised by Prof. **Yeong Min Jang**
Co-supervised by Prof. **Sunwoong Choi**

A dissertation submitted in fulfillment of the requirement for the degree of Doctor of Philosophy (Electronics Engineering)
June 2012

Graduate School, Kookmin University
Department of Electronics Engineering
Mostafa Zaman Chowdhury
2011

Mostafa Zaman Chowdhury 의 博士學位 請求論文을 認准함

**2012 年 06 月**

<u>審査委員長　　金 基 斗</u>㊞

<u>審 査 委 員　　張 暎 民</u>㊞

<u>審 査 委 員　　李 尙 煥</u>㊞

<u>審 査 委 員　　鄭 成 鎬</u>㊞

<u>審 査 委 員　　李 承 珩</u>㊞

國民大學校大學院

*Dedicated to my beloved mother, father, and wife.*

# Acknowledgement

I am grateful to Almighty Allah, the most beneficial and merciful, for giving me the opportunity to complete my PhD study. I would like to express my sincere gratitude to my advisor, Prof. Yeong Min Jang, for his valuable guidance and support throughout my study. Besides providing excellent academic guidance, he has been incredibly encouraging, supportive, and patient. It has been a great pleasure to have Prof. Sunwoong Choi as my co-supervisor. I would also like to convey my heartiest gratitude to my course teachers. My sincere acknowledgement is for the dissertation committee members who have granted me their valuable time and effort to review my dissertation. Their valuable comments have helped me a lot to improve my dissertation. I would like to express my special acknowledgements to Prof. Ki-Doo Kim, Prof. Youngil Park, and Prof. Zygmunt J. Haas (Cornell University, USA) for their valuable suggestions during my PhD study.

I would like to express my thanks to the authority of Kookmin University to provide me sufficient facilities during my study period. Special thanks go to International affair team and Electronics Engineering department of Kookmin University for their endless help. I am also thankful to Korean Government for giving me the precious opportunity of study here in Republic of Korea. I would like to thanks to the authority of KUET for giving me the leave to study in Kookmin University.

I would like to express my love and affection to my dear wife Fahmida. My wife's endless inspiration helped me a lot to go ahead towards my goal. I am grateful to my parents and parents-in-law for their blessings. My sister, sisters-in-law, nephews, every one of my family had been supportive to me, I won't forget their kindness. I would like to thank my relatives, teachers, and my friends for their continuous love, sacrifice, encouragement, and faith on me. I am grateful to all my friends who are living in Korea. For their company, I never felt alone here.

I am grateful to all the members of the Wireless Networks and Communication Laboratory, Young Min Seo, Tuan, Hyunh, Anh Tuan, Shahin, Nirzhar, Ratan, and Sung Hun for their friendship and help during last four years. They have been so much supporting in every step, which is really unimaginable.

# Table of Contents

# List of Figures

# List of Tables

# Acronyms

| | |
|---|---|
| AQoS | Adaptive QoS |
| ARPU | Average revenue per user |
| BB | Bandwidth broker |
| BS | Base station |
| CAC | Call admission control |
| CAPEX | Capital expenditure |
| CATV | Cable TV |
| CN | Core network |
| CR | Cognitive radio |
| FAP | Femto-access point |
| FGW | Femto-gateway |
| FRA | Flexible resource-allocation |
| FMC | Fixed mobile convergence |
| FMS | Femtocell management system |
| GPS | Global positioning system |
| GBR | Guaranteed bit rate |
| HCDP | Handover call dropping probability |
| IMS | IP multimedia subsystem |
| ISP | Internet service provider |
| MBS | Multicast/broadcast services |
| NGMN | Next generation mobile network |
| NGBR | Non-guaranteed bit rate |
| OPEX | Operational expenditure |
| PDF | Probability density function |
| QoS | Quality of service |
| RNC | Radio network controller |
| RRC | Radio resource control |
| RSSI | Received signal strength indicator |
| SIP | Session initiated protocol |
| SVC | Scalable video coding |
| SeGW | Security GateWay |
| SIR | Signal-to-interference ratio |
| SON | Self-organizing network |
| UE | User equipment |
| UMA | Unlicensed mobile access |
| WiMAX | Worldwide interoperability for microwave access |

# Abstract

# Adaptive Resource Management for Multimedia Applications in Femtocellular and Macrocellular Networks

*by* Mostafa Zaman Chowdhury

*Dept. of Electronics Engineering*
*Graduate School, Kookmin University*
*Seoul, Korea*

The increasing demands of various high data rate wireless applications have been seen in the recent years and it will continue in the future. To fulfill these demands, the limited existing wireless resources should be utilized properly or new wireless technology should be developed. Therefore, we propose some novel idea to manage the wireless resources and deployment of femtocellular network technology. The study was mainly divided into two parts: (a) femtocellular network deployment and resource allocation and (b) resource management for macrocellular networks. The femtocellular network deployment scenarios, integrated femtocell/macrocell network architectures, cost-effective frequency planning, and mobility management schemes are presented in first part. In the second part, we provide a CAC based on adaptive bandwidth allocation for the wireless network in. The proposed CAC relies on adaptive multi-level bandwidth-allocation scheme for non-real-time calls. We propose video service provisioning over wireless networks. We provide a QoS adaptive radio resource allocation as well as popularity based bandwidth allocation schemes for scalable videos over wireless cellular networks. All the proposed schemes are verified through several numerical and simulation results. The research results presented in this dissertation clearly imply the advantages of our proposed schemes.

# Chapter 1
# Introduction

A notable trend of growing demand of high data rate multimedia traffic was found during the last couple of years. It is well forecasted that the demand of these data traffic will be increased as well in future. Also, users are more interested with the wireless connectivity compared to the wired connectivity. The load on the wireless network technologies is increasing even the wireless capacity is increasing. However, the capacity increasing rate for the wireless networks is always lower than the increasing demand rate. Moreover, many wired services are converted to wireless services (e.g., mobile IPTV). Therefore, the limited wireless resources should be utilized properly. New wireless technologies e.g., femtocells are developing to support mobile users in different environments. The new indoor wireless technologies will divert huge amount of users from the expensive macrocellular networks to the indoor wireless connectivity.

This dissertation contributes to deploy the femtocellular networks, Quality of Service (QoS) provisioning for the macrocellular users, and video service provisioning through wireless networks. This dissertation represents the main contributions of the author's research studies during the past four years.

## 1.1 Problem Statement

Integrated femtocell/macrocell networks, comprising a conventional cellular network overlaid with femtocells, offer an economically appealing way to improve coverage, quality of service, and access network capacity. There are several technical, business, and regulatory issues of the femtocellular technology that remain to be addressed [1]-[9] for the dense deployment of femtocells. As the femtocellular network co-exists with the macrocellular infrastructure, the technical issues relate to radio resource management, end-to-end QoS support for network architecture, and mobility management. In particular, the main technical issues that affect the performance of the integrated network and influence the QoS level of the femtocell users are: frequency and interference management and handover between a macrocell and femtocells.

In wireless systems, supporting QoS requirements of different traffic types is more challenging due to the need to minimize two performance metrics - the probability of dropping a handover call and the probability of blocking a new call. Since QoS requirements are not as stringent for non-real-time traffic types, as opposed to real-time traffic, more calls can be accommodated by releasing some bandwidth from the already admitted non-real-time traffic calls.

Good quality video services always require higher bandwidth. Hence, to provide the video services e.g., multicast/broadcast services (MBS) and unicast services along with the existing voice, internet, and other background traffic services over the wireless cellular networks, it is required to efficiently manage the wireless resources in order to reduce the overall forced call termination probability, to maximize the overall service quality, and to maximize the revenue. Fixed bandwidth allocation for the MBS sessions either reduces the quality of the MBS videos and bandwidth utilization or increases the overall forced call termination probability and of course the handover call dropping probability as well.

## 1.2 Research Objectives

This research aim is to efficiently manage the radio resource for femtocellular and macrocellular networks.

### 1.2.1 Femtocellular Network Deployment

*Femto-access points (FAPs)* are low-power, small-size home-placed Base Stations (also known as *Home NodeB* or *Home eNodeB*) that create islands of increased capacity in addition to the capacity provided by the cellular system. These areas of increased capacity are referred to as *femtocells*. Femtocells operate in the spectrum licensed for cellular service providers. The key feature of the femtocell technology is that users require no new equipment (UE). However, to allow operation in the licensed cellular spectrum, coordination between the femtocells and the macrocell infrastructure is required. Such coordination is realized by having the femtocells connected to the local mobile operator's network using one or more of the following backhaul network technologies: residential xDSL, cable TV (CATV), Metro Ethernet, or WiMAX. There are several technical, business, and regulatory issues of the femtocellular technology that remain to be addressed. As the femtocellular network co-exists with the macrocellular infrastructure, the technical issues relate to radio resource management, end-to-end QoS support for network architecture, and mobility management. In particular, the main technical issues that affect the performance of the integrated network and influence the QoS level of the femtocell users are: frequency and interference management, service provisioning in capacity limited xDSL backhaul, and handover between a macrocell and femtocells.

In this dissertation, we provide the efficient management of femtocellular networks co-existing with the macrocellular networks. The femtocellular network configurations, integrated femtocell/macrocell network architecture based on the density of femtocells, cost-effective frequency planning, detail mobility management including the call admission control, and traffic model are performed in this dissertation for the femtocellular network deployment.

### *1.2.2 CAC based on Adaptive Bandwidth Allocation for Wireless Networks*

Provisioning of QoS is a key issue in any multi-media system. However, in wireless systems, supporting QoS requirements of different traffic types is more challenging due to the need to minimize two performance metrics - the probability of dropping a handover call and the probability of blocking a new call. Since QoS requirements are not as stringent for non-real-time traffic types, as opposed to real-time traffic, more calls can be accommodated by releasing some bandwidth from the already admitted non-real-time traffic calls. If we require that such a released bandwidth to accept a handover call ought to be larger than the bandwidth to accept a new call, then the resulting probability of dropping a handover call will be smaller than the probability of blocking a new call. In this paper we propose an efficient Call Admission Control (CAC) that relies on adaptive multi-level bandwidth-allocation scheme for non-real-time calls. The scheme allows reduction of the call dropping probability along with increase of the bandwidth utilization.

Numerous prior research works have been published that allow higher priority for handover calls over new calls (e.g., [10], [11]). Most of these proposed schemes are based on the notion of "guard band," where a number of channels are reserved for the exclusive use of handover calls. Although schemes based on guard bands are simple and capable of reducing the HCDP, these schemes also result in reduced bandwidth utilization.

Another approach to reduce HCDP is handover-queuing schemes, which allow handover calls to queue and wait for a certain time for resources to become available. However, the handover-queuing schemes are not practical approaches for real-time multimedia services, because of the limited queuing time that could be allowed for real-time traffic.

We propose the "*Prioritized bandwidth-allocation scheme*," a multi-level bandwidth-allocation scheme for non-real-time traffic, which supports negligible HCDP without reducing the resource utilization. (We will also often refer to this scheme simply as "adaptive *bandwidth-allocation scheme.*) The proposed scheme reserves some releasable bandwidth to accept handover calls. The prioritized adaptive bandwidth-allocation scheme allows to reclaim more bandwidth in the case of handover calls, thus increasing the probability of accepting a handover call, as opposed to new calls. And even though our scheme blocks more new calls, still the bandwidth utilization is not reduced, because the scheme accepts new calls for which it expects to be able to provide sufficient resources until the call ends.

### *1.2.3 Scalable Video Services over Wireless Cellular Networks*

Good quality video services always require higher bandwidth. Hence, to provide the video services e.g., MBS and unicast services along with the existing voice, internet, and other background traffic services over the wireless cellular networks, it is required to efficiently manage the wireless resources in order to reduce the overall forced call termination probability, to maximize the overall service quality, and to maximize the revenue. Fixed bandwidth allocation for the MBS sessions either reduces the quality of the MBS videos and bandwidth utilization or increases the overall forced call termination probability and of course the handover call dropping probability as well. Scalable Video Coding (SVC) technique allows the variable bit rate allocation for the video services. In this study, we propose a bandwidth allocation scheme that efficiently allocates bandwidth among the MBS sessions and the non-MBS traffic calls (e.g., voice, unicast, internet, and other background traffic). The proposed scheme reduces the bandwidth allocation for the MBS sessions during the congested traffic condition only to accommodate more calls in the system. Instead of allocating fixed bandwidths for the MBS sessions and the non-MBS traffic, our scheme allocates variable bandwidths for them. However, the minimum quality of the videos is guaranteed by allocating minimum bandwidth for them.

### *1.2.4 Popularity based Bandwidth Allocation for Scalable Video over Wireless Networks*

With the rapid improvement of various wireless network technologies, now it is possible to provide high quality video transmission over the wireless networks. The high quality video streams need higher bandwidth. Hence, during the video transmission through wireless networks, it is very important to make the best utilization of the limited bandwidth. Therefore, when many video sessions are active, the bandwidth per video session can be allocated based on popularity of the video sessions (programs). Instead of allocating equal bandwidth to each of them, our proposed scheme allocates bandwidth per video session based on popularity of the video program. When the system bandwidth is not sufficient to allocate the demanded bandwidth for all the active video sessions, our scheme efficiently allocates the total system bandwidth among all the scalable active video sessions in such a way that more bandwidth is allocated to higher popularity one.

## 1.3 Dissertation Outline

The Fig. 1 highlights the structure of this dissertation.

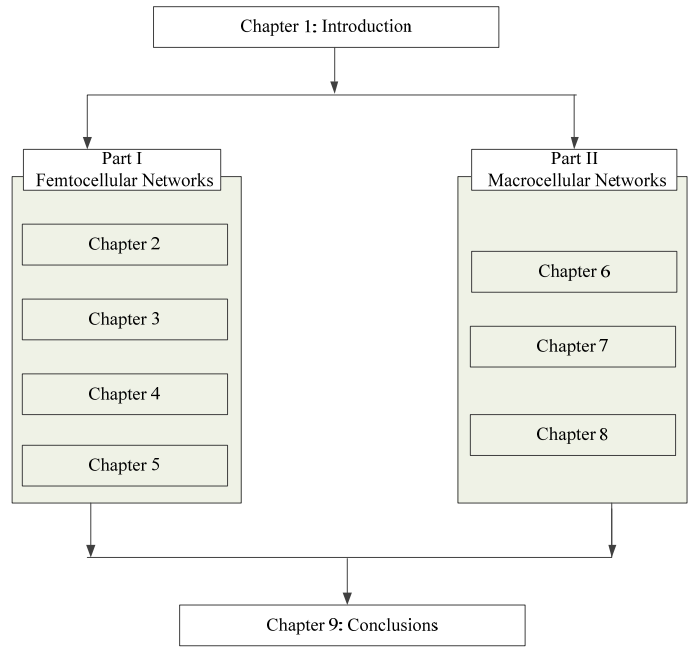

**Fig. 1.1:** Organization of the dissertation chapter.

The dissertation is organized as follows. Chapter 2 discusses about the femtocellular network deployments. Chapter 3 provides the integrated femtocell/macrocell network architectures. Cost-effective frequency planning for capacity enhancement of femtocellular networks is presented in Chapter 4. Chapter 5 discusses the mobility management issues for the dense femtocellular network deployment. We provide a CAC based on adaptive bandwidth allocation for the wireless network in Chapter 6. In Chapter 7, we provide a QoS adaptive radio resource allocation for scalable videos over wireless cellular networks. Popularity based bandwidth allocation for scalable video over wireless networks is proposed in Chapter 8. Finally, Chapter 9 summarizes the research results and contributions.

Part I:

Resource Management for

Femtocellular Networks

# Chapter 2
# Femtocellular Network Deployments

*FAPs* are low-power, small-size home-placed Base Stations (also known as *Home NodeB* or *Home eNodeB*) that create islands of increased capacity in addition to the capacity provided by the cellular system. These areas of increased capacity are referred to as *femtocells*. Femtocells operate in the spectrum licensed for cellular service providers. The key feature of the femtocell technology is that users require no new UE. However, to allow operation in the licensed cellular spectrum, coordination between the femtocells and the macrocell infrastructure is required. Such coordination is realized by having the femtocells connected to the local mobile operator's network using one or more of the following backhaul network technologies: residential xDSL, CATV, Metro Ethernet, or WiMAX. Fig. 2.1 shows a simple example of femtocellular network deployment. Femtocells are normally deployed in indoor environments. FAP can be connected to core network through either xDSL networks or cable networks. Several FAPs are connected to femto-gateway (FGW). The subscribers receive the wireless services through macrocellular networks whenever they stay outside indoor environment. However, when the users move to indoor environments, they use the femtocellular networks.

It is envisioned that a FAP would be typically designed to support simultaneous cellular access of two to six mobile users in residential or small indoor environments. Predictions show that in the near future about 60% of voice traffic and about 90% of data traffic will originate from indoor environments, such as a home, an office, an airport, and a school [12]. Therefore, there is the need for improved indoor coverage with larger data rates. However, due to the limited cellular capacity, it might be difficult and too expensive to accommodate this increased traffic demand using the current macrocellular coverage. Femtocells are a new candidate technology that is capable of providing expanded coverage with increased data rates. Due to interest from operators, such as the NGMN (Next Generation Mobile Network) Alliance, and standardization bodies, such as 3GPP, 3GPP2, Femto Forum, IEEE 802.16m, WiMAX Forum,

Broadband Forum, ITU-T, and ITU-R WP5D, the integrated femtocell/macrocell will be a major part of network evolution and revolution for IMT-Advanced architectures [13]. ABI Research expects cellular-based femtocells to outpace UMA and SIP-based Wi-Fi solutions by 2013, grabbing 62 % of the market.

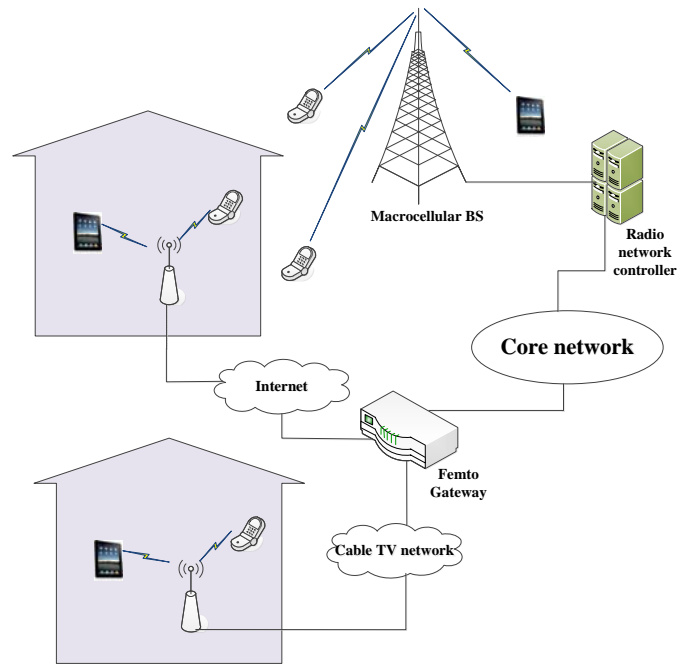

**Fig. 2.1:** A simple example of femtocellular network deployment.

There are several technical, business, and regulatory issues of the femtocellular technology that remain to be addressed [1]-[9]. As the femtocellular network co-exists with the macrocellular infrastructure, the technical issues relate to radio resource management, end-to-end QoS support for network architecture, and mobility management. In particular, the main technical issues that affect the performance of the integrated network and influence the QoS level of the femtocell users are: frequency and interference management, service provisioning in capacity limited xDSL backhaul, and handover between a macrocell and femtocells.

## 2.1 Key Advantages of Femtocellular Network Architecture

The advantages of the femtocellular technology can be examined from multiple perspectives, such as the users, the manufacturers, the application developers, the network operators, and the service and content providers. For example, from the user's perspective, customers typically expect high data-rate wireless indoor access at low cost and with good QoS. Of course, the key advantage of the femtocells for users is that there is no need for an expensive dual-mode handset; rather, the same single-mode handset is used to access the FAPs and the macrocellular network.

From the perspective of a network operator, the followings are the key advantages of the integrated femtocell/macrocell network:

- **Improved coverage:** Providing extensive in-building coverage has long been a challenge for mobile operators. This is even a more difficult problem for communication at higher frequencies, where radio propagation loss is larger. Femtocells can provide better signal reception within the indoor environment, as FAPs are also located inside the building. Thus, using of the basic concept of spectrum re-use, femtocells can improve the network coverage and increase the network capacity. Especially, femtocells extend the service coverage into remote or indoor areas, where access to a macrocellular network is unavailable or is limited. Of course, improved coverage and access capacity enhances customer satisfaction, allowing the network operator to retain and expand its customer pool.

- **Reduced infrastructure and capital costs:** Femtocells use the existing home broadband connectivity for backhauling the femtocells' traffic. Thus, by steering users' traffic into their own FAPs and away from the macrocells, femtocells reduce the expensive backhaul costs of macrocellular networks.

- **Power saving:** Since femtocells target indoor coverage, the transmission power is significantly smaller, as compared to the transmission power of the macrocellular network that is required to penetrate into buildings. Smaller transmission power results in decreased battery drainage of the mobile devices, prolonging the devices' lifetime. Furthermore, decrease in the transmission power reduces inter-cell interference, increasing the signal-to-interference-ratio. This, in turn, improves the reception, increasing capacity and coverage.
- **Provisioning of QoS:** The radio path loss close to the fringe of a macrocell can be quite severe. Femtocells can in particular improve the QoS for users with poor macrocell reception.

## 2.2 Comparison with Some Other Network Technologies

Even though there are several technical approaches to improve the indoor coverage, femtocell appears to be the most attractive alternative. Compared with the *Fixed Mobile Convergence (FMC)* framework, femtocells allow servicing large numbers of indoor users. In contrast with the 3G/Wi-Fi UMA(*Unlicensed Mobile Access*) technology, femtocell do not require dual-mode handset. Furthermore, another drawback of Wi-Fi is its use of the increasingly crowded unlicensed ISM band that causes significant interference. Finally, repeaters (or signal booster) [14] improve the wireless access coverage, but not the wireless capacity. Repeaters need new backhaul connections and only solve the poor coverage problem in remote areas, where fixed broadband penetration is low. Table 2.1 shows the comparison between the Wi-Fi and FAP technologies. Table 2.2 explains the comparison between the repeater, FAP, and Wi-Fi technologies.

**Table 2.1:** Comparison between the Wi-Fi and FAP

|  | Wi-Fi | FAP (Home Base Station) |
| --- | --- | --- |
| **Transmitting Power** | ~50 mW | ~15 mW |
| **Radio Signal Robustness** | More readily fades | More robust |
| **Voice Optimization** | Wi-Fi is not optimized for voice | Femtocell is optimized for voice |
| **Spectrum** | Use un-license spectrum | Use license spectrum |
| **Handset** | Costly dual-mode | Single mode |
| **Complexity of Service Delivery** | High | Relatively Low |
| **Required Residential Networking Equipment** | Wi-Fi router (accessible to the handset) | Consumer Broadband (DSL, Cable, or Wireless…) |

**Table 2.2:** Technical solutions for indoor poor coverage

| Solutions / Feature | FAP | Repeater (or signal booster) | 3G/Wi-Fi UMA |
|---|---|---|---|
| **Installing infrastructure** | Existing xDSL or CATV connection | Needs new connection | Existing Wi-Fi connection |
| **Coverage** | In indoor, hot spot or remote area where cellular voice or data services are required | Solves poor coverage issues in remote areas where fixed broadband penetration is low. | In indoor, hot spot or remote area where cellular data services are required |
| **Terminal mode** | Single mode | Single mode | Dual-mode |
| **Data speed** | High | Low | High |

## 2.3 Femtocell Network Configurations

There is a fundamental shift in the architecture of the femtocell networks, with smaller cells closer to the user. We briefly analyze some of the benefits expected in its deployment, for both operators and end-users. Fig. 2.2 shows three possible scenarios for femtocell configuration based on the coverage of existing ISP and macrocellular networks:

**Type A - a single stand-alone femtocell**: This could be the case of a remote area with no macrocellular coverage or a poor coverage area (such as an indoor or macrocell edge), and when no other neighboring femtocells are available. This type of a configuration extends the service coverage into remote areas.

**Type B - a network of stand-alone femtocells:** In this scenario, multiple FAPs are situated within an area in such a way that a radio signal from one FAP overlaps with other FAPs' signals. There is no macrocellular coverage or the coverage is poor. Femtocell-to-femtocell handovers are present and need to be handled by the femtocellular network. As the Type A configuration, the Type B configuration is also able to extend the service coverage into remote areas.

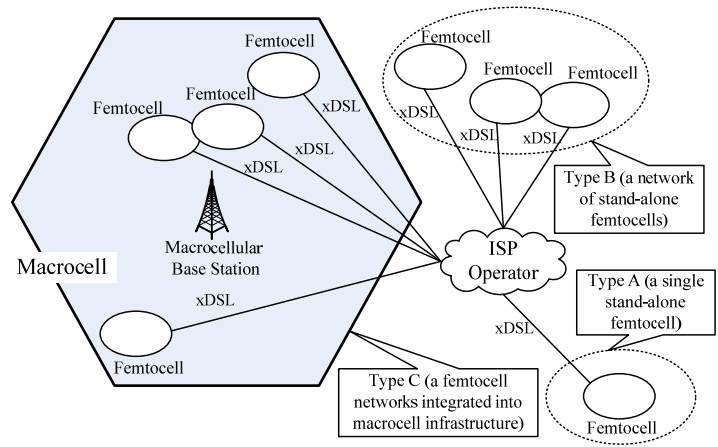

**Fig. 2.2:** The three types of the femtocellular network configurations.

**Type C - a femtocell network integrated with a macrocellular infrastructure:** This scenario can be viewed as a two-tier hierarchical network, where the macrocells create the upper tier and the femtocells the lower tier. Handover between macrocells and femtocells, as well as handover between femtocells, are common occurrence in this scenario. This configuration improves the indoor service quality and reduces the traffic load of the macrocells by diverting traffic to femtocells.

# Chapter 3

# Integration of Femtocellular and Macrocell Networks

Integrated femtocell/macrocell networks, comprising a conventional cellular network overlaid with femtocells, offer an economically appealing way to improve coverage, QoS, and access network capacity. The key element to successful femtocells/macrocell integration lies in its self-organizing capability. Provisioning of quality of service is the main technical challenge of the femtocell/macrocell integrated networks, while the main administrative challenge is the choice of the proper evolutionary path from the existing macrocellular networks to the integrated network. From the wireless operator point of view, the most important advantage of the integrated femtocell/macrocell architecture is the ability to offload a large amount of traffic from the macrocell network to the femtocell network. This will not only reduce the investment capital, the maintenance expenses, and the operational costs, but will also improve the reliability of the cellular networks. Dense deployment of femtocells will offload huge traffic from the macrocellular network to femtocellular network by the successful integration of macrocellular and femtocellular networks.

Fig. 3.1 shows an example of macrocellular network integration with femtocells. Macrocells are operated by a mobile wireless operator, while femtocells are privately owned and connected to a broadband service provider, such as an Internet Service Provider (ISP). Thousands of femtocells may co-exist in a a macrocellular coverage area. The provision of QoS in femtocellular networks is more difficult than for the existing macrocellular networks due to the large number of neighboring FAPs and the possible interference conditions among the femtocells and between macrocells and femtocells [15]. There are a number of technical challenges to support satisfactory QoS to the users in a femtocellular network, solutions of which are being studied. The QoS of femtocellular networks is influenced by procedures such as resource allocation, network architecture, frequency and interference management, power control, handover control, security assurance, and QoS provisioning in backhaul networks.

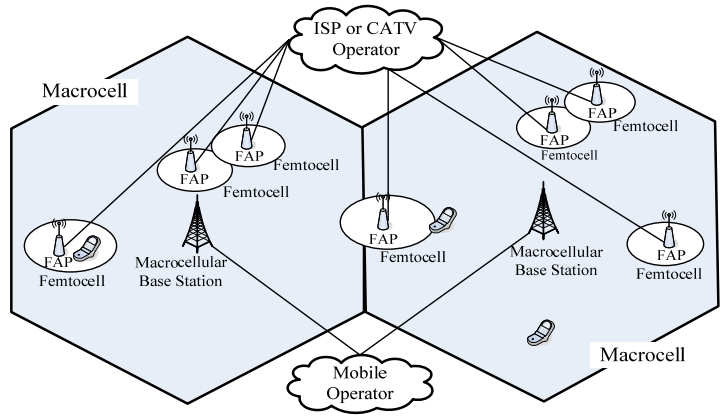

**Fig. 3.1:** Integration of a macrocellular network with femtocells.

## 3.1 Overview of Integrated Femtocell/Macrocell Networks

In the proposed three types of the femtocell deployment in Fig. 2.2, the number of FAPs, their locations, and identification parameters change dynamically. Integration with the core network (CN) should be possible using existing standardized interfaces. Some issues that should be considered for the network integration are low capital expenditure (CAPEX) and low operational expenditure (OPEX), coexistence with other wireless networks, higher spectral efficiency, improved cell-edge performance, scalability of the provisioning and the management processes, self-organizing network (SON) architecture, and QoS support. Fig. 3.2 depicts the basic device-to-CN connectivity for the femtocell network deployment [3], [16]-[18]. The femtocell access networks are connected to CN through backhaul networks. A femtocell management system (FMS) is used to control and to manage the FAPs within a regional area. The FMS functionalities include configuration, fault detection, fault management, monitoring, and software upgrades.

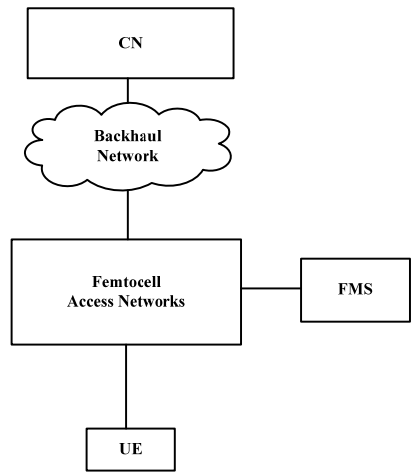

**Fig 3.2:** The basic femtocellular network architecture

There are several deployment options for the femtocell/macrocell network integration. For a particular situation, the choice of architecture depends on numerous factors, such as the size of the network, the existing network features, the future rollout and convergence plans of the network operator, capacity planning, the ability to co-exist and interact with the existing network, and the predicted network evolution. We propose three methods for integration of the femtocells with the macrocellular infrastructure:

- **Small-scale** deployment using the existing macrocell *Radio Network Controller* (RNC) interface
- **Medium- and large-scale** deployment using a concentrator and *IP Multimedia Subsystem (IMS)*
- **Large-scale and highly-dense** integration using *SON* and cognitive radio *(CR)* along with concentrator and *IMS*

The "small-scale," "medium- and large-scale," and "large-scale and highly-dense" femtocell deployment network architectures differ in terms of network entities, connecting procedures, and management systems. The size of the scale refers to the number and density of FAPs that are to be installed. For small-scale femtocell deployment, to reduce the implementation costs, the existing macrocellular RNC interface is used for FAPs connection to the macrocellular infrastructure. For medium-and large-scale femtocell deployment, the existing network infrastructure is insufficient to support the number of FAPs, and introduction of new communication elements, or modification of the existing infrastructure is necessary [19]. One of such new elements is the *FGW*, which is able to control many FAPs. IMS provides an efficient way to control a large number of signaling messages using the *Session Initiated Protocol* (*SIP*). As the number of FAPs within an area increases, the interference from neighbor FAPs increases as well, as does the rate of handovers. This is where SON [20] architectures can be useful in reducing the interference by the auto-configuration of frequency allocations and by self-adapting the transmission power of the neighbor FAPs. A SON architecture is also able to improve handover performance.

## 3.2 The Small-Scale Deployment using the Existing Macrocellular RNC Interface

Fig. 3.3 shows the femtocell/macrocell integrated network architecture for small-scale deployment. Such architecture could be useful in a rural or a remote area, where only a few femtocells are required. This architecture is similar in its control part to the existing UMTS-based 3G network; each FAP is corresponding to a macrocellular Base Station and is connected to the RNC. The *Security GateWay* (*SeGW*) implements a secure communication tunnel between FAP and RNC, providing mutual authentication, encryption, and data integrity functions for signalling, voice, and data traffic [3]. Security of communications between the FAP and the *SeGW* can also be handled by the *IPsec* protocol. The RNC is responsible for controlling and managing the macrocellular Base Stations, together with the newly added FAPs. The pros of this architecture are that it is simple and cost-effective for a small number of FAPs and that it uses the existing RNC interface. The disadvantages of the architecture are that it is not scalable to a larger number of FAPs and that the RNC capacity to support the macrocellular infrastructure is reduced.

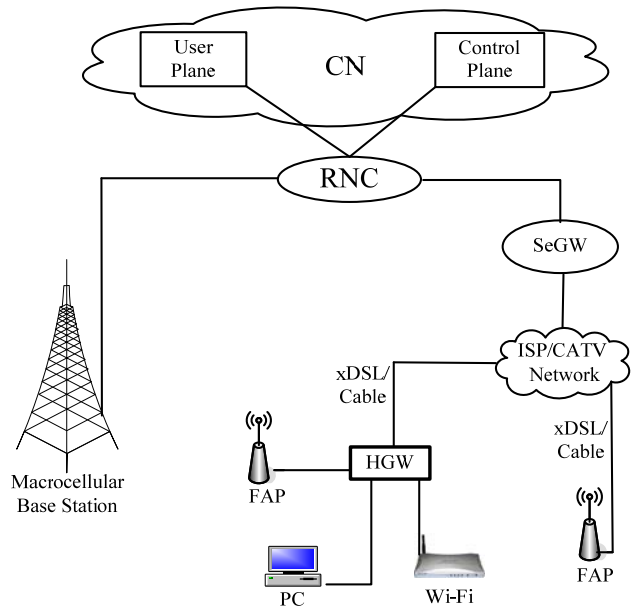

**Fig. 3.3:** The proposed small-scale deployment of the integrated femtocell/macrocell network architecture using the existing macrocellular RNC interface.

## 3.3 The Medium- and Large-Scale Deployment using a Concentrator and IMS

HSDPA/HSUPA cellular networks utilize centralized RNC to control their associated Base Stations. A single RNC is in charge of radio resource management of approximately 100 macrocellular Base Stations. However, since it is envisioned that thousands of femtocells may exist within an area of a single macrocell, a single RNC is incapable of controlling such a large number of femtocells. Therefore, a different and more efficient femtocell control scheme is necessary. One such a scheme in Fig. 3.4 implements a concentrator-based UE-to-CN connectivity for integrated femtocell/macrocell networks [3], [9], [19]. The FGW acts as a concentrator and several FAPs are connected to FGW through a broadband network, such as an ISP, for example. There is no direct interface between the RNC and FGW, so the FGW communicates with RNC through CN. The FGW manages thousands of femtocells and appears as a legacy RNC to the existing CN. Inter-operability between the mobile operator and the ISP network and among the mobile operators is required to connect the femtocell users to other users. Whenever a FAP is installed, a mobile operator stores its location information gathered from the macrocellular networks using femtocell searching (sniffing) for management purpose. Other alternatives to find location information are using the customer contractual billing address information of the ISP operator and the GPS (Global Positioning System) module, which could be optionally implemented on a FAP, but is likely to suffer from poor satellite coverage.

The main advantages of IMS-supported integrated femtocell/macrocell networks are scalability and the possibility of rapid standardization ([4]). IMS also supports seamless mobility, which is required by cellular operators [21]. A policy and an IMS-based *Bandwidth Broker (BB)* [22] is required to control the end-to-end bandwidth allocation. The advantages of this architecture are that it supports a large number of FAPs, that it supports better QoS than the small-scale deployment architecture, and that the RNC capacity is not affected by FAPs' deployment. The disadvantage of this architecture is that it is not cost effective for a small number of FAPs.

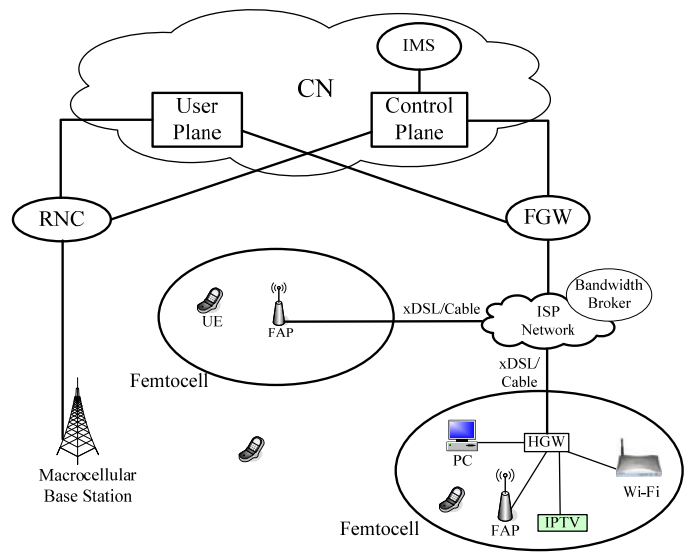

**Fig. 3.4:** The proposed medium- and large-scale deployment of the integrated femtocell/macrocell network architecture using concentrator and IMS

## 3.4 The Large-Scale and Highly-Dense Deployment Architecture using SON and CR along with Concentrator and IMS

Traditionally, macrocellular networks require complex and expensive manual planning and configuration. Currently, 3GPP LTE-Advanced and IEEE 802.16m are standardizing the SON concept for IMT-Advanced networks. The main functionalities of SON for integrated femtocell/macrocell networks are self-configuration, self-optimization, and self-healing [20]. The self-configuration function includes intelligent frequency allocation among neighboring FAPs; self-optimization attribute includes optimization of transmission power among neighboring FAPs, maintenance of neighbor cell list, coverage control, and robust mobility management; and self-healing feature includes automatic detection and resolution of most failures. The sniffing function is required to integrate femtocell into a macrocellular network, so that a FAP can scan the air interface for available frequencies and other network resources. Self-organization of radio network access is regarded as a new approach that reduces OPEX/CAPEX. It enables cost-effective support of a range of high-quality mobile communication services and applications for affordable prices in a dense femtocell deployment. An advanced self-organizing mechanism enables deployment of dense femtocell clusters [23] and the integrated femtocell/macrocell networks should incorporate SON capabilities. This way, neighboring FAPs can communicate with each other to reconfigure resources, transmission powers, and frequency assignments. Network operators may need to deploy the hierarchical femtocell network architecture based on centralized or distributed SON. The centralized SON architecture may be necessary for cooperative hotspot coverage, but the distributed and flat SON architecture may be required for individual, ad-hoc, and random femtocell coverage. Therefore, further enhancements of SON will be an important element of a future femtocell deployment. The ultimate deployment architectures of the IMT-Advanced network will rely on concentrator-based IMS/SIP and SON capable all-IP network architecture. Fig. 3.5 shows the basic features and framework of the proposed SON-capable integrated femtocell/macrocell network architecture of the large-scale and highly-dense femtocell network. Next, we briefly discuss three examples of operations of a femtocell network.

- **Scenario A - frequency configuration and power optimization:** If large number of FAPs are deployed in an indoor building or femto-zone area, signals from different FAPs will interfere with each other. The FAPs need to coordinate each other to configure frequency and optimize transmission power.
- **Scenario B - interference mitigation and cell size adjustment:** If an UE connected to a FAP that is designated as a master FAP receives interfering signal from other FAPs, then the master FAP requests these FAPs to reconfigure their transmission power, so that interference at the UE is reduced. The neighboring femtocells re-adjust their cell size to reduce the interference.

- **Scenario C - a seamless handover:** When an UE moves from a macrocell to a femtocell or from a femtocell to another femtocell and detects more than one FAP, the FAPs and macrocellular Base Station (e.g., *eNodeB*) coordinate with each other to facilitate fast and seamless handover.

The advantages of this architecture are that it supports large number and high density of FAPs, that it supports better QoS than the other two architectures, that it can reduce interference effects, and that it improves spectral efficiency. The disadvantage of this architecture is that it is not economically suitable for a deployment of a small number of FAPs. Future all-IP CN will provide a common IP-based network platform for heterogeneous access networks, advanced mobility management, enhanced session management, and network extension/composition [24]. These features will enhance the performance of the integrated femtocell/macrocell networks.

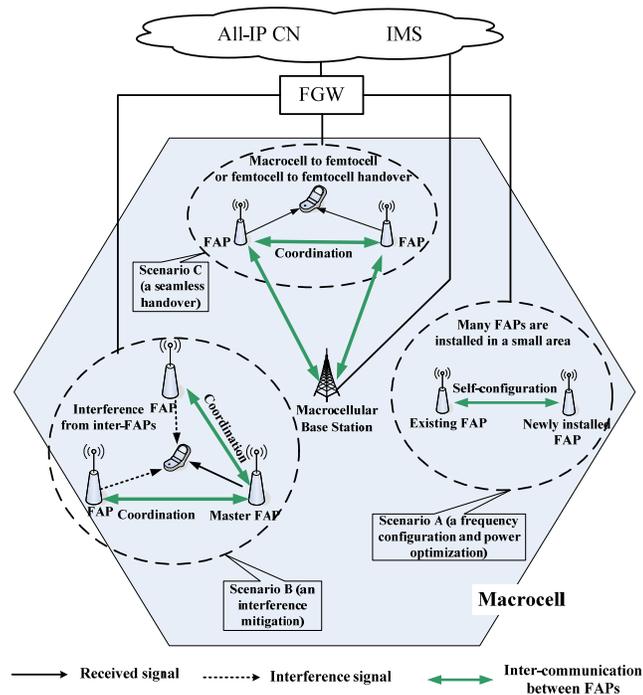

**Fig. 3.5:** The proposed SON features for the large-scale and highly-dense deployment of an integrated femtocell/macrocell network architecture.

# Chapter 4
# Cost-Effective Frequency Planning for Capacity Enhancement of Femtocellular Networks

As *femtocellular* networks will co-exist with *macrocellular* networks, mitigation of the interference between these two network types is a key challenge for successful integration of these two technologies. To support QoS and increased capacity in integrated macrocell/femtocell networks, dynamic frequency and interference management, similar to those in macrocellular networks, is necessary. At the femtocell edges, users experience significantly more interference than users located closer to the FAP [4]. Intelligent and automated radio-frequency planning is needed to minimize interference for random and unknown installation of femtocells. In 3GPP [3], a number of different deployment configurations have been considered for FAP. In particular, there are several interference mechanisms between the femtocellular and the macrocellular networks, and the effects of the resulting interference depend on the density of *femtocell*s and the overlaid *macrocell*s in a particular coverage area. While improper interference management can cause a significant reduction in the system capacity and can increase the outage probability, effective and efficient frequency allocation among femtocells and macrocells can result in a successful co-existence of these two technologies. Furthermore, highly dense femtocellular deployments – the ultimate goal of the femtocellular technology – will require significant degree of self-organization in lieu of manual configuration. In this chapter presents various femtocellular network deployment scenarios, and proposes a number of frequency-allocation schemes to mitigate the interference and to increases the spectral efficiency of the integrated network. These schemes include: *shared frequency band*, *dedicated frequency band*, *sub-frequency band*, *static frequency-reuse*, and *dynamic frequency-reuse*.

## 4.1 Interference Scenarios in a Femtocellular Network

There are different scenarios in which interference, created due to the co-existence of macrocells and femtocells in the same geographical area, may affect the performance of the femtocellular network. The amount of interference depends on the network architecture, location of femtocells, and density of femtocells. Based on these factors, Fig. 4.1 depicts four scenarios for femtocellular deployment.

- *Scenario A (Single femtocell without overlaid macrocells):* In this case, there is no interference effect from other cells. This scenario is typical of remote areas and interference is not an important issue in this case.
- *Scenario B (Single stand-alone femtocells overlaid by a macrocell):* In this case, there is a single femtocells overlaid by macrocell, so there is no femtocell-to-femtocell interference. However, a significant amount of femtocell-to-macrocell interference could exist in this case.
- *Scenario C (Multi-femtocells overlaid by a macrocell):* In this case, there are few neighboring femtocells in addition to the overlaid by macrocell. The transmissions of the four basic network entities: the macrocellular base station (BS), the macrocellular user equipment (UE), the FAPs, and the femtocellular user equipment, all potentially affect each other.

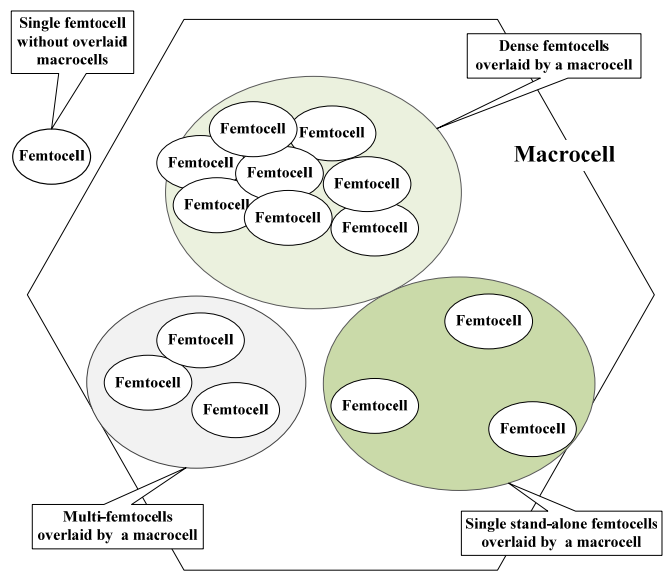

**Fig. 4.1:** Various scenarios for femtocellular/macrocellular interference.

- *Scenario D (Dense femtocells overlaid by a macrocell):* In this case, many femtocells are deployed in a relatively small geographical area. Although this is the scenario that is the goal of a successful femtocellular deployment, it also presents the worst case of interference. As in scenario *C*, the four basic network entities are all potentially affected by mutual interference.

Hence, the deployment scenario *B* or *C* or the combination of deployment scenarios *B* and *C* are termed "non-dense femtocellular network deployment," while the deployment scenario *D* is referred to as "dense femtocellular network deployment."

The identity of the offenders (entities generating the interference) and the victims (entities affected by the interference) depend on the relative positions of the four basic network entities: the FAPs, the macrocellular UE, the femtocellular UE, and the macrocellular BS. The four different link types, the macrocellular downlink and uplink, and the femtocellular downlink and uplink, can potentially create harmful interference affecting the other basic network entities [25].

- *Macrocell downlink:* The femto UEs within a macrocell coverage area receive interference from the macrocell downlink if both the macrocell and femtocell are allocated the same frequency. The situation is of particular concern when the location of the femtocell is close to the macrocellular BS and the femtocell UE is located at the edge of the femtocell, so that the transmitted power from the macrocellular BS can potentially cause severe interference to the femto UE receiver. This situation can occur in every one of the deployment scenarios *B, C, and D,* and is demonstrated in Fig. 4.2 by the macrocell downlink causing interference to the femto UE-1, femto UE-2, and femto UE-3.

- *Macrocell uplink:* Whenever a macro UE is inside the femtocell coverage area or just close to a femtocell, the uplink signal from the macro UE to the macrocellular BS can cause interference to the FAP receiver. This situation can occur in the deployment scenarios *B, C,* and *D*. Fig. 4.2 shows an example where the macrocell uplink of macro UE-1 causes interference to FAP-2.

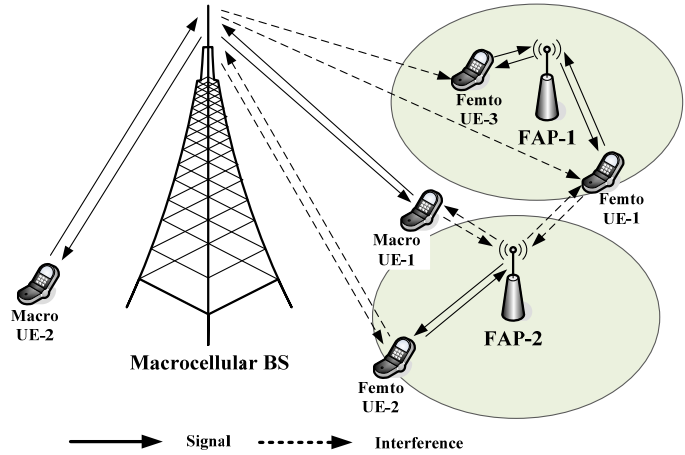

**Fig. 4.2:** Example of interference scenarios in integrated femtocell/macrocell networks.

- *Femtocell downlink:* In this case, the femtocell downlink causes interference to the macro UE receivers and to the nearby femto UE receivers. Whenever a macro user is inside or near a femtocell coverage area, the macro UE is subjected to interference from the femtocell downlink (deployment scenarios *B, C,* and *D*). Similarly, when two (or more) femtocells are in close proximity to each other, then the femto UE of one femtocell is affected by the interference from the neighbor femtocell downlink (scenarios *C* and *D*). Fig 4.2 depicts a situation where the downlink of the FAP-2 femtocell causes interference to macro UE-1 and femto UE-1.

- *Femtocell uplink:* When a femtocell is close to the macrocellular BS, the transmitted uplink signal from the femto UE causes interference to the macrocell receivers (deployment scenarios *B, C,* and *D*). Similarly, when two (or more) femtocells are close to each other, one femtocell uplink causes interference to the neighbor FAP receiver (scenarios *C* and *D*). Fig. 4.2 presents an example where the uplink from femto UE-2 causes interference to macrocellular BS receiver and to UE-1 femtocell uplink at FAP-2.

## 4.2 The Proposed Frequency Allocation Schemes

In this section, we use the parameters as defined in Table 4.1 below.

**Table 4.1:** Basic Nomenclature

| Symbol | Definition |
|---|---|
| $B_T$ | The total system-wide spectrum of frequencies (frequency band) allocated to for macrocells and femtocells |
| $B_m$ | The total frequency spectrum (frequency band) allocated to macrocells |
| $B_f$ | The total frequency spectrum (frequency band) allocated to femtocells |
| $B_{f1}, B_{f2},$ and $B_{f3}$ | The total frequency spectrum (frequency band) allocated to all the femtocells in the macrocells #1, #2, and #3, respectively, of a macrocell cluster |
| $B'_{f1}, B'_{f2},$ and $B'_{f3}$ | The actual frequency band allocation to a femtocells in the macrocell #1, #2, and #3, respectively, of a macrocellular cluster |
| $B_{fnc}$ | The frequency spectrum (frequency band) allocated to the center of a newly installed femtocell |
| $B_{fne}$ | The frequency spectrum (frequency band) allocated to the edge of a newly installed femtocell |
| $B_{foc(k)}$ | The frequency spectrum (frequency band) allocated to the center of a *k-th* overlapping interfering femtocell |
| $B_{foe(k)}$ | The frequency spectrum (frequency band) allocated to the edge of a *k-th* overlapping interfering femtocell |

The cellular spectrum is a quite limited and expensive resource, so spatial reuse of radio spectrum has been a well-known technique for cost reduction. Though, allocation of the same frequencies among neighboring femtocells and overlaid macrocells can potentially cause serious interference effects especially for the dense femtocells deployment case. However, a properly designed frequency allocation scheme can mitigate the interference effects, while improving the utilization of the frequency spectrum. The frequency allocation for the different femtocellular network deployment scenarios should differ to achieve better utilization of the spectrum. In this section, we propose possible efficient and cost effective frequency planning for different femtocellular network deployment scenarios. The total system-wide cellular frequency spectrum, the frequency spectrum allocated for macrocells, and the frequency spectrum allocated to femtocells, are denoted as $B_T$, $B_m$, and $B_f$ respectively.

### *4.2.1 The Dedicated Frequency Band Allocation Scheme*

The *dedicated frequency band* allocation is the case where the same frequency band is shared by all the femtocells and a different frequency band is allocated to the macrocells; i.e., in this scheme, femtocells and macrocells use totally separate frequency bands. In the example in Fig. 4.3, the femtocells use the frequency band from $f_1$ to $f_3$, whereas the frequency band allocation for the macrocells is $f_3$ to $f_2$. This scheme is not suitable to support dense femtocells deployment (scenarios *D*), as the use of the same frequencies by the densely located femtocells of scenario *D* would cause severe interference problem. On the other hand, use of this allocation scheme in the deployment scenarios *A* would cause significant inefficiency. Thus, this scheme could be used only for initial and relatively small-scale deployment of a femtocellular network. For this scheme, we can write:

$$\left.\begin{aligned} B_m &= f_2 - f_3 \\ B_f &= f_3 - f_1 \end{aligned}\right\} \tag{1}$$

$$\left.\begin{aligned} B_T &= B_m + B_f \\ B_m \cap B_f &= \emptyset \\ B_m \cup B_f &= B_T \end{aligned}\right\} \tag{2}$$

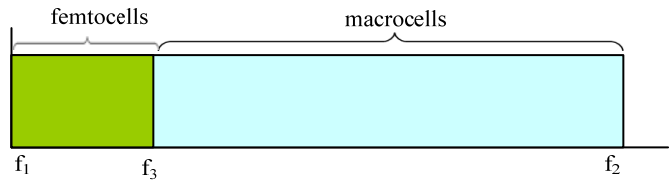

**Fig. 4.3:** Frequency allocation using the *dedicated frequency band* scheme.

### 4.2.2 The Shared Frequency Band Allocation Scheme

In the *shared frequency band* allocation scheme, the frequencies from the same spectrum can be allocated for the femtocells and the macrocells, as shown in Fig. 4.4. This scheme is very efficient for the deployment scenarios *A* resulting in the best utilization of the spectrum because in this deployment scenario there are no interfering overlapping femtocells and/or overlaying macrocell. This scheme can also be used for the deployment scenario *B* with a small numbers of discrete femtocells overlaid by a macrocell. However, this scheme is inappropriate for other deployment scenarios due to the amount of interference that could be present. For this scheme, we can write:

$$B_m = B_f = B_T = f_2 - f_1 \tag{3}$$

$$\left.\begin{aligned} B_m \cap B_f &= B_T \\ B_m \cup B_f &= B_T \end{aligned}\right\} \tag{4}$$

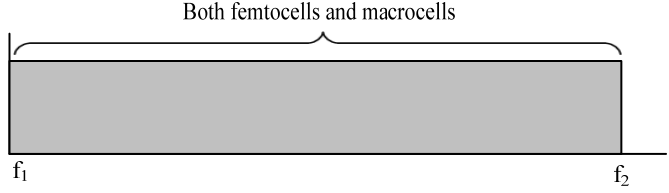

**Fig. 4.4:** Frequency allocation using the *shared frequency band* scheme

### *4.2.3 The Sub-Frequency Band Scheme*

In the *sub-frequency band* allocation scheme, the macrocells use the total system spectrum, while only part of this total frequency band can be used by the femtocells. This is depicted in Fig. 4.5, where the femtocells use frequency band $f_1$ to $f_4$, while the total frequency allocation for the macrocells is $f_1$ to $f_2$. This scheme cannot support dense femtocells deployment (e.g., deployment scenario *D*) and can also cause significant interference in deployment scenarios *C*. However, in the deployment scenario *B*, the amount of femtocell-to-femtocell interference is limited. This scheme is mostly useful for the deployment scenario *A*. For this scheme, we can write the following set of equations:

$$\left.\begin{array}{l} B_m = B_T = f_2 - f_1 \\ B_f = f_4 - f_1 \end{array}\right\} \tag{5}$$

$$\left.\begin{array}{l} B_m \cap B_f = B_f \\ B_m \cup B_f = B_T \end{array}\right\} \tag{6}$$

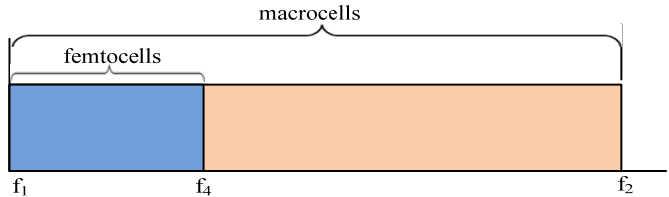

**Fig. 4.5:** Frequency allocation using *sub-frequency band* scheme.

### *4.2.4 Reuse Frequency among Femtocells and Macrocells*

In a large scale deployment of femtocellular networks, the implementation of frequency-reuse can be performed in a number of ways. Here, we present two alternatives for frequency reuse. In both the proposed schemes, we assume that the reuse factor in the macrocellular network is 3 (i.e., the number of macrocells in a macrocellular cluster).

### 4.2.4.1 Scheme 1 (the *static frequency-reuse* scheme)

In the *static frequency-reuse* scheme, the set of all cellular frequencies is divided into three equal bands: $B_{m1}$, $B_{m2}$, and $B_{m3}$ and each one of the three macrocells in a macrocellular cluster uses one of these three different frequency bands. If a macrocell uses a particular frequency band, then the femtocells within that macrocell use the other two frequency band. Figs. 4.6 and 4.7 show an example of the *static frequency-reuse* scheme. Macrocell #1 of the macrocell cluster uses frequency band $B_{m1}$ and each femtocell within this macrocell uses either $B_{m2}$ or $B_{m3}$. The use of two different frequency bands for neighboring femtocells reduces the femtocell-to-femtocell interference. The deployment scenarios *B* and *C* are quite effective for use with the *static frequency-reuse* scheme. While deployment scenario *D* could also be used with this scheme, but cannot mitigate interference well. SON-based FAPs can be optionally used for automatic assignment of frequencies to femtocells. For this scheme, we can write the following set of equations:

$$\left. \begin{array}{l} B_{m1} = B_{m2} = B_{m3} = \dfrac{B_m}{3} \\ B'_{f1} = a.B_{m2} + a'.B_{m3} \\ B'_{f2} = b.B_{m3} + b'.B_{m1} \\ B'_{f3} = c.B_{m1} + c'.B_{m2} \\ B_{f1} = B_{m2} + B_{m3} \\ B_{f2} = B_{m3} + B_{m1} \\ B_{f3} = B_{m1} + B_{m2} \end{array} \right\} \quad (7)$$

$$\left. \begin{array}{l} B_{m1} \cap B_{f1} = \emptyset, B_{m2} \cap B_{f2} = \emptyset, B_{m3} \cap B_{f3} = \emptyset \\ B_{m1} \cup B_{f1} = B_T, B_{m2} \cup B_{f2} = B_T, B_{m3} \cup B_{f3} = B_T \end{array} \right\}, \quad (8)$$

where *a, b,* and *c* are binary variables and $a'$, $b'$, and $c'$ are their respective complements. $B_{m1}$, $B_{m2}$, and $B_{m3}$ refer the frequency band allocations for macrocell #1, #2, and # of macrocell cluster, respectively. $B_{f1}$, $B_{f2}$, and $B_{f3}$ refer the total frequency band allocations for all the femtocells in the macrocells #1, #2, and #3, respectively, of a macrocellular cluster. $B'_{f1}$, $B'_{f2}$, and $B'_{f3}$ refer to a particular frequency band allocated to a femtocell in the macrocell #1, #2, and #3 of a macrocellular cluster, respectively.

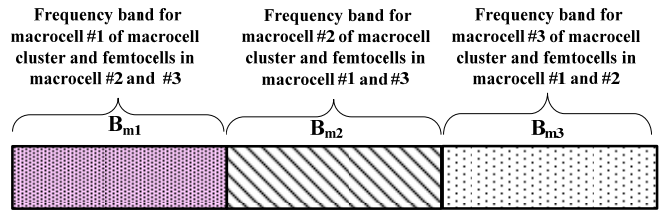

**Fig. 4.6:** Frequency allocation for macrocells using the *static frequency-reuse* scheme.

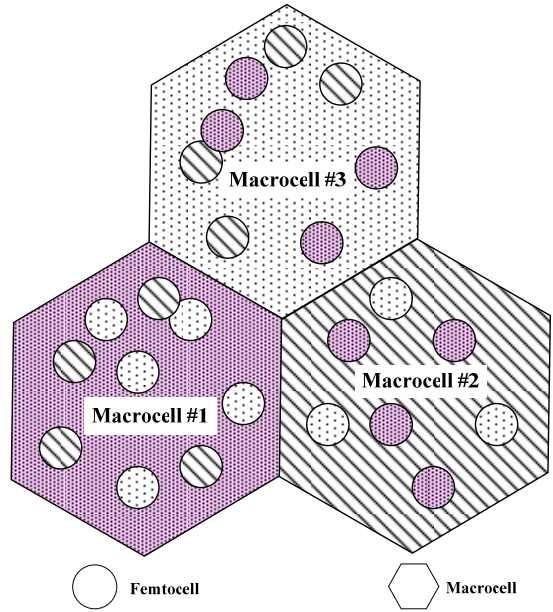

**Fig. 4.7:** An example of frequency allocation to macrocells of a macrocell cluster and to femtocells within the macrocells for the *static frequency-reuse* scheme.

**4.2.4.2 Scheme 2 (the *dynamic frequency-reuse* scheme)**

The total frequency allocation for the three macrocells in a macrocell cluster and the total frequency allocation for the femtocells in each of these three macrocells are similar to the *static frequency-reuse* scheme. However, as opposed to the *static frequency-reuse* scheme where each femtocell uses only one of the two frequency bands which is not assigned to its underlying macrocell, in the *dynamic frequency-reuse* scheme, each femtocell uses two frequency bands. In each femtocells, one band is used in the center of the femtocell, while the other band is used at the edge of the femtocell. The frequency band used in the center of all the femtocells of the same macrocell is the same. However, the frequency bands used in the edges of the various femtocells are, in general, different, to avoid the interference. Figs. 4.8 and 4.9 show an example of the frequency allocation for the *dynamic frequency-reuse* scheme. For clarity, only the femtocells of one macrocell (macrocell #1) are shown in the example. The overlapping of femtocells and the interfering signals from other femtocells in this proposed *dynamic frequency-reuse* scheme is mitigated through the use of the different bands at the edges of the femtocells. For the *dynamic frequency-reuse* scheme, we can write the following set of equations:

$$\left.\begin{aligned}
B_{m1} = B_{m2} = B_{m3} &= \frac{B_m}{3} \\
B_1 = B_2 = B_3 &= \frac{B_{m3}}{3} \\
B_4 = B_5 &= \frac{B_{m3}}{2} \\
B_{f1} &= B_{m2} + B_{m3} \\
B_{f2} &= B_{m3} + B_{m1} \\
B_{f3} &= B_{m1} + B_{m2}
\end{aligned}\right\} \quad (9)$$

$$B_{f1}' = \begin{cases} B_{m2} + a.B_1 + b.B_2 + c.B_3 & \text{for 3 interfering femtocells} \\ B_{m2} + x.B_4 + x'.B_5 & \text{for 2 interfering femtocells} \\ B_{m2} + B_{m3} & \text{for no interfering femtocell} \end{cases} \quad (10)$$

$$\left.\begin{aligned} B_{m1} \cap B_{f1} = \emptyset,\ B_{m2} \cap B_{f2} = \emptyset,\ B_{m3} \cap B_{f3} = \emptyset \\ B_{m1} \cup B_{f1} = B_T,\ B_{m2} \cup B_{f2} = B_T,\ B_{m3} \cup B_{f3} = B_T \end{aligned}\right\}, \quad (11)$$

where $a, b, c$ are the binary parameters and $x'$ is the complement of $x$. The value of $a$, $b$, and $c$ depend on the neighboring femtocell's edge frequency band. $B_{m1}$, $B_{m2}$, and $B_{m3}$ refers the frequency band allocation for macrocell #1, #2, and #3, respectively, of the macrocell cluster. $B_{f1}$, $B_{f2}$, and $B_{f3}$ refers the total frequency band allocation for all the femtocells in macrocell #1, #2, and #3, respectively, of the macrocell cluster. $B_{f1}'$ refers to the actual frequency band allocation of a femtocells in macrocell #1 (of the macrocell cluster).

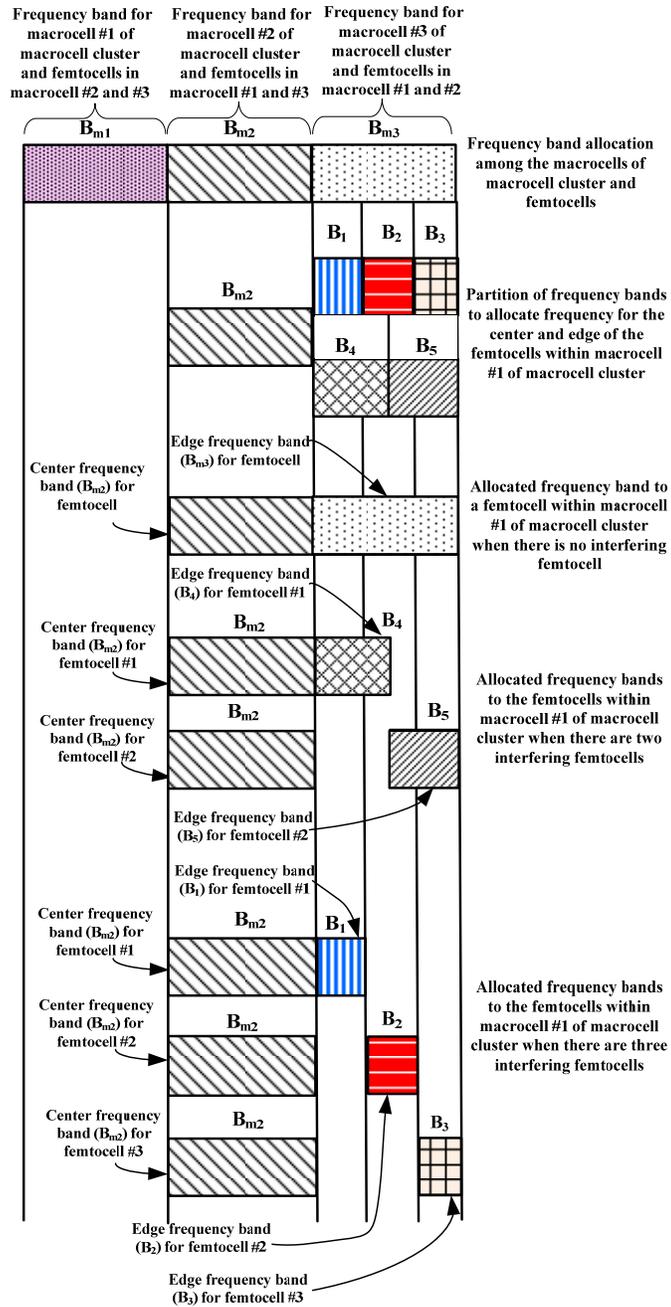

**Fig. 4.8:** The division of frequency spectrum for the *dynamic frequency-reuse* scheme.

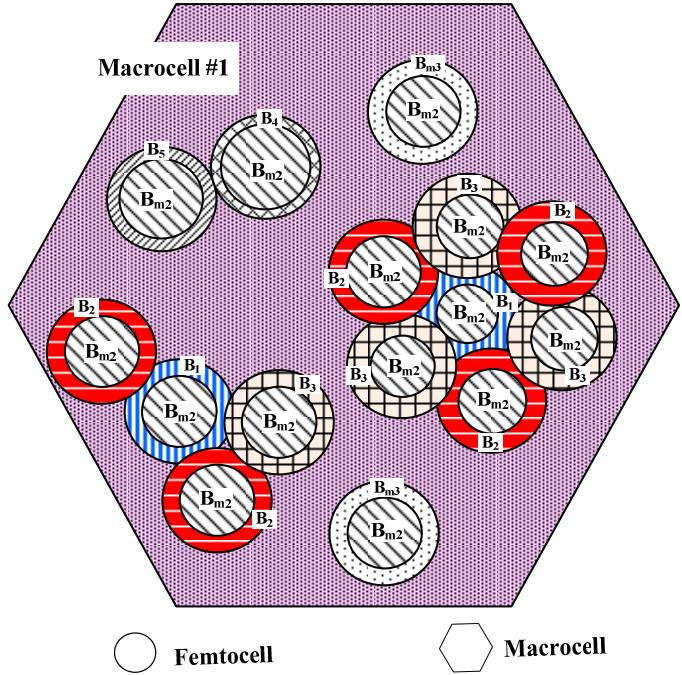

**Fig. 4.9:** Frequency allocation in the macrocell #1 and the femtocells within this macrocell using the *dynamic frequency-reuse* scheme.

The radius of the inside circle of a femtocell can vary based on the number of neighbor femtocells and their relative distance. Deployment scenarios *C* and *D* can be quite effectively used with this scheme. Although deployment scenario *B* could also be used with this scheme, however, it would be inefficient. The use of SON-based network feature is essential for the *dynamic frequency-reuse* scheme; e.g., using the SON functionalities, the transmitted power can be automatically adjusted and the edge frequencies can be automatically assigned [23]. The frequency allocation among macrocells in a macrocellular cluster can follow similar procedure. The algorithm to configure the frequency for a newly installed femtocell (consider as an example only one macrocell within a macrocellular cluster of size 3) is proposed below.

**Algorithm to select the frequency band for a newly installed femtocell:**

```
1:   while the detected frequency band of overlaid macrocell = B_{m1}
2:      the total frequency band allocation for all femtocells in the macrocell = B_{m2} + B_{m3}
3:      if a newly installed femtocell does not detect any interfering femtocell then
4:         B_{fnc} = B_{m2};
5:         B_{fne} = B_{m3};
6:      end if
7:      if a newly installed femtocell detects one interfering femtocell then
8:         B_{fnc} = B_{m2};
9:            if B_{foe(1)} = B_5 then
10:              B_{fne} = B_4;
11:           else if B_{foe(1)} = B_4 then
12:              B_{fne} = B_5;
13:           else if B_{foe(1)} = B_1 then
14:              B_{fne} = B_2;
15:           else if B_{foe(1)} = B_2 then
16:              B_{fne} = B_3;
17:           else if B_{foe(1)} = B_3 then
18:              B_{fne} = B_1;
19:           end if
20:     end if
21:     if a newly installed femtocell detects two common interfering femtocells then
22:        B_{fnc} = B_{m2};
23:           if B_{foe(1)} = B_4 and B_{foe(2)} = B_5 then
24:              B_{fne} = B_3;
25:              B_{foe(1)} = B_1 ;
26:              B_{foe(2)} = B_2 ;
27:           else if B_{foe(1)} = B_1 and B_{foe(2)} = B_2 then
28:              B_{fne} = B_3;
29:           else if B_{foe(1)} = B_2 and B_{foe(2)} = B_3 then
30:              B_{fne} = B_1;
31:           else if B_{foe(1)} = B_3 and B_{foe(2)} = B_1 then
32:              B_{fne} = B_2;
33:           end if
34:     end if
```

Whenever a FAP is removed from a femtocellular network, the existing femtocells re-configure their frequencies in a similar way to the case when a new femtocell is installed. In the proposed scheme, maximum three overlapped femtocells are considered. In a practical case where more than three femtocells overlap, then the size of the femtocells will need to be reduced or re-adjusted automatically to reduce the inter-femtocell interference effect. In a very dense femtocells area, if the best frequency reconfiguration is not possible using the available combination of frequency bands, then the femtocell size will need to be reduced or re-adjusted automatically to mitigate the inter-femtocell interference effect. Thus, the SON-based network architecture is essential for the *dynamic frequency-reuse* scheme in the femtocellular network environment.

As mentioned before, the SON architecture is required to support the *dynamic frequency-reuse* scheme. The basic features, of an integrated SON femtocell/macrocell architecture to support the *dynamic frequency-reuse* scheme are as follows:

- ***Case 1 (Frequency configuration among neighboring femtocells):*** When a FAP is newly installed, it configures its center and edge frequencies according to the detected frequencies of the neighboring femtocells. The FAPs of the entire neighborhood coordinate this frequency allocation.
- ***Case 2 (Cell size re-adjustment):*** If a number of femtocells interfere with each other even after the frequency configuration procedure was performed, the affected FAPs coordinate among themselves to re-adjust their center and edge areas, as to reduce the interference effect.
- ***Case 3 (Frequency configuration between a newly installed FAP and macrocellular BS):*** When a FAP is newly installed in a macrocellular coverage area, the FAP and the respective macrocellular BS communicate with each other to configure the frequencies of the newly installed FAP.

In addition to frequency allocation schemes, there are other techniques that are also used to mitigate interference in a SON-based femtocellular network. Power optimization technique can be quite effective in areas of dense femtocell deployment and for femtocells located on the fringe of a macrocell. Based on the presence of femto users and their operating modes, the FAP can change its mode of operation to mitigate the interference among the neighboring femtocells. For instance, the users on the edge of a macrocell typically transmit higher power that causes interference to the neighboring FAPs. A FAP may increase its cell size to accept such macro users into its femtocell. Thus by adjusting the cell size, the interference to the neighboring femtocells can be reduced. As another example, power control of the UE is needed when the distance between a FAP and the respective macrocellular BS is short.

## 4.3 Outage Probability Analyses

For the analysis in this section, we use the parameters as defined in Table 4.2.

**Table 4.2:** Nomenclature for Outage Probability Analysis

| Symbol | Definition |
|---|---|
| $S_o$ | The received signal from the associated (reference) FAP |
| $I_{m(j)}$ | The received interference from the *j-th* interfering macrocellular BS |
| $K$ | The total number of neighboring femtocells |
| $N$ | The total number of interfering macrocells |
| $I_{n(i)}$ | The received interference from the *i-th* neighboring femtocell |
| $I_f$ | The total received interference from all the neighboring femtocells |
| $P_T$ | The transmitted signal power |
| $P_R$ | The signal power received at the receiver |
| $d$ | The distance between the transmitter and the receiver |
| $\xi$ | The slow-fading random variable |
| $Z$ | The fast-fading random variable |
| $\eta$ | The path loss exponent |

There are various interference mechanisms present in the macrocell/femtocell integrated network architecture; in particular, between macrocells and femtocells, and among femtocells. Inadequate interference management system reduces system capacity for both the macrocellular and the femtocellular networks. More specifically, interference reduces users' QoE, and causes higher outage probability. On the other hand, appropriate interference management can increase the capacities of both, the femtocellular and the macrocellular networks.

Assuming that the spectrums of the transmitted signals are spread, we can approximate the interference as AWGN. Then, following the Shannon Capacity Formula, $C = W \log_2(1 + SNR)$ [bits/sec], we can state that the capacity of a wireless channel decreases with decreased signal-to-interference (*SIR*) level. The received *SIR* level of a femtocell user in a macrocell/femtocell integrated network can be expressed as:

$$SIR = \frac{S_o}{\sum_{j=0}^{N-1} I_{m(j)} + \sum_{i=1}^{K} I_{n(i)}}, \tag{12}$$

where $S_o$ is the power of the signal from the associated (reference) femtocell, $I_{m(j)}$ is the power of the interference signal from the *j-th* interfering macrocell from among the *N* interfering macrocells, and $I_{n(i)}$ is the received interference signal from the *i-th* femtocell from among the *K* neighboring femtocells. The indices *0, i,* and *j* refer to the reference femtocell, the *i-th* neighboring femtocell, and the *j-th* interfering macrocell, respectively.

The outage probability of a femtocell user can be calculated as:

$$P_{out} = P_r(SIR < \gamma), \tag{13}$$

where $\gamma$ is a threshold value of *SIR* below which there is no acceptable reception. Alternatively, equation (13) can be rewritten as:

$$\begin{aligned} P_{out} &= P_r\left(\frac{S_o}{\sum_{j=0}^{N-1} I_{m(j)} + \sum_{i=1}^{K} I_{n(i)}} < \gamma\right) \\ &= P_r\left\{\sum_{i=1}^{K} I_{n(i)} > \left(\frac{S_o}{\gamma} - \sum_{j=0}^{N-1} I_{m(j)}\right)\right\} \\ &= P_r\left\{I_f > \left(\frac{S_o}{\gamma} - I_m\right)\right\}, \end{aligned} \tag{14}$$

where $I_f$ and $I_m$ are the total received interference from the neighbor femtocells and macrocells, respectively. We assume that the probability density function (PDF) of $I_f$ is Gaussian [26].

As per equation (14), to reduce the probability of outage, techniques to mitigate the interference from the interfering macrocells and the neighboring femtocells need to be employed. The choice of such a technique varies according to the macrocell coverage area and the density of the femtocells. For example, in remote areas, where macrocell coverage is sparse or not available at all, or when the macrocell and the femtocells use non-overlapping frequency bands, the interference from macrocells can be assumed negligible and $\forall_j, I_{m(j)} \approx 0$. In a dense femtocellular deployment, the interference form neighboring femtocells is of main concern and proper inter-femtocell interference management can suffice to increase the signal-to-interference ratio and to reduce the outage probability.

The received signal power $P_R$ from a transmitter located at distance *d* from the receiver, when the transmitted signal is of power $P_T$ is:

$$P_R = P_T P_0 \, d^{-\eta} \xi Z, \tag{15}$$

where $P_0$ is a function of the carrier frequency, the antenna height, and the antenna gain; $\xi$ is a random variable that accounts for slow-fading (the so-called "shadowing") of the radio channel; $Z$ is a random variable that represents the effect of fast fading (i.e., multi-path effect) of the communication channel; and, finally, $\eta$ is the path loss exponent. The distribution of $\xi$ is typically assumed to be log-normal [27]. We assume that the envelope of a fast-faded channel is Rayleigh [27] and, therefore, the distribution of the power attenuation factor, Z, due to Rayleigh fading is modeled as exponential [28].

When the femtocell user receives signal from its FAP that is situated indoors, then typically the shadowing is negligible and the slow-fading can be neglected. Thus, using equation (15), the received signal by a femtocell user from its associated FAP, $S_o$, can be expressed by:

$$S_o = P_{Rf(0)} = P_{Tf(0)} P_{0f} d_0^{-\eta_1} Z_0 = \bar{S} Z_0 \tag{16}$$

while the interference $I_{n(i)}$ and $I_{m(j)}$ from the *i-th* neighbor femtocells and *j-th* interfering macrocell, respectively, can be expressed as:

$$I_{n(i)} = P_{Rf(i)} = P_{Tf(i)} P_{0f} d_i^{-\eta_2} \xi_i Z_i \tag{17}$$

$$I_{m(j)} = P_{Rm(j)} = P_{Tm} P_{0m} d_j^{-\eta_3} \xi_{m(j)} Z_{m(j)} , \tag{18}$$

where $i = 0$ and $j = 0$ denote the reference (associated) femtocell and the reference (underlying) macrocell, respectively. $P_{Rm(j)}$ and $P_{Rf(i)}$ stand for the received power at UE from the *j-th* macrocell and the *i-th* femtocell, respectively. $P_{Tm}$ and $P_{Tf(i)}$ represent the transmitted power from each of the macrocells and the *i-th* femtocell respectively. $d_j$ and $d_i$ are the distances from the reference UE to the *j-th* macrocellular BS and to the *i-th* FAP respectively.

We denote by $I_f$ and $I_m$ the power of the total interference from the *K* femtocell neighbors and the power of the total interference from the *N* interfering macrocells BSs, respectively. Using equations (17) and (18), $I_f$ and $I_m$ are derived as:

$$I_f = \sum_{i=1}^{K} I_{n(i)} = \sum_{i=1}^{K} P_{Tf(i)} P_{0f} d_i^{-\eta_2} \xi_i Z_i X_i \tag{19}$$

$$I_m = \sum_{j=0}^{N-1} I_{m(j)} = \sum_{j=0}^{N-1} P_{Tm} P_{0m} d_j^{-\eta_3} \xi_{m(j)} Z_{m(j)} Y_j \tag{20}$$

where $X_i$ and $Y_j$ are binary indication functions which take the value of 1 when the reference femtocell and the $i$-th neighboring femtocell ($X_i$) or the $j$-th macrocell ($Y_j$) use same frequency. Otherwise, $X_i = 0$ or $Y_j = 0$.

Using equations (14) and (16), $P_{out}$ can be written as:

$$P_{out} = P_r\left\{Z_0 < \frac{\gamma}{S}(I_f + I_m)\right\}. \tag{21}$$

The PDF of $Z_0$ is exponentially distributed. Thus, the solution of equation (21) can be computed as:

$$\begin{aligned}
P_{out} &= \int_0^{\frac{\gamma}{S}(I_f+I_m)} \exp(-Z_0)\, dZ_0 \\
&= 1 - \exp\left\{-\frac{\gamma}{S}(I_f + I_m)\right\} \\
&= 1 - \exp\left[-\frac{\gamma}{S}\left\{\sum_{i=1}^{K} X_i I_{n(i)} + \sum_{j=0}^{N-1} Y_j I_{m(j)}\right\}\right] \\
&= 1 - \left\{\prod_{i=1}^{K} e^{\left\{-\frac{\gamma}{S} I_{n(i)} X_i\right\}}\right\}\left\{\prod_{j=0}^{N-1} e^{\left\{-\frac{\gamma}{S} I_{m(j)} Y_j\right\}}\right\}.
\end{aligned} \tag{22}$$

We used equations (16) − (21) in equation (22) to calculate $P_{out}$. In the equation (22), $Y_j = 1$ if the *j-th* macrocell and the reference femtocell are allocated the same frequency, otherwise $Y_j = 0$ (in other words, $B_m \cap B_f = \emptyset$). Thus, for the *shared frequency band* and the *sub-frequency band* schemes, $Y_j = 1$. For the *frequency-reuse* schemes introduced in this paper, $Y_0 = 0$ and for some macrocells in the first and the higher tiers $Y_j = 1$ and for the others $Y_j = 0$. However, the macrocells in the second and in the higher tiers cause only negligible interference. The value of $X_i$ is related to the allocated frequency for the reference femtocell and the *i-th* neighbor femtocell. $X_i = 1$ for the *shared frequency band*, the *dedicated frequency band*, and the *sub-frequency band* schemes, where all the $K$ neighboring femtocells use the same frequency as the reference femtocell, For the *static frequency-reuse* scheme, $X_i = 1$ for almost 50% of the neighboring femtocells and $X_i = 0$ for the remaining neighboring femtocells. Similarly, for the *dynamic frequency-reuse* scheme, $X_i = 0$ for more than 66% of the neighboring femtocells. Thus, the outage probability for the *dynamic frequency- reuse* scheme is lower than other schemes.

## 4.4 Numerical Results

In this section, we evaluate the throughput and the outage probability of the proposed *static* and *dynamic frequency- reuse* schemes and compare with the *shared frequency band* scheme and the *dedicated frequency band* scheme for various femtocells densities. In our evaluation, we define as dense femtocells deployment (scenario *D*) as more than 1000 femtocells within a macrocell; otherwise, we consider it as non-dense femtocellular deployment (scenarios *B* and *C*). Table 4.3 summarizes the values of the parameters that we used in our numerical analysis. For simplicity, we used the macrocell propagation model from [29] and the femtocell propagation model from [30]. We neglect the effects of the macrocells in the second and the higher tiers, as the contributed interference is minimal. Only the reference macrocell and the other six macrocells in the first tier are considered. Furthermore, for the purpose of calculation of inter-femtocell interference, we assume that there exists one wall between two femtocells. We consider two femtocells as neighbors, if their FAPs are within 60 meter of each other. The femtocells are placed randomly within the macrocell coverage area and the number of femtocells within the neighbor area is randomly generated according to the Poisson distribution. For the *dedicated frequency band* scheme, we assume 33.3% of total cellular frequency band is allocated to femtocells and the remaining 66.7% of the total frequency band is allocated to the macrocells. For capacity analysis we use the Shannon Capacity Formula.

**Table 4.3:** Summary of the parameter values used in our analysis

| Parameter | Value |
|---|---|
| Macrocell radius | 1 [km] |
| Femtocell radius | 10 [m] |
| Distance between the reference macrocellular BS and the reference FAP | 200 [m] |
| Carrier frequency | 900 [MHz] |
| Transmit signal power by the macrocellular BS | 1.5 [kW] |
| Maximum transmitted signal power by a FAP | 10 [mW] |
| Height of a macrocellular BS | 50 [m] |
| Height of a FAP | 2 [m] |
| Threshold value of SIR ($\gamma$) | 9 [dB] |

Fig. 4.10 demonstrates that the proposed *static frequency-reuse* scheme provides better throughput than the *dedicated frequency band* and the *shared frequency band* schemes for non-dense femtocellular networks deployment. The throughput of the *dynamic frequency-reuse* scheme and of the *static frequency-reuse* scheme are almost same. However, *dynamic frequency-reuse* scheme is inferior for non-dense femtocells deployment, because of the increased implementation cost of the SON features that are needed for the *dynamic frequency-reuse* scheme. Fig. 4.11 shows that the *static frequency-reuse*-scheme also reduces the outage probability within the considered range of parameters. The outage probability of the *dynamic frequency-reuse* scheme is almost same as that of the *static frequency-reuse* scheme for non-dense femtocellular network deployment. Thus, the *static frequency-reuse* scheme is recommended for the non-dense femtocellular network deployment.

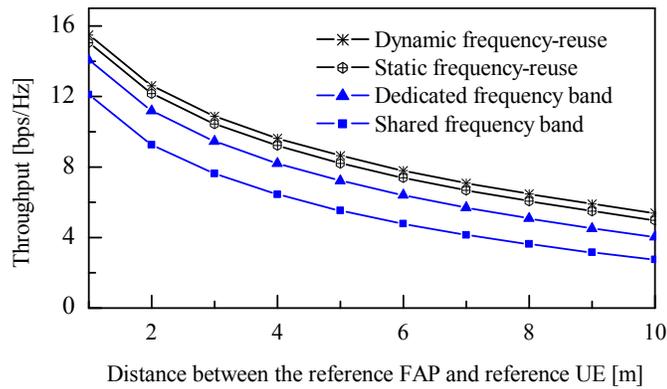

**Fig. 4.10:** Throughput comparison of non-dense femtocellular network deployment (scenarios *B* and *C*)

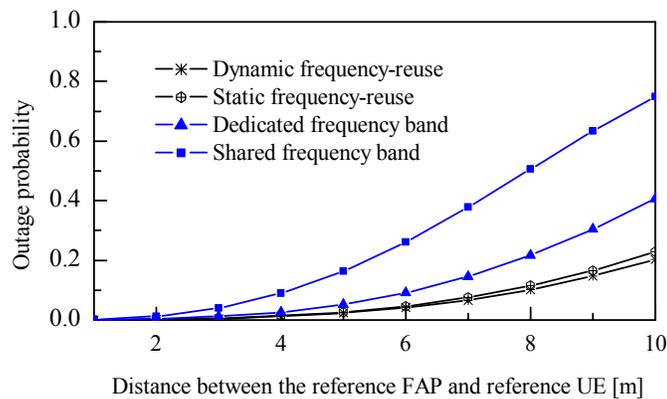

**Fig. 4.11:** Outage probability comparison of non-dense femtocellular network deployment (scenarios *B* and *C*).

Fig. 4.12 depicts the throughput performance of the *dynamic frequency-reuse* scheme in dense femtocells environment. The throughput of the *dynamic frequency-reuse* scheme is larger than that of the other schemes for dense femtocellular deployment. However, the throughput quickly degrades to small values for the *shared frequency band* scheme and for the *dedicated frequency band* scheme. Fig. 4.13 illustrates the fact that that the outage probability of the *dynamic frequency-reuse* scheme is significantly smaller compared with the other schemes even for highly dense femtocellular network deployment. The results in Figs. 4.12 and 4.13 were obtained when the distance between the femto UE and the reference FAP was maintained at 5 meters.

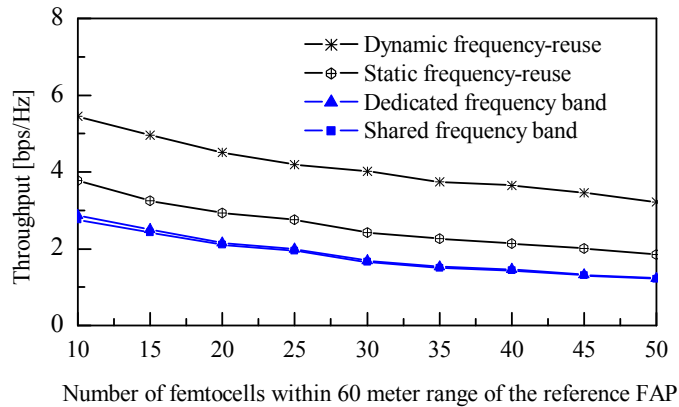

**Fig. 4.12:** Throughput comparison of the dense femtocells scenario

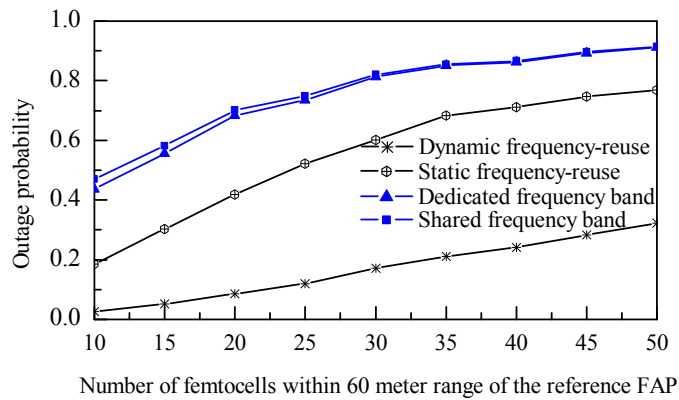

**Fig. 4.13:** Outage probability of the dense femtocells scenario.

*4.4.1 Summary of Numerical Results*

The *dynamic frequency-reuse* scheme outperforms the other schemes investigated in this paper for all the femtocellular network environments in terms of throughput, outage probability, and spectral use. However, due to the overhead associated with the implementation of the SON features, the *dynamic frequency-reuse* scheme is not appropriate for small-scale femtocellular network deployment. On the other hand, SON-based network architecture is not essential for the *static frequency-reuse* scheme. Therefore, since in the non-dense femtocellular network deployment the performance of the *static frequency-reuse* scheme and the *dynamic frequency-reuse* scheme are essentially equal, the *static frequency-reuse* scheme is preferable for the non-dense deployment cases (scenarios *B* and *C*).

For the scenario *A*, the *shared frequency band* should be used to increase the spectral efficiency. We recommend that for the dense femtocellular network deployment (scenario *D*) the use of *dynamic frequency-reuse* scheme should be adopted. Finally, we point out that if the implementation of the SON features in the femtocellular network architecture is not expensive or if cost is not a major factor (as may be the case in some military installations), then the *dynamic frequency-reuse* scheme is, indeed, the best choice for all types of femtocellular network deployments.

# Chapter 5
# Mobility Management for Highly Dense Femtocellular Networks

Dense femtocells are the ultimate goal of the femtocellular network deployment. Dense deployment of femtocells will offload huge traffic from the macrocellular network to femtocellular network by the successful integration of macrocellular and femtocellular networks. Efficient handling of handover calls is the key for the successful macrocell/femtocell integration. For the efficient handling of handover calls and to effectively transfer the macrocell traffic to the femtocellular network, an intelligent integrated macrocell/femtocell network architecture, neighbor cell list with minimum number of femtocells, and handover process with proper signaling are the important elements. An appropriate traffic model for the integrated macrocell/femtocell network is also important for the performance analysis measurement.

The femtocells are deployed under the macrocellular network coverage or they are deployed in a separate non-macrocellular coverage area. In the overlaid macrocell coverage area, femtocell-to-femtocell, femtocell-to-macrocell, and macrocell-to-femtocell handovers are occurred due to the deployment of femtocells. The frequency of these handovers increases as the density of femtocells is increased. Thus, effective handover mechanisms are essential to support these handovers. The efficient femtocell-to-femtocell and femtocell-to-macrocell handovers result in seamlessly movement of the femtocell users. Even though the macrocell-to-femtocell handover is not essential

## 5.1  Neighbor Femtocell List

Finding the neighboring FAPs and determining the appropriate FAP for the handover are challenging for optimum handover decision [5]. Macrocell-to-femtocell and femtocell-to-femtocell handovers in dense femtocellular network environment suffers from some additional challenges because of dense neighbor femtocells. In these handovers, MS needs to select the appropriate target FAP among many neighbor FAPs. These handovers create significant problem if there is no minimum number of femtocells in the neighbor femtocell list. The MSs use much more power consumption for scanning many FAPs, and the MAC overhead becomes significant. This increased size of neighbor FAP list message and broadcasting of large information occurs too much overhead. So, an appropriate and optimal neighbor FAP list is essential for the dense femtocellular network deployment.

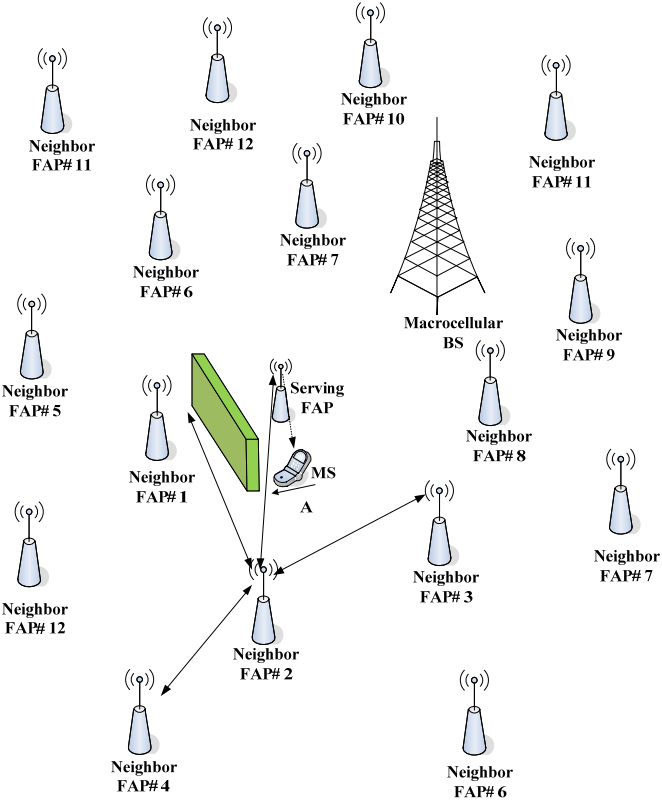

**Fig. 5.1:** A scenario of dense femtocellular network deployment where several hidden FAPs and other FAPs are situated as a neighbor femtocell.

The FAPs and the macrocellular BS coordinate with each other to facilitate a smooth handover. If large number of FAPs are deployed in an indoor building or femto zone area, signals from different FAPs will interfere with each other. Thus during the handover phase it is quite difficult to sense the actual FAP for which the user is going to be handed over. The need of reduced size of neighbor femtocell list is essential to make minimum numbers of scanning and signal flowing during the handover. Large neighbor femtocell list causes many unnecessary scanning for the handover. Also missing of some hidden femtocells in the neighbor femtocell list causes the failure of handover. Our main goal is to build such a neighbor femtocell list for the femtocell-to-femtocell and macrocell-to-femtocell handovers so that the list contains minimum number of femtocells as well as the list consider all the hidden femtocells. Fig. 5.1 shows a scenario of dense femtocellular network deployment where several FAPs are situated as a neighbor femtocell. In this Fig., for the MS at position "A" shown here, the optimized neighbor femtocell list must contain FAP# 1, FAP# 2, FAP# 3, and FAP# 8. However, due to a wall and other obstacle between the MS and the FAP# 1, the MS cannot receive sufficient signal level from this FAP. Serving FAP and FAP# 1 also cannot coordinate each other. Thus a neighbor femtocell list only based on the RSSI (received signal strength indicator) measurement cannot include FAP# 1 in the neighbor femtocell list. In this situation, using the SON features, FAP# 2 and FAP# 1 coordinate each other. The FAP# 2 gives the location information of FAP# 1 to the serving FAP. Thus, getting this location information, the neighbor femtocell list includes FAP# 1.

Figs. 5.2 and 5.3 show the flow mechanism for the design of the optimal neighbor femtocells list. $N_f$ and $N_c$, respectively, represents the total number of femtocells and cells included in the neighbor cell list. Our proposed scheme considers initially the received RSSI level. For the dense femtocellular network deployment, the frequency for each FAP is allocated based on the neighbor overlapping femtocells. Thus, the overlapping two femtocells do not use same frequency [6]. Only same frequency is used by apart femtocells. Thus, we deduct those FAPs from the neighbor femtocell list those use same frequency as the serving FAP. Finally we added the hidden femtocells in the neighbor femtocell list using the location information coordinated among neighbor FAPs or among neighbor FAPs and macrocellular BS.

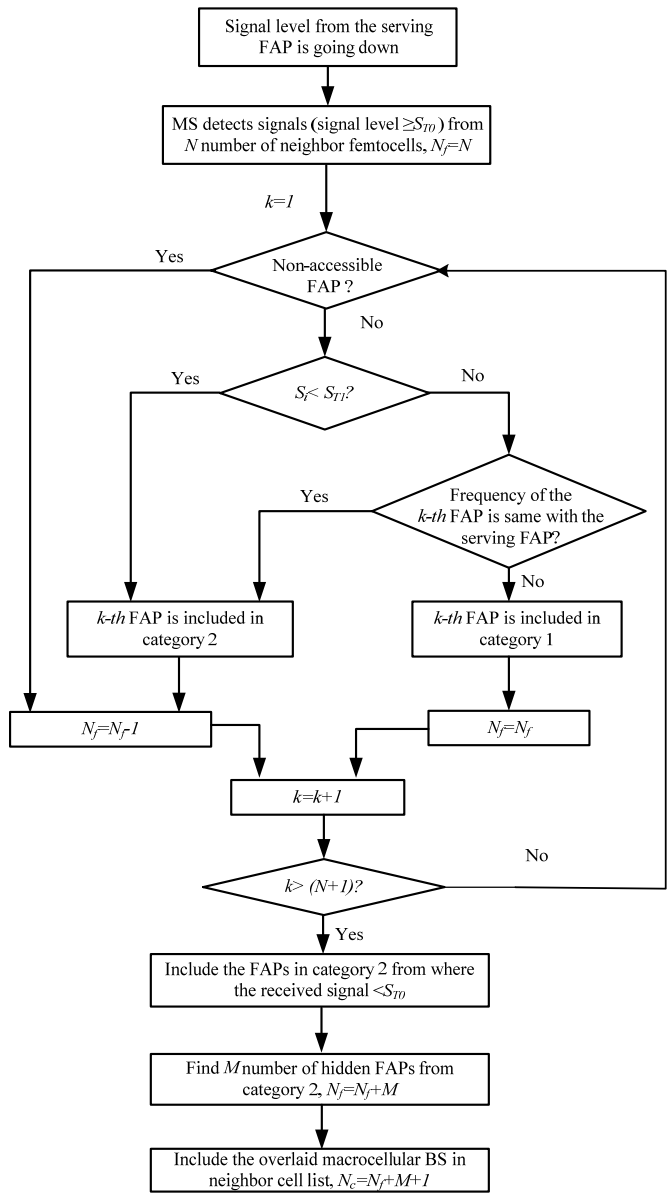

**Fig. 5.2:** The flow mechanism for the design of the optimal neighbor cell list for the handover when the MS is connected with a FAP.

Fig. 5.2 describes the flow mechanism for the design of the optimal neighbor cell list for the handover when the MS is connected with a FAP. Fig. 5.3 describes the flow mechanism for the design of the optimal neighbor cell list for the handover when the MS is connected with the overlaid macrocellular network..

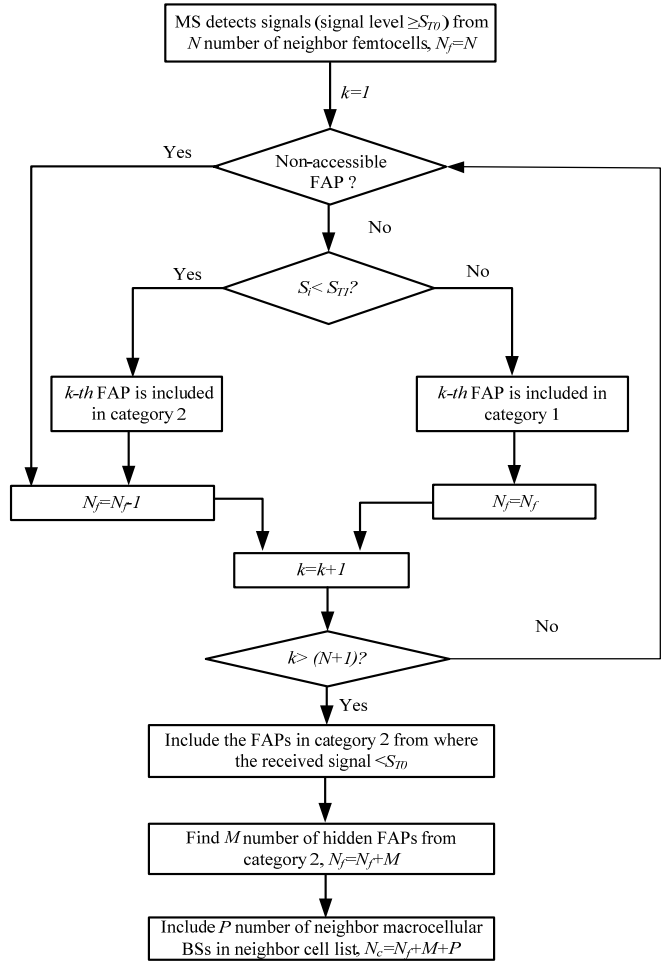

**Fig. 5.3:** The flow mechanism for the design of the optimal neighbor cell list for the handover when the MS is connected with the overlaid macrocellular BS.

We use two threshold levels of signal to design the flow mechanisms. The first threshold signal level, $S_{T0}$ is the minimum level of RSSI that must be needed to connect a FAP. The second signal level, $S_{T1}$ which is higher than $S_{T0}$. After checking the open/close access [31] system, the *i-th* FAP will be directly added in the neighbor cell list if the received signal, $S_i$, from the *i-th* FAP is greater than or equal to a the second threshold $S_{T1}$. All the $N$ number of FAPs from where the MS receives signals are initially considered to build the neighbor cell list. Then for the close access case, all the non-accessible FAPs are deducted. The frequency allocations are considered to find out the nearest FAPs for the possible handover. The coordination among the neighbor FAPs as well as among the FAPs and macrocellular BS are performed to find out the hidden FAPs. Hidden FAPs are those from where the received signals are less than the second signal level, $S_{T1}$, but these FAPs are very close to the serving FAP. Even though these FAPS are very near to the MS, it receives low level of signals from these FAPs due to some obstacles between the MS and the FAPs. Thus the addition of these hidden FAPs in the neighbor cell list reduces the chance that the MS fail to perfectly handover to the target FAP

The FAPs that are initially listed as the neighbor femtocell list based on the received RSSI level can expressed as a set $A$:

$$A = \{...FAP\#i(RSSI_i),... : 1 \leq i,\ RSSI_i \geq S_{T0})  \qquad (1)$$

where FAP#i(RSSI$_i$) represents that *i-th* neighbor FAP from which the received RSSI level at the MS is greater than or equal to $S_{T0}$. $S_{T0}$ is the minimum level of received signal from a FAP that can be detected by a MS.

The number of FAPs listed based on the minimum level of received signal level, $S_{T0}$ can be calculated as:

$$N = |\{...FAP\#i(RSSI_i),... : 1 \leq i,\ RSSI_i \geq S_{T0}\}| \qquad (2)$$

Instead of considering only the RSSI level, we consider RSSI level; frequency used by the serving FAP and *i-th* neighbor FAP; and the location information to construct an appropriate neighbor femtocell list.

We consider little higher RSSI level $S_{T1}$, compared to $S_{T0}$, to select better signal quality FAPs. Some hidden femtocells are picked for the neighbor femtocell list from where the received signal levels are less than $S_{T1}$. The FAPs considering the RSSI level of $S_{T1}$, the neighbor femtocell list can be expressed as:

$$B = \{...FAP\#j(RSSI_j),... : 1 \leq j,\ RSSI_j \geq S_{T1}) \qquad (3)$$

The number of FAPs listed based on the minimum level of received signal level, $S_{T1}$ can be calculated as:

$$N_1 = |\{...FAP\#j(RSSI_j),... : 1 \leq j,\ RSSI_j \geq S_{T1}\}| \qquad (4)$$

In the dense femtocell deployment, same frequency is not used for the overlapped femtocells [5], [6]. Therefore same frequencies are used by two femtocells those are little far away. So from the neighbor femtocell list we can deduct those femtocells which use same frequency as the serving femtocells. The femtocells those can be categorized in this group are:

$$C = \{...FAP\#k(f_k),... : 1 \leq k,\ C \in B, f_s \cup f_i = f_s\} \quad (5)$$

$$N_2 = |\{...FAP\#k(f_k),... : 1 \leq k, C \in B, f_s \cup f_i = f_s\}| \quad (6)$$

where $FAP\#k(f_k)$ represents that *k-th* neighbor femtocell that use frequency $f_k$. Whereas $f_s$ is the frequency used by the serving femtocell. $N_2$ indicates the number of femtocells in this group.

Now, we apply the location information for the neighbor femtocell list to include the hidden FAPs in the neighbor femtocell list. The hidden femtocells are chosen from the category 2 femtocells. The femtocells included in this category are (a) the femtocells from where the received RSSI level is less than $S_{T1}$, or (b) which femtocells use the same frequency as the serving femtocell. As the serving FAP can coordinate with some nearest FAPs, [6], [32] the nearest FAPs can inform the location of some hidden FAPs. Thus the hidden FAPs within a range of distance can be included in the neighbor femtocell list. The femtocells those are included in this group can be expressed as:

$$D = \{...FAP\#m(RSSI_m, f_m, d),... :\ 1 \leq m, (RSSI_m < S_{T1}\ or\ f_s \cup f_m = f_s)\ and\ d \leq d_{max}\} \quad (7)$$

$$M = |\{...FAP\#m(RSSI_m, f_m, d),... :\ 1 \leq m, (RSSI_m < S_{T1}\ or\ f_s \cup f_m = f_s)\ and\ d \leq d_{max}\}| \quad (8)$$

where *d* is the distance between the MS and the *m-th* neighbor femtocell that use frequency $f_m$. The *m-th* femtocell will be included in this group only if the distance between the MS and the *m-th* neighbor FAP is less than or equal to a pre-defined threshold distance $d_{max}$.

Considering the above three facts (RSSI level, frequency, and location information) the femtocells included in the final neighbor femtocell list are:

$$E = (B/C) \cup D \quad (9)$$

The total number of femtocells in the neighbor femtocell list is thus:

$$N_f = N_1 - N_2 + M \qquad (10)$$

## 5.2 Handover Call Flow

The ability to seamlessly switch between the femtocells and the macrocell network is a key driver for femtocell network deployment. However, until now there is no effective and complete handover scheme for the femtocell network deployment. The handover procedures for existing 3GPP networks are presented in [33]-[39]. This section proposes the complete handover call flows for the integrated femtocell/macrocell network architecture in a dense femtocellular network deployment. The proposed handover schemes optimize the selection/reselection/RRC management functionalities in the femtocell/macrocell handover.

### *5.2.1 Femtocell-to-Macrocell Handover*

Fig. 5.4 shows the detail handover call flow procedures for femtocell-to-macrocell handover in dense femtocellular network deployment. If femto user detects that femto signal is going down, MS sends this report to the connected FAP (steps 1, 2). The MS searches for the signals from the neighbor FAPs and the macrocellular BS (step 3). The MS, serving FAP (S-FAP), neighbor FAPs, and the macrocellular BS together perform the SON configuration to accomplish the optimized neighbor cell list for the handover (steps 4 and 5). The MS performs the pre-authentication to all the access networks that are included in the neighbor cell list (step 6). Based on the pre-authentication and the signal levels, the MS and S-FAP together decide for handover to the macrocellular BS (step 7). FAP starts handover (HO) procedures by sending a handover request to macrocellular BS through the CN (steps 8, 9, 10, and 11). The CAC and radio resource control (RRC) are performed to check whether the call can be accepted or not (step 12). Then the macrocellular BS responses for the handover request (steps 13, 14, 15, and 16). Steps 17, 18, 19, 20, and 21 are used to setup a new link between target-RNC (T-RNC) and the macrocellular BS. The packet data are forwarded to macrocellular BS (step 22). The MS re-establishes a channel with the macrocellular BS and detached from the S-FAP, and also synchronized with the macrocellular BS (steps 23, 24, 25, 26, and 27). MS sends a handover complete message to FGW to inform that, the MS already completed handover and synchronized with the target macrocellular BS (steps 28, 29, and 30). Then the FAP deletes the old link with the S-FAP (steps 31, 32, and 33). Now the packets are sent to MS through the macrocellular BS.

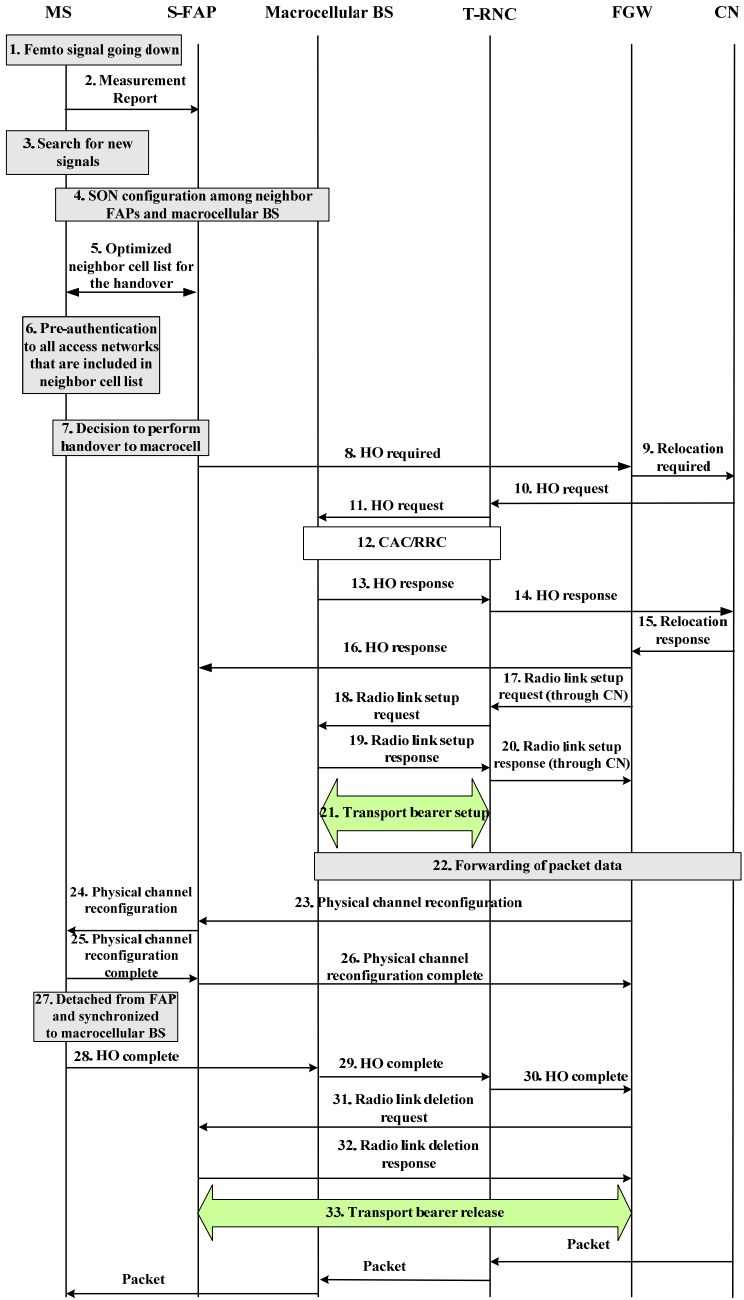

**Fig. 5.4:** Call flow for the femtocell-to-macrocell handover for a dense femtocellular network deployment.

## 5.2.2 Macrocell-to-Femtocell Handover

In this handover MS needs to select the appropriate target-FAP (T-FAP) among many candidate FAPs. Also, interference level should be considered for handover decision. The authorization should be checked during the handover preparation phase. Fig. 5.5 shows the detail call flow procedures for macrocell-to-femtocell handover in dense femtocellular network deployment. Whenever the MS in the macrocell network detects a signal from femtocell, it sends a measurement report to the connected macrocellular BS (steps 1, 2). The MS, macrocellular BS, and neighbor FAPs combine perform the SON configuration to accomplish the optimized neighbor cell list for the handover (steps 3 and 4). The MS performs the pre-authentication to all the access networks that are included in the neighbor cell list (step 5). Based on the pre-authentication and the signal levels, the MS decides for handover to target-FAP (T-FAP) (step 6). The macrocellular BS starts handover procedures by sending a handover request to the serving-RNC (S-RNC) (step 7). The handover request is forwarded from the macrocellular BS to T-FAP through the CN and FGW (steps 8, 9, and 10). The FAP checks the user's authorization (steps 11 and 12). The T-FAP performs CAC, RRC, and also compares the interference levels to admit a call (steps 13). Then the T-FAP responses for the handover request to macrocellular BS through the CN (step 14, 15, 16, and 17). A new link is established between the FGW and the T-FAP (steps 18, 19, 20, 21, and 22). Then the packet data are forwarded to the T-FAP (step 23). Now the MS re-establishes a channel with the T-FAP, detached from the source macrocellular BS, and synchronized with the T-FAP (steps 24, 25, 26, 27, and 28). MS sends a handover complete message to S-RNC to inform that, the MS already completed handover and synchronized with the T-FAP (steps 29, 30, and 31). Then the macrocellular BS deletes the old link with the RNC (steps 32, 33, and 34). Now the packets are forwarded to MS through the FAP.

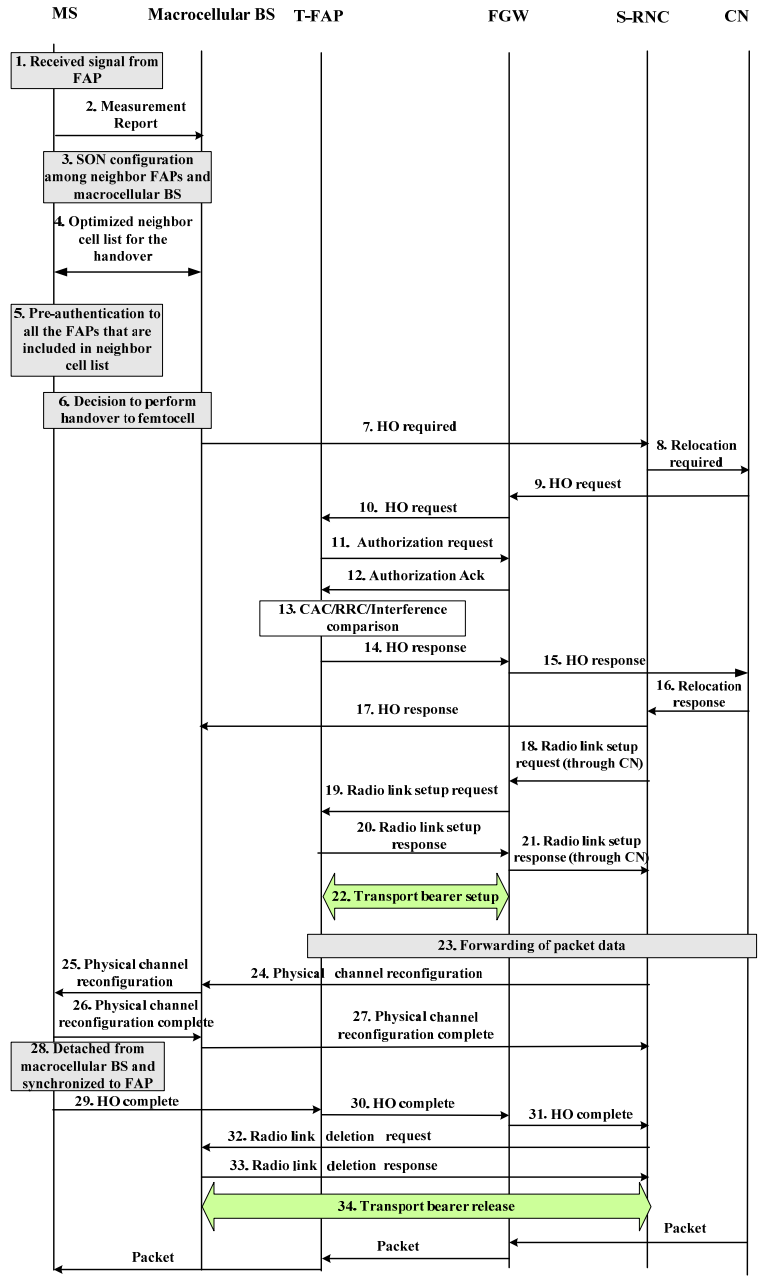

**Fig. 5.5:** Call flow for the macrocell-to-femtocell handover for dense femtocellular network deployment.

### 5.2.3 Femtocell-to-Femtocell Handover

Femtocell-to-femtocell handover is also challenging issue. In this handover MS needs to select the appropriate T-FAP among many neighbor FAPs. The authorization should be checked during the handover preparation phase. Fig. 5.6 shows the detail call flow procedures for femtocell-to-femtocell handover in dense femtocellular network environment. If femto user detects that femto signal is going down, MS send this report to the connected FAP (steps 1, 2). The MS searches for the signals from the neighbor FAPs and the macrocellular BS (step 3). The MS, serving-FAP (S-FAP), neighbor FAPs, and the macrocellular BS together perform the SON configuration to accomplish the optimized neighbor cell list for the handover (steps 4 and 5). The MS performs the pre-authentication to all the access networks that are included in the neighbor cell list (step 6). Based on the pre-authentication and the signal levels, the MS and S-FAP together decide for handover to the T-FAP (step 7). FAP starts handover procedures by sending a handover request to T-FAP through the FGW (steps 8 and 9). The FAP checks the user's authorization (steps 10 and 11). The T-FAP performs CAC and RRC to admit the handover call (steps 12). Then the T-FAP responses for the handover request to S-FAP through the FGW (step 13 and 14). A new link is established between the FGW and the T-FAP (steps 15, 16, and 17). Then the packet data are forwarded to the T-FAP (step 18). Now the MS re-establishes a channel with the T-FAP, detached from the S-FAP, and synchronized with the T-FAP (steps 19, 20, 21, 22, and 23). MS sends a handover complete message to FGW to inform that, the MS already completed handover and synchronized with the T-FAP (steps 24, 25, and 26). Then the S-FAP deletes the old link with FGW (steps 27, 28, and 29). Now the packets are forwarded to MS through the T-FAP.

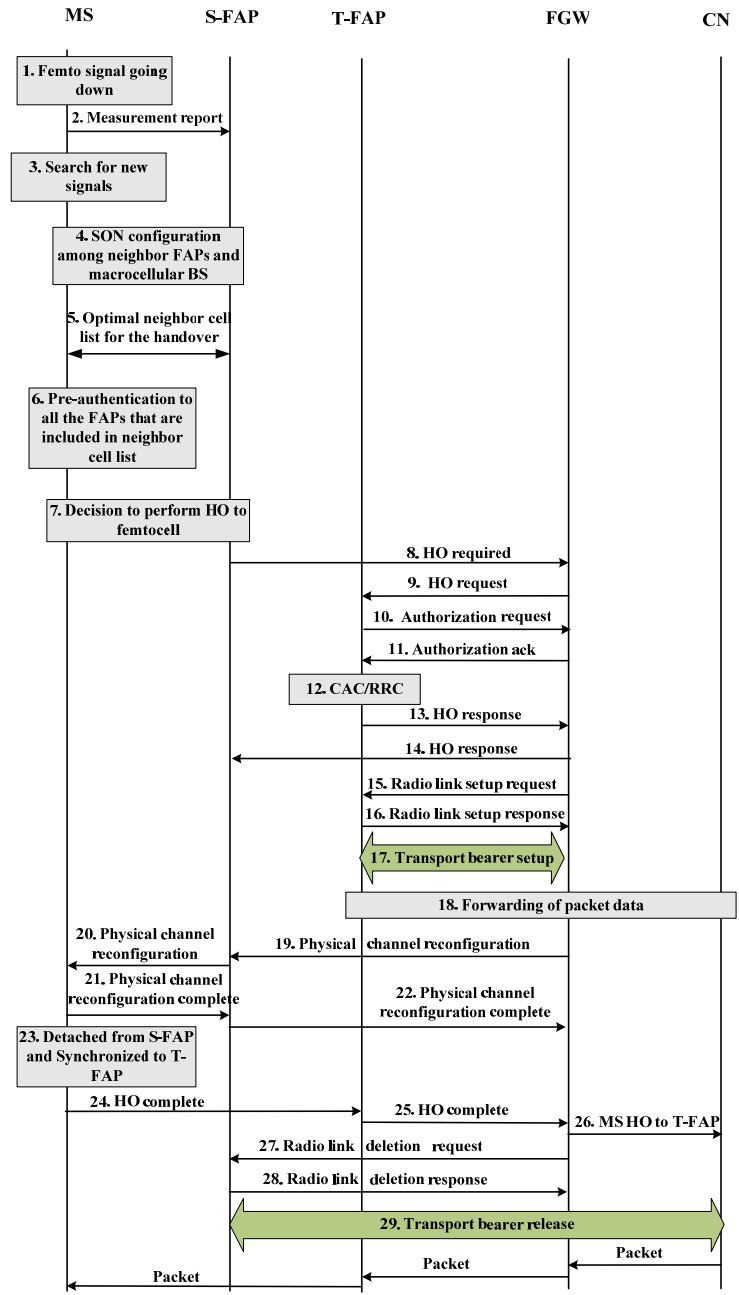

**Fig. 5.6:** Call flow for the femtocell-to-femtocell handover for dense femtocellular network deployment.

## 5.3 CAC for Femtocell/Macrocell Overlaid Networks

For the macrocell/femtocell integrated networks, the CAC can play a vital role to maximize the resource utilization especially for the macrocellular networks by efficiently controlling the admission of various traffic calls inside the macrocell coverage area. The main goal of our proposed scheme is to transfer macrocell calls to femtocellular networks as many as possible. We divided the proposed CAC into three parts. The first one for the new originating call, the second one for the calls that are originally connected with the macrocellular BS, and the last one for the calls that are originally connected with the FAP. We also used two threshold levels of SNIR to admit a call in the system. The first threshold level, $\Gamma_1$ is the minimum level of received SNIR that must be needed to connect a FAP. The second signal level, $\Gamma_2$ which is higher than $\Gamma_1$. The second threshold, $\Gamma_2$ is used in the CAC to reduce the unnecessary macrocell-to-femtocell handovers. We offer the QoS degradation [40], [41] of the QoS adaptive multimedia traffic to accommodate femtocell-to-macrocell and macrocell-to-macrocell handover calls. The existing QoS adaptive multimedia traffic in macrocellular network releases $C_{release,m}$ amount of bandwidth to accept the handover calls in the macrocellular network. This releasable amount depends on the number of running QoS adaptive multimedia calls and their maximum level of allowable QoS degradation and total number of existing calls in the macrocellular network. $\beta_{r,m}$ and $\beta_{min,m}$, respectively, the requested and the minimum required bandwidth for each of the requesting calls in macrocell. Hence each of the QoS adaptive calls can release maximum $\beta_{r,m} - \beta_{min,m}$ amount of bandwidth to accept a handover call in the macrocell system. If $C$ and $C_{occupied,m}$, respectively, the macrocell system bandwidth capacity and the occupied bandwidth by the existing macrocell calls, then the available empty bandwidth, $C_{availale,m}$ in the macrocellular network is $C - C_{occupied,m}$.

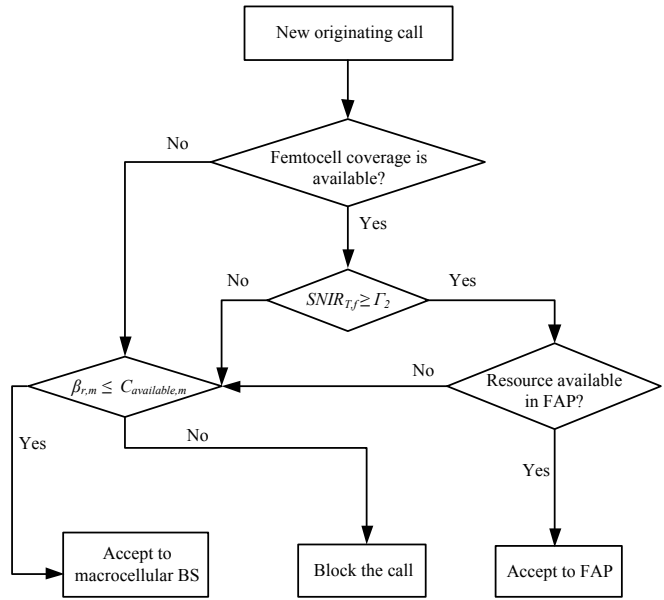

**Fig. 5.7:** CAC policy for the new originating calls.

### 5.3.1 New Originating Call

Fig. 5.7 shows the CAC policy for the new originating calls. Whenever a new call arrives, the CAC initially checks whether the femtocell coverage is available or not. If the femtocell coverage is available then FAP is the first choice to connect a call. A FAP accepts a new originating call if received SINR level, $\varGamma_2$ is satisfied and resource in the FAP is available. $SNIR_{T,f}$ is the received SINR level considering the target FAP. If the above conditions are not satisfied, then the call tries to connect with the overlaid macrocellular network. The macrocell system does not allow QoS degradation policy to accept any new originating call. A call is rejected if the requested bandwidth, $\beta_{r,m}$ is not available in the overlaid macrocellular network.

### 5.3.2 Calls are Originally Connected with the Macrocellular BS

Fig. 5.8 shows the CAC policy for the calls that are originally connected with the macrocellular BS. Whenever the moving MS detects signal from a FAP, the CAC policy checks the received SNIR level, $SNIR_{T,f}$ for the target FAP. A macrocell call is handed over to femtocell if $SNIR_{T,f}$ meets minimum $\varGamma_2$ or currently received SNIR level considering macrocellular BS, $SNIR_m$ is less than or equal to $SNIR_{T,f}$. If any one of the above conditions is satisfied, then the CAC policy checks the resource availability in the target FAP. We prefer higher level of threshold, $\Gamma_2$ to avoid some unnecessary macrocell-to-femtocell handovers.

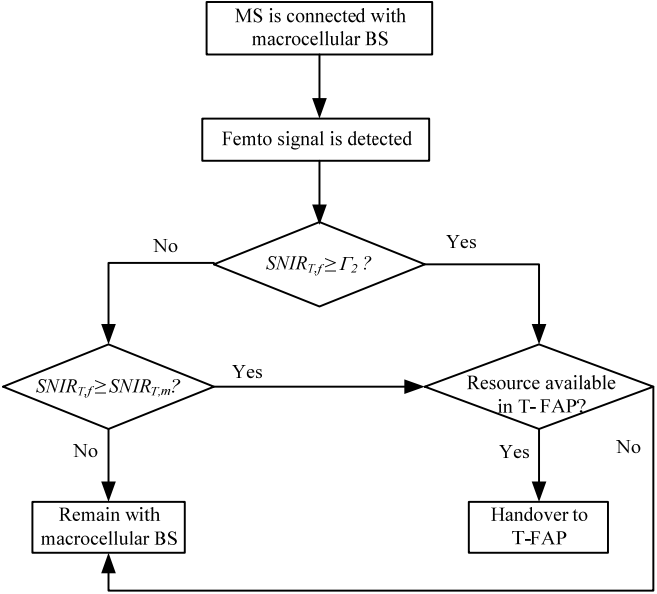

**Fig. 5.8:** CAC policy for the calls that are originally connected with the macrocellular BS.

### *5.3.3 Calls are Originally Connected with the FAP*

Fig. 5.9 shows the CAC policy for the calls that are originally connected with the FAP. Femtocell-to-femtocell and femtocell-to-macrocell handover calls are controlled by this CAC policy. Whenever the signal level from the serving FAP (S-FAP) is going down, the MS initiates to handover to other femtocell or overlaid macrocell. Whenever another target FAP (T-FAP) is not available for handover, the call tries to connect with the macrocellular network. If the empty resource in the macrocell system is not enough to accept the call, the CAC policy allows releasing of some bandwidth from the existing calls by degrading the QoS level of them. The CAC policy also permits the reduction of required bandwidth from $\beta_{r,m}$ to $\beta_{min,m}$ for the handover call request. If the minimum required bandwidth $\beta_{min,m}$ is not available in the macrocell system after releasing of some bandwidth from the existing calls, then the call will be dropped. If the received SNIR considering the T-FAP is greater than or equal to $\Gamma_2$, the MS firstly tries to handover to T-FAP. If the received SNIR considering the T-FAP is in between $\Gamma_1$ and $\Gamma_2$, then the MS tries to connect with the macrocellular BS. If resource in the macrocell is not available in the macrocell system, then the MS is handed over to the T-FAP even the received SNIR considering the T-FAP is less than or equal to $\Gamma_2$. However, during this condition, QoS degradation policy is not applicable. QoS degradation policy is only applicable when the received SNIR considering the T-FAP is less than $\Gamma_1$ or the resource in the T-FAP is not available.

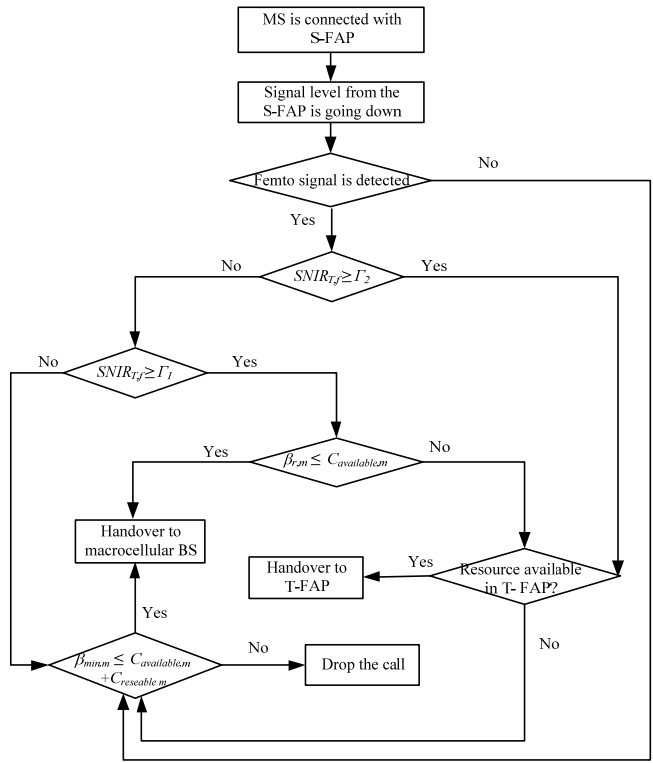

**Fig. 5.9:** CAC policy for the calls that are originally connected with the FAP.

## 5.4 Traffic Model

The proposed CAC schemes can be modeled by Markov Chain. The Markov Chain for the queuing analysis of a femtocell layer is shown in Fig. 5.10, where the states of the system represent the number of calls in the system. The maximum number of calls that can be accommodated in a femtocell system is $K$. As the call arrival rate in a femtocell is very low and the data rate of a femtocellular network is high, there is no need of

- 60 -

handover priority scheme for the femtocellular networks. The calls that are arrived in a femtocellular network are new originating calls, macrocell-to-femtocell handover calls, and the femtocell-to-femtocell handover calls. Femtocell-to-femtocell handover calls are divided into two types. The first type of calls are those for which the received SNIR considering the T-FAP is greater than or equal to $\Gamma_2$. The second type of calls are those for which the received SNIR considering the T-FAP is in between $\Gamma_1$ and $\Gamma_2$, and these calls are rejected by the macrocellular BS. We define $\mu_m$ ($\mu_f$) as the channel release rate of macrocell (femtocell).

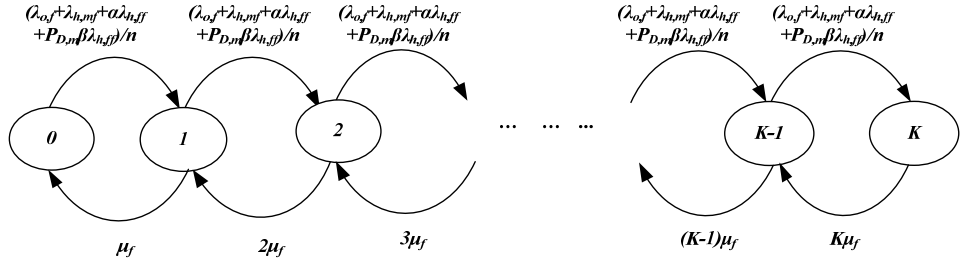

**Fig. 5.10:** The Markov Chain of a femtocell layer.

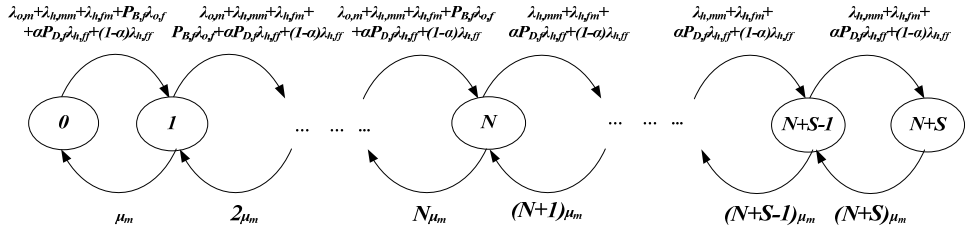

**Fig. 5.11:** The Markov Chain of a macrocell layer.

Fig. 5.11 shows the Markov Chain for the queuing analysis of the overlaid macrocell layer, where the states of the system represent the number of calls in the system. In Figs. 5.10 and 5.11, the symbols $\lambda_{o,f}$ and $\lambda_{o,m}$, respectively, represents total originating call arrival rate considering all $n$ number of femtocells within a macrocell coverage area and only macrocell coverage area. $\lambda_{h,mm}$, $\lambda_{h,ff}$, $\lambda_{h,fm}$, and $\lambda_{h,mf}$, respectively, indicates the total macrocell-to-macrocell, femtocell-to-femtocell, femtocell-to-macrocell, and macrocell-to-femtocell handover call arrival rate within the macrocell coverage area. $P_{B,m}$ ($P_{B,f}$) is the call blocking probability in the macrocell (femtocell) system. $P_{D,m}$ ($P_{D,f}$) is the call dropping probability in the macrocell (femtocell) system. We assume that for a femtocell-to-femtocell handover, the probability that the received SNIR considering the T-FAP is greater $\Gamma_2$ is $\alpha$ and the received SNIR considering the T-FAP is in between $\Gamma_2$ and $\Gamma_2$ is $\beta$. The consideration of single level of *SNIR* threshold, i.e., $\Gamma_1=\Gamma_2$ results $\alpha=1$ and $\beta=0$.

Fig. 5.11 also shows that the macrocell system provides $S$ number of additional states to support handover calls by the proposed adaptive QoS policy. The state $N$ is the maximum number of calls that can be accommodated by the macrocell system without QoS adaptation policy. Hence, the system provides QoS adaptation policy only to accept handover calls in the macrocell system. These handover calls include macrocell-to-macrocell and femtocell-to-macrocell handover calls. Femtocell-to-macrocell handover calls are divided into two types. The first types of calls are those which are directly arrived to the macrocell system. The second types of calls are those for which the calls are firstly arrived to femtocell but these calls are not accepted to femtocell due to lagging of resources or due to poor SNIR level.

The average channel release rates for the femtocell layer and the macrocell layer are calculated as follows,

For the macrocell layer it is:

$$\mu_m = \eta_m(\sqrt{n}+1)+\mu \qquad (11)$$

and for the femtocell layer it is:

$$\mu_f = \eta_f + \mu \qquad (12)$$

where $1/\mu$, $1/\eta_m$, and $1/\eta_f$, respectively, the average call length (exponentially distributed), average cell dwell time for the macrocell (exponentially distributed), and the average cell dwell time for femtocell (exponentially distributed).

The handover call arrival rates are calculated as follows,

The macrocell-to-macrocell handover call arrival rate is:

$$\lambda_{h,mm} = P_{h,mm}\frac{\left(1-P_{B,m}\right)\left(\lambda_{m,o}+\lambda_{f,o}P_{B,f}\right)+\left(1-P_{D,m}\right)\left\{\lambda_{h,fm}+\lambda_{h,ff}\left(1-\alpha+\alpha P_{D,f}\right)\right\}}{1-P_{h,mm}\left(1-P_{D,m}\right)} \qquad (13)$$

the macrocell-to-femtocell handover call arrival rate is:

$$\lambda_{h,mf} = P_{h,mf} \frac{(1-P_{B,m})(\lambda_{m,o}+\lambda_{f,o}P_{B,f})+(1-P_{D,m})\{\lambda_{h,fm}+\lambda_{h,ff}(1-\alpha+\alpha P_{D,f})\}}{1-P_{h,mm}(1-P_{D,m})} \quad (14)$$

the femtocell-to-femtocell handover call arrival rate is:

$$\lambda_{h,ff} = P_{h,ff} \frac{\lambda_{f,o}(1-P_{B,f})+\lambda_{h,mf}(1-P_{D,f})}{1-P_{h,ff}(1-P_{D,f})\{\alpha+(1-\alpha)P_{D,m}\}} \quad (15)$$

and the femtocell-to-macrocell handover call arrival rate is:

$$\lambda_{h,fm} = P_{h,fm} \frac{\lambda_{f,o}(1-P_{B,f})+\lambda_{h,mf}(1-P_{D,f})}{1-P_{h,ff}(1-P_{D,f})\{\alpha+(1-\alpha)P_{D,m}\}} \quad (16)$$

where $P_{h,mm}$, $P_{h,mf}$, $P_{h,ff}$, and $P_{h,fm}$, respectively, the macrocell-to-macrocell handover probability, macrocell-to-femtocell handover probability, femtocell-to-femtocell handover probability, and femtocell-to-macrocell handover probability.

The macrocell-to-macrocell handover probability, macrocell-to-femtocell handover probability, femtocell-to-femtocell handover probability, and femtocell-to- macrocell handover probability are calculated as follows,

The macrocell-to-macrocell handover probability is:

$$P_{h,mm} = \frac{\eta_m}{\eta_m + \mu} \quad (17)$$

the femtocell-to-macrocell handover probability is:

$$P_{h,fm} = \left[1 - n\left(\frac{r_f}{r_m}\right)^2\right]\frac{\eta_f}{\eta_f + \mu} \quad (18)$$

the femtocell-to-femtocell cell handover probability is:

$$P_{h,ff} = (n-1)\left(\frac{r_f}{r_m}\right)^2 \frac{\eta_f}{\eta_f + \mu} \quad (19)$$

and the macrocell-to-femtocell handover probability is:

$$P_{h,mf} = n \left(\frac{r_f}{r_m}\right)^2 \frac{\eta_m \sqrt{n}}{\eta_m \sqrt{n} + \mu} \tag{20}$$

where $r_f$ and $r_m$, respectively, the radius of a femtocell and radius of a macrocell coverage areas.

There is no guard channel for the handover calls in femtocell layer in our proposed scheme. For the femtocell layer, the average call blocking probability, $P_{B,f}$ and the average call dropping probability, $P_{D,f}$ can be calculated as,

$$P_{D,f} = P_{B,f} = P_f(K) = \frac{\left(\frac{\lambda_{T,f}}{n}\right)^K \frac{1}{K!\mu_f^K}}{\sum_{i=0}^{K}\left(\frac{\lambda_{T,f}}{n}\right)^i \frac{1}{i!\mu_f^i}} \tag{21}$$

where $\lambda_{T,f} = \lambda_{f,o} + \lambda_{h,mf} + \alpha \lambda_{h,ff} + P_{D,m} \beta \lambda_{h,ff}$

There QoS adaptation/degradation policy is allowed for the handover calls of macrocell layer in our proposed scheme. For the macrocell layer, the average call blocking probability, $P_{B,m}$ and the average call dropping probability, $P_{D,m}$ can be calculated as,

$$P_{B,m} = \sum_{i=N}^{N+S} P(i) = \sum_{i=N}^{N+S} \frac{(\lambda_{m,0} + \lambda_{h,m})^N (\lambda_{h,m})^{i-N}}{i!\mu_m^i} P(0) \tag{22}$$

$$P_{D,m} = P(N+S) = \frac{(\lambda_{m,0} + \lambda_{h,m})^N \lambda_{h,m}^S}{(N+S)!\mu_m^{N+S}} P(0) \tag{23}$$

where $\lambda_{h,m} = \lambda_{h,mm} + \lambda_{h,fm} + \alpha P_{D,f} \lambda_{h,ff} + (1-\alpha)\lambda_{h,ff}$

and $P(0) = \left[\sum_{i=0}^{N} \frac{(\lambda_{m,0} + \lambda_{m,h})^i}{i!\mu_m^i} + \sum_{i=N+1}^{N+S} \frac{(\lambda_{m,0} + \lambda_{m,h})^N (\lambda_{m,h})^{i-N}}{i!\mu_m^i}\right]^{-1}$

## 5.5 Performance Analysis

In this section we performed the effect of integrated macrocell/femtocell networks as well as the performance analysis of our proposed schemes. All the call arriving processes are assumed to be Poisson. The positions of the deployed femtocells within the macrocell coverage area are random. Table 5.1 shows the basic parameters that are used for the performance analysis. We also assume random manner about the hidden femtocells. We consider both the open access and close access randomly in the simulation.

**Table 5.1.** Summary of the parameter values used in our analysis

| Parameter | Value |
|---|---|
| First threshold value of received signal (RSSI) from a FAP $(S_{T0})$ | -90 (dBm) |
| Second threshold value of received signal (RSSI) from a FAP $(S_{T1})$ | -75 (dBm) |
| Bandwidth capacity of macrocell (C) | 6 (Mbps) |
| Required/allocated bandwidth for each of the QoS non-adaptive call | 64 (Kbps) |
| Maximum required/allocated bandwidth for each of the QoS adaptive calls | 56 (Kbps) |
| Minimum required/allocated bandwidth for each of the QoS adaptive calls | 28 (Kbps) |
| Ratio of arrival traffic (QoS non-adaptive calls: QoS adaptive calls) | 1:1 |
| First SNIR threshold $(\Gamma_1)$ | 10 (dB) |
| Second SNIR threshold $(\Gamma_2)$ | 12 (dB) |
| Number of deployed femtocells in a macrocell coverage area | 1000 |
| Average call duration time $(1/\mu)$ considering all calls (exponentially distributed) | 120 (sec) |
| Average cell dwell time $(1/\eta_f)$ for femtocell (exponentially distributed) | 360 (sec) |
| Average cell dwell time $(1/\eta_m)$ for macrocell (exponentially distributed) | 240 (sec) |
| Density of call arrival rate (at femtocell coverage area: at macrocell only coverage area ) | 20:1 |
| Penetration loss | 20 (dB) |

Fig. 5.12 shows the probability comparison that the target femtocell is missing from the neighbor femtocell list. In a traditional neighbor cell list based on the received signal strength only cannot include the hidden femtocells in the neighbor cell list. Thus there is a possibility that the target femtocell is not included in the neighbor femtocell list. This causes a failure of the handover to the target femtocell. With the increased number of deployed femtocells within an area increases the possibility that the neighbor femtocells coordinate with the serving femtocell and keep informed about the location of the hidden neighbor femtocells. As a consequence, increased number of deployed femtocells results in the reduction of probability that the hidden femtocell is out of the neighbor femtocell list. Also missing the appropriate neighbor femtocell from the neighbor femtocell list may cause a handover failure. Thus the handover failure rate decreases with the increase of the number of deployed femtocells for our scheme. Fig. 5.13 shows the comparison of the numbers of neighbor femtocells in the neighbor femtocell list for different schemes based on different parameters matrix. The result shows that the neighbor femtocell list based on the proposed scheme contains very small number of femtocells during the handovers. Thus the signal flow for the handover process became very small. Therefore, the results in Figs. 5.12 and 5.13 demonstrate that the proposed neighbor femtocell list algorithms for the femtocell-to-femtocell and the macrocell-to-femtocell handovers offer optimal number of femtocells in neighbor femtocell list. However, the reduced number of femtocells in the neighbor femtocell list does not increase the handover failure probability. Instead of increasing the handover failure probability due to hidden femtocell problem, our scheme reduces the handover failure probability with the increase of deployed femtocells.

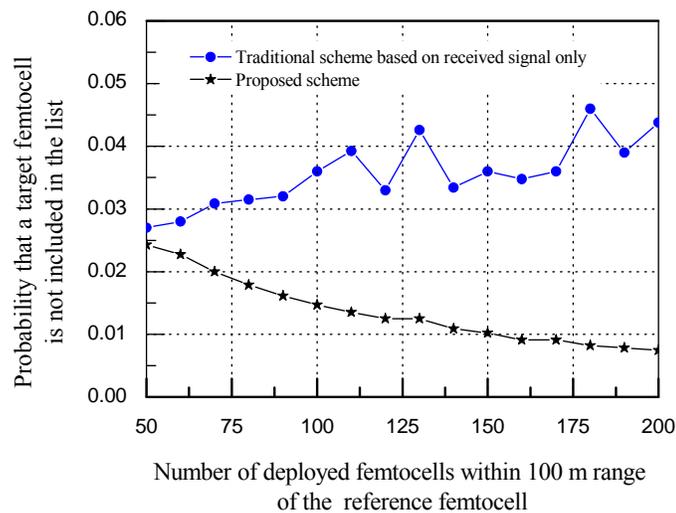

**Fig. 5.12:** The probability comparison that the target femtocell is missing from the neighbor femtocell list.

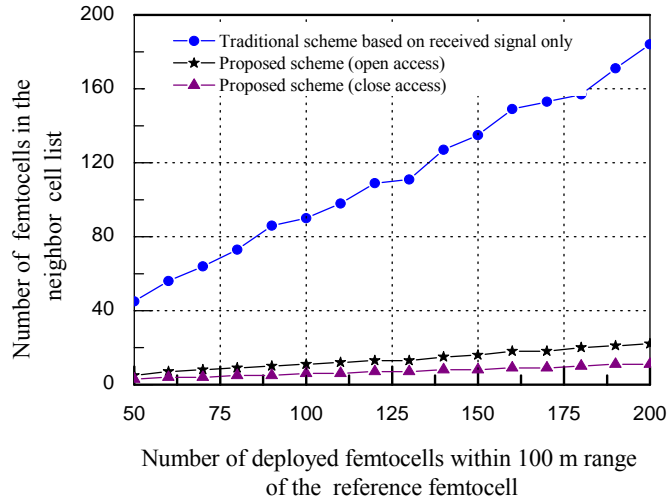

**Fig. 5.13:** A comparison of the numbers of neighbor femtocells in the neighbor femtocell list for different schemes based on different parameters matrix.

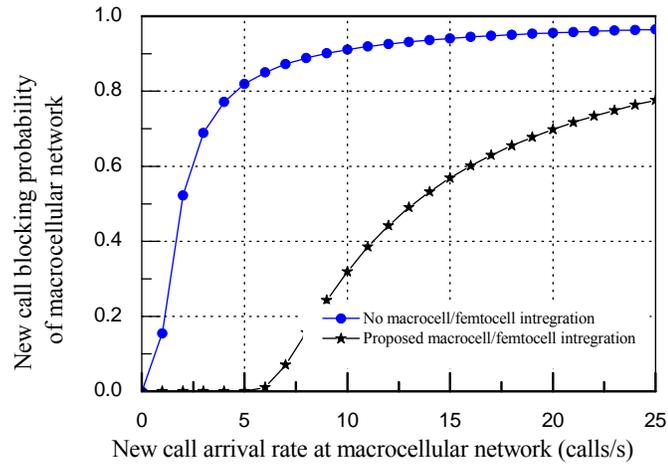

**Fig. 5.14:** A comparison of new call blocking probability in macrocellular network.

Whenever the macrocell and the femtocells are integrated, a huge number of macrocell calls are diverted to femtocells by macrocell-to-femtocell handover. As a result, the macrocell system can accommodate more number of new calls. Fig. 5.14 shows that the integrated macrocell/femtocell network system significantly reduces the new call blocking probability of the macrocellular networks even the macrocell system capacity for both the cases are same. For the same reason, the integrated macrocell/femtocell networks significantly reduces the overall forced call termination probability in the macrocellular networks. Fig. 5.15 shows the performance improvement of macrocellular networks in terms of overall forced call termination probability. Due to movement of the users, the femto users go out from the serving femtocell coverage area. If there is no integrated macrocell/femtocell network system, femtocell-to-macrocell handover calls will be surely dropped. As a result the handover call dropping probability for the femto users will be very high.

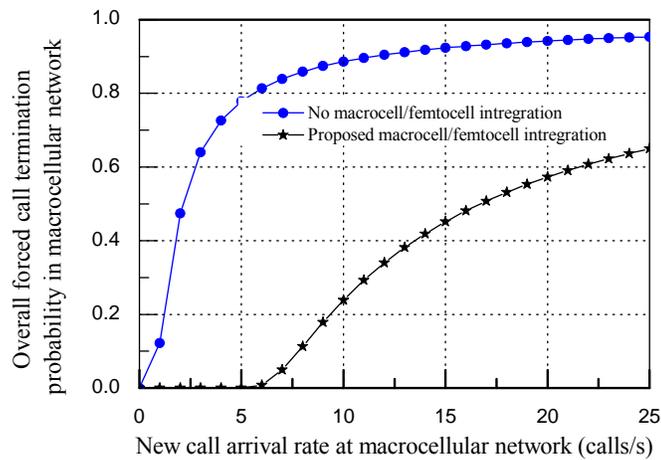

**Fig. 5.15:** A comparison of overall forced call termination probability in the macrocell system.

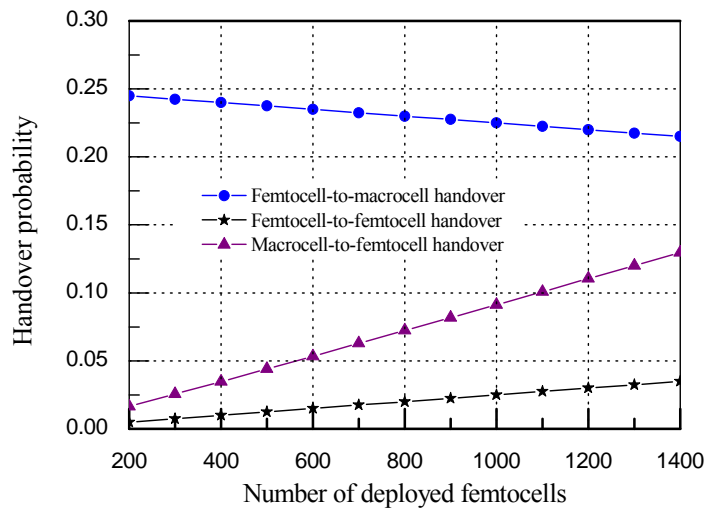

**Fig. 5.16.** A comparison of handover probability.

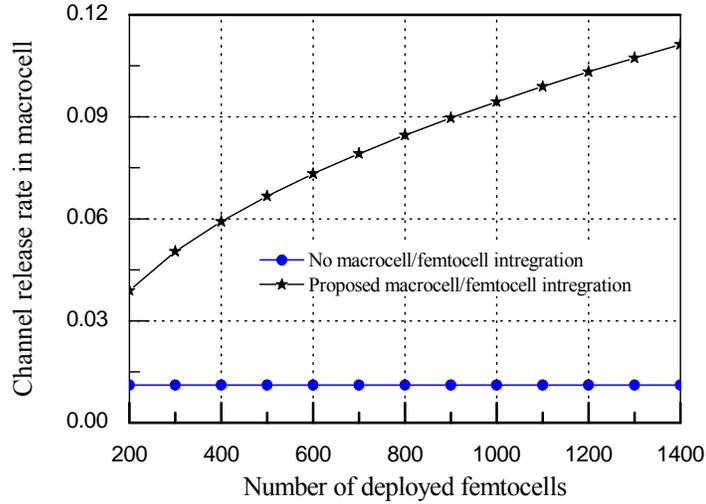

**Fig. 5.17.** A comparison of channel release rate in overlaid macrocellular network.

Fig. 5.16 shows the effect of different handover probabilities with the increase of the number of deployed femtocells within a macrocellular network coverage. With the increase of the number of deployed femtocells, the femtocell-to-femtocell handover and macrocell-to-femtocell handover probabilities are significantly increased. Also, the femtocell-to-macrocell handover probability is very high. Thus the management of these large numbers of femtocell-to-femtocell, femtocell-to-macrocell, and macrocell-to-femtocell handover calls is the important issue for the dense femtocellular network deployment. Fig. 5.17 shows the effect of the macrocell/femtocell integrated network in terms of channel release rate of the macrocellular network. Due to the integration, a huge number of macrocell users are handed over to femtocellular networks. Thus the channel release rate is increased with the increase of number of deployed femtocells. As a consequence, the macrocellular network can significantly reduce the overall forced call termination probability

# Part II:

# Resource Management for Macrocellular Networks

# Chapter 6
# Call Admission Control based on Adaptive Bandwidth Allocation for Wireless Networks

Provisioning of QoS is a key issue in any multimedia system. However, in wireless systems, supporting QoS requirements of different traffic types is more challenging due to the need to minimize two performance metrics - the probability of dropping a handover call and the probability of blocking a new call. Since QoS requirements are not as stringent for non-real-time traffic types, as opposed to real-time traffic, more calls can be accommodated by releasing some bandwidth from the already admitted non-real-time traffic calls. If we require that such a released bandwidth to accept a handover call ought to be larger than the bandwidth to accept a new call, then the resulting probability of dropping a handover call will be smaller than the probability of blocking a new call

In recent years, a notable trend in the design of wireless cellular systems is the decrease in the cell size; from macrocells, to microcells, to femtocells, and to picocells. Furthermore, user mobility has been increasing as well. These two factors result in more frequent handovers in wireless communication system. But when a handover occurs, there is a possibility that, due to limited resources in the target cell, the handed over connection will be dropped. From a user's point of view, blocking a new connection (e.g., the "busy" tone in phone communication) is more preferable than dropping the connection after it has already begun. Therefore, of interest are mechanisms that would allow reduction in the *HCDP*, even if this reduction comes at the expense of increasing the call blocking probability.

## 6.1  Background and Related Works of Handover Priority Scheme

Numerous prior research works have been published that allow higher priority for handover calls over new calls (e.g., [10], [11]). Most of these proposed schemes are based on the notion of "guard band," where a number of channels are reserved for the exclusive use of handover calls. Although schemes based on guard bands are simple and capable of reducing the HCDP, these schemes also result in reduced bandwidth utilization. Other approaches to reduce HCDP are handover-queuing schemes, which allow handover calls to queue and wait for a certain time for resources to become available. However, the handover-queuing schemes are not practical approaches for real-time multimedia services, because of the limited queuing time that could be allowed for real-time traffic ([42]).

The QoS adaptability of some multimedia traffic types has been used by several schemes (e.g., [11], [42], [47]-[50]) to reduce the call blocking probability. The adaptive QoS schemes proved more flexible and efficient in guaranteeing QoS than the guard channel schemes [11]. D. D. Vergados *et al.* [11] proposed an adaptive resource allocation scheme to prioritize particular traffic classes over others. Their scheme is based on the QoS degradation of low priority traffic to accept higher priority traffic call requests. W. Zhuang *et al.* [42] proposed an adaptive QoS (AQoS) scheme which reduces the QoS levels of calls that carry adaptive traffic to accept the handover call requests. F. A. Cruz-Pérez *et al.* [49] proposed flexible resource-allocation (FRA) strategies that prioritizes the QoS of particular service types over the others. Their scheme releases bandwidth from the low priority calls based on the prioritized call degradation policy to accept the higher priority call requests. I. Habib *et al.* [50] presented an adaptive QoS channel borrowing algorithm. A cell can borrow channels from any neighboring cell to reduce the call blocking probability.

A naïve bandwidth-adaptive scheme would be to merely reclaim bandwidth from the non-real-time traffic calls to accept a handover call or a new call <u>without differentiating between the two types of calls.</u> We refer to such a scheme as the "*Non-prioritized bandwidth-allocation scheme.*" In this non-prioritized bandwidth-adaptive scheme, when a handover or a new call request arrives, to accommodate this call, the system permits release of (up to some maximum allowable) bandwidth from non-real-time calls in progress. However, since the bandwidth release operation does not differentiate between handover and new calls, it cannot increase the priority of the former type of calls compared to the latter one. Indeed, in heavy traffic condition, the number of handover call requests increases faster than the increase in new originating call requests. Hence, the existing non-real-time traffic cannot release sufficient bandwidth to accept large number of handover calls. Consequently, the non-prioritized bandwidth-adaptive scheme cannot significantly reduce the HCDP, even though it reduces the new call blocking probability.

The AQoS handover priority scheme [42] allows reclaiming some of the allocated bandwidth from already admitted non-real-time traffic calls only to accept handover call requests. Therefore, this scheme can reduce the HCDP, but it cannot maximize the bandwidth utilization. This scheme also cannot significantly reduce the overall forced call termination rate (new originating calls plus handover calls).

As compared to our proposed scheme, the adaptive QoS schemes in [11], [49], [50] do not differentiate between handover calls and new calls. Hence, these schemes only ensure the QoS levels of the calls of higher priority traffic classes, but cannot reduce the overall HCDP of the system. Indeed, for the medium and heavy traffic conditions, these schemes cause very high HCDP and very large delays in transmission of the low priority traffic calls. The channel borrowing scheme [50] results in increased signaling overhead due to communication with the neighboring cells.

Therefore, we propose the "*Prioritized bandwidth-allocation scheme*," a multi-level bandwidth-allocation scheme for non-real-time traffic, which supports negligible HCDP without reducing the resource utilization. (We will also often refer to this scheme simply as "adaptive *bandwidth-allocation scheme.*) The proposed scheme reserves some releasable bandwidth to accept handover calls. In particular, the scheme is based on $M$ traffic classes, where two bandwidth-degradation thresholds are defined for each traffic class. Both thresholds signify the maximum portion of the allocated bandwidth that can be reclaimed from a non-real-time call of a particular traffic class. The first threshold is defined for the case when the arrival is a new call, while the second threshold is defined for a handover call.[1] By setting the first threshold to be smaller than the second threshold, the proposed prioritized adaptive bandwidth-allocation scheme allows to reclaim more bandwidth in the case of handover calls, thus increasing the probability of accepting a handover call, as opposed to new calls. And even though the proposed scheme blocks more new calls, still the bandwidth utilization is not reduced, because the scheme accepts new calls for which it expects to be able to provide sufficient resources until the call ends.

---

[1] Also, the minimum required bandwidth to accept a non-real-time handover call is less than that of a non-real-time new call.

## 6.2 The System Model

Contemporary and future wireless network are required to serve different multimedia traffic types, which are classified by standardization bodies. The QoS parameters of the various traffic types can be significantly different ([43]-[46]). Bit rate is one such a parameter ― some traffic types require guaranteed bit rate (GBR), while others are categorized as "best effort" delivery only. Delay is another QoS parameter. For example, according to 3GPP, the delay of real-time conversational services is characterized by the round trip time, which is required to be short, because of the interactive nature of such services. On the other hand, streaming services are limited to the delay variation of the end-to-end flow, and background services are delay insensitive [45]. Typically, real-time services necessitate GBR, while for non-real-time services non-guaranteed bit rate (NGBR) suffices. Thus, under heavy traffic condition, the QoS of non-real-time services can be purposely degraded (e.g., by restricting bandwidth allocation), so that the QoS of real-time services is preserved (e.g., by maintaining low probability of blocking new calls or low probability of dropping handover calls).

There are various considerations that affect the tradeoffs of such bandwidth-allocation schemes. For example, as mentioned before, it would be reasonable to commit larger amount of bandwidth to handover calls than to new originating call. Similarly, while in progress, non-real-time calls could be subject to some bandwidth reduction, alas by increasing the duration (i.e., the lifetime) of such connections. Hence, to analyze the QoS of the various traffic types with the proposed scheme, an appropriate system model is proposed in this paper.

### *6.2.1 The Bandwidth Allocation/Degradation Model*

Fig. 6.1 shows the multi-level bandwidth-allocation model for non-real-time services of the traffic of class *m.* The bandwidth-allocation scheme is characterized by bandwidth-degradation factors $\gamma_m$, $\gamma_{m,n}$, and $\gamma_{m,h}$, which are defined for each class *m* traffic, respectively, as: the fraction of the bandwidth that has been already degraded of an admitted non-real-time call, the maximum fraction of the bandwidth of an admitted non-real-time call that can still be degraded to accept a new call, and the maximum fraction of the bandwidth of an admitted non-real-time call that can still be degraded to accept a handover call. The values of $\gamma_{m,h}$ for different classes of traffic types ensure the minimum QoS requirements. With increasing the values of $\gamma_{m,n}$, the delay and the HCDP are increased, while the new call blocking probability is decreased. The parameters $\beta_{m,a}$, $\beta_{m,n}$, and $\beta_{m,h}$ represent the per-call bandwidth allocations of the traffic of class *m,* respectively, as: the allocated bandwidth of already admitted calls, the minimum allocated bandwidth to accept a new call, and the minimum allocated bandwidth to accept a handover call. Since the bandwidth of real-time traffic classes cannot be degraded at all, the bandwidth degradation factor of all the real-time traffic classes equals zero. However, the system can release bandwidth from the existing non-real-time traffic calls (i.e., degrade the QoS of the non-real-time calls) to accept non-real-time and real-time traffic calls. Though, the level of bandwidth degradation to accept a new call and a handover call are not, necessarily, equal.

The bandwidth-degradation factors relate to the bandwidth allocations as follows:

$$\gamma_m = \frac{\beta_{m,r} - \beta_{m,a}}{\beta_{m,r}}, \tag{1}$$

$$\gamma_{m,h} = \frac{\beta_{m,r} - \beta_{m,h}}{\beta_{m,r}}, \tag{2}$$

$$\gamma_{m,n} = \frac{\beta_{m,r} - \beta_{m,n}}{\beta_{m,r}}, \tag{3}$$

where $\beta_{m,r}$ represents the bandwidth requested by a call of the *m-th* class traffic. A new call can be accepted only if the condition $\beta_{m,a} \geq \beta_{m,n}$ (for all the traffic classes of *m*=1… *M*) holds after a new call is accepted. A handover call (any class of traffic) can be accepted only if the condition $\beta_{m,a} \geq \beta_{m,h}$ (for all the traffic classes of *m*=1…*M*) holds after a handover call is accepted. Due to the above definitions, the scheme is more likely to accept handover calls over new calls.

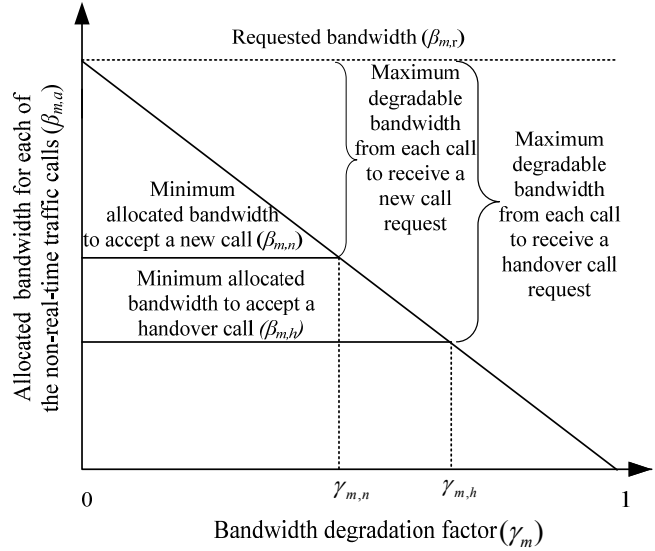

**Fig. 6.1**: The model of the proposed multi-level bandwidth allocation scheme for non-real-time traffic of class *m*

The non-prioritized bandwidth-adaptive scheme represents a particular limiting case of the proposed scheme in which $\gamma_{m,n} = \gamma_{m,h}$ for each class of traffic. It means that the non-prioritized bandwidth-adaptive scheme does not differentiate between the handover calls and the new calls. The AQoS handover priority scheme [42] is also a special case of the proposed scheme in which $\gamma_{m,n} = 0$ for all traffic classes. It implies that the AQoS handover priority scheme does not allow the bandwidth degradation to accept a new call. The key advantages of our proposed prioritized bandwidth-adaptive scheme are that it provides a system operator with the ability to adjust the parameters $\gamma_{m,n}$ and $\gamma_{m,h}$ in order to achieve the desired new call blocking probability and HCDP, as well as to satisfy the minimum expected QoS level for each class of traffic calls. The only disadvantage of the proposed scheme is that it increases the average call duration of the non-real-time traffic calls. However, the increased call duration is less than in the non-prioritized bandwidth-adaptive scheme. Compared to the non-prioritized bandwidth-adaptive and AQoS handover priority schemes, our proposed scheme does not have significant limitations in terms of implementation and complexity. Furthermore, our proposed scheme is based on the QoS adaptation mechanism, a mechanism that is already well accepted in the field of wireless communications.

## 6.2.2 The Traffic Model

Fig. 6.2 shows the relation of the new-call-arrival rate ($\lambda_n$), the handover-call-arrival rate ($\lambda_h$), and the average channel release rate ($\mu_c$). In the figure, $P_B$ and $P_D$ represent the blocking probability of new calls and the dropping probability of handover calls, respectively. All call arriving processes are assumed to be Poisson.

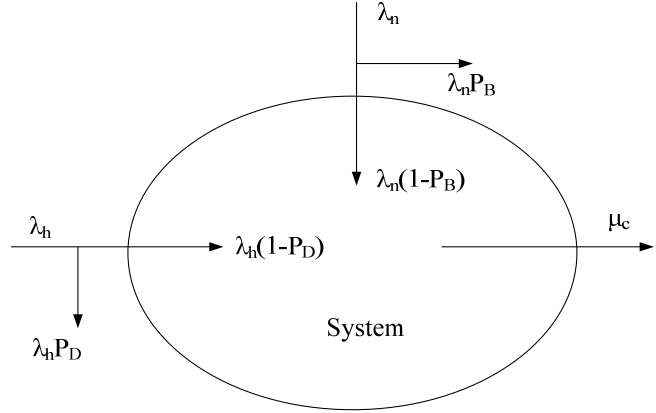

**Fig. 6.2:** The system model: new-call arrival rate ($\lambda_n$), handover-call arrival rate ($\lambda_h$), and service rate ($\mu_c$)

A new call that arrives in the system may either complete within the original cell or may handover to another cell or cells before completion. The probability of a call handover depends on two factors, (a) the average cell dwell time[2] ($1/\eta$) and (b) the average call duration ($1/\mu$). We note that the average duration of non-real-time calls (e.g., file download) depends on the amount of allocated bandwidth. The average channel release rate ($\mu_c$), also depends on the above two parameters (a) and (b).

Since both the call duration and the cell dwell time are assumed to be exponential, the handover probability of a call at a particular time is given by:

$$P_h = \frac{\eta}{\eta + \mu}. \qquad (4)$$

The average call duration, ($1/\mu$), is a weighted sum of the call durations of the $q$ real-time traffic classes and the $M$-$q$ non-real-time traffic classes. However, since the bandwidth allocated to a real-time traffic is fixed (i.e., $\beta_{m,a} = \beta_{m,r}$), while the bandwidth allocated to a non-real-time traffic of class *m* can be degraded (i.e., $\beta_{m,a} \leq \beta_{m,r}$), the average call duration of a real-time call is independent of bandwidth adaptation,[3] while the average call duration of non-real-time traffic strongly depends on the bandwidth-degradation factors. Thus, if we label $T_m(\beta_m)$ as the duration of a call of class *m*, where $\beta_m$ is the bandwidth allocated to calls of class *m*, then:

$$\frac{1}{\mu} = \frac{\sum_{m=1}^{q} N_m \cdot T_m(\beta_{m,r}) + \sum_{m=q+1}^{M} N_m \cdot T_m(\beta_{m,a})}{\sum_{m=1}^{M} N_m} \ . \tag{5}$$

---
[2] Also referred to as "sojourn time"

The handover-call arrival rate into a cell is calculated as:

$$\lambda_h = \frac{P_h(1 - P_B)}{[1 - P_h(1 - P_D)]} \lambda_n \ . \tag{6}$$

where the equation follows from balancing the rates of handover calls into and out of a cell (see Fig. 2.)

## 6.3 Bandwidth Adaptation and the Optimal CAC

Efficient allocation of bandwidth is a key element of the adaptive bandwidth-allocation scheme to guarantee the QoS of different classes of traffic and to ensure the best utilization of the bandwidth. This section presents the bandwidth allocation rules, the bandwidth release rules, and the *CAC* policy.

The bandwidth allocated to the traffic of class *m* (among the total *M* traffic classes) is represented by $\beta_{m,a}$. Among the *M* traffic classes, *q* traffic classes are bandwidth non-adaptive (e.g., conversational non-compressed voice), whereas the remaining *(M-q)* traffic classes are bandwidth-adaptive (e.g., file transfer) [51]. We label the total number of real-time and of non-real-time calls in the system, respectively, as:

$$N_R = \sum_{m=1}^{q} N_m \quad \text{and} \quad N_{nR} = \sum_{m=q+1}^{M} N_m \ . \tag{7}$$

Suppose that C and $N_m$ represent the total bandwidth (i.e., capacity) of the system and the total number of current calls in the system of the traffic of class *m*, respectively. We define the "residual fractional non-real-time capacity" as *X*:

---

[3] For calls which complete without being dropped

$$X = \frac{C - \sum_{m=1}^{q} N_m \beta_{m.a}}{\sum_{m=q+1}^{M} N_m \beta_{m.r}}, \qquad N_{nR} \geq 1 \tag{8}$$

where, the allocation of bandwidth for each of bandwidth-adaptive traffic classes is based on the value of $X$.

The allocated bandwidth for each of the bandwidth non-adaptive (real-time) calls is:

$$\beta_{m,a} = \beta_{m,r}, \qquad 1 \leq m \leq q \tag{9}$$

If $X \geq 1$, then:

$$\beta_{m,a} = \beta_{m,r}, \qquad (q+1) \leq m \leq M \tag{10}$$

and if $X \leq 1$, then:

$$\beta_{m.a} = \frac{C - \sum_{k=1}^{q} N_k \beta_{k.a}}{\sum_{k=q+1}^{M} N_k (1-\gamma_{k,h}) \beta_{k.r}} (1 - \gamma_{m,h}) \beta_{m,r}, \qquad (q+1) \leq m \leq M \text{ and } N_{nR} \geq 1 \tag{11}$$

Next, we show how to calculate the maximum bandwidth that can be released from non-real-time calls, the occupied bandwidth by all the existing calls, and the available bandwidth to accept a call.

If $\beta_{m,a} > \beta_{m,h}$ for the traffic of class *m*, then bandwidth could be released from the calls of class *m* to accommodate an arrival of a handover call. The overall releasable bandwidth from the non-real-time calls to accept a handover call is:

$$C_{releasable,hand} = \sum_{m=q+1}^{M} N_m (\beta_{m,a} - \beta_{m,h}). \tag{12}$$

If $\beta_{m,a} > \beta_{m,n}$ for the traffic of class *m*, then bandwidth could be released from the calls of class *m* to accommodate an arrival of a new call. The overall releasable bandwidth from the non-real-time calls in the system to accept a new call is:

$$C_{releasable,new} = \sum_{m=q+1}^{M} N_m (\beta_{m,a} - \beta_{m,n}). \tag{13}$$

The bandwidth occupied by all the calls in the system is:

$$C_{occupied} = \sum_{m=1}^{M} N_m \beta_{m,a}. \tag{14}$$

The maximum possible available bandwidth to accept a handover call is:

$$C_{available,hand} = C - \sum_{m=1}^{q} N_m \beta_{m,r} - \sum_{m=q+1}^{M} N_m \beta_{m,h}. \tag{15}$$

and the maximum possible available bandwidth to accept a new call is:

$$C_{available,new} = C - \sum_{m=1}^{q} N_m \beta_{m,r} - \sum_{m=q+1}^{M} N_m \beta_{m,n}. \tag{16}$$

The required minimum bandwidth to accept the $(N_m+1)^{th}$ call of class $m$, for which the requested bandwidth is $\beta_{m,r}$, can be calculated as follows:

For a handover call it is:

$$C_{m,h(required)} = \begin{cases} \beta_{m,r}, & 1 \leq m \leq q \\ (1 - \gamma_{m,h})\beta_{m,r}, & (q+1) \leq m \leq M \end{cases} \tag{17}$$

and for a new call it is:

$$C_{m,n(required)} = \begin{cases} \beta_{m,r}, & 1 \leq m \leq q \\ (1 - \gamma_{m,n})\beta_{m,r}, & (q+1) \leq m \leq M \end{cases} \tag{18}$$

A call (of any class of traffic) can be accepted only if the required bandwidth for that call is less than or equal to the unused bandwidth plus releasable bandwidth. The CAC policy for the proposed scheme, shown in Fig. 6.3, determines whether a call can be accepted or not based on the following rules. After the arrival of the $(N_m+1)^{th}$ call of class $m$, the input to the CAC algorithm includes: the total capacity ($C$) of the system, the bandwidth occupied by all the system calls ($C_{occupied}$), the call type (new or handover), and the amount of requested bandwidth ($\beta_{m,r}$). A new call is rejected if $\beta_{m,a}$ is less than or equal to $\beta_{m,n}$. It means that for this condition, the existing non-real-time calls are not allowed to release any bandwidth to accept a new call; i.e., only handover calls can be accepted.

Whenever the requested bandwidth is strictly less than the total available bandwidth ($C - C_{occupied}$), the system accepts the call. Otherwise, the system calculates the minimum required bandwidth to accept the call and the maximum available bandwidth if all the existing non-real-time calls release the maximum allowable bandwidth (i.e., $C_{releasable,new}$ to accept a new call and $C_{releasable,hand}$ to accept a handover call). For the proposed CAC, $C_{releasable,new} < C_{releasable,hand}$ to reserve more releasable bandwidth for handover calls, so that $P_D < P_B$. The CAC then determines whether it is possible to admit the call or not after reducing the requested bandwidth and releasing the bandwidth from the existing calls. If the condition is satisfied, the system releases the required bandwidth from the existing non-real-time calls to accept the call. In summary, the proposed CAC policy results in higher priority to handover calls than to new calls.

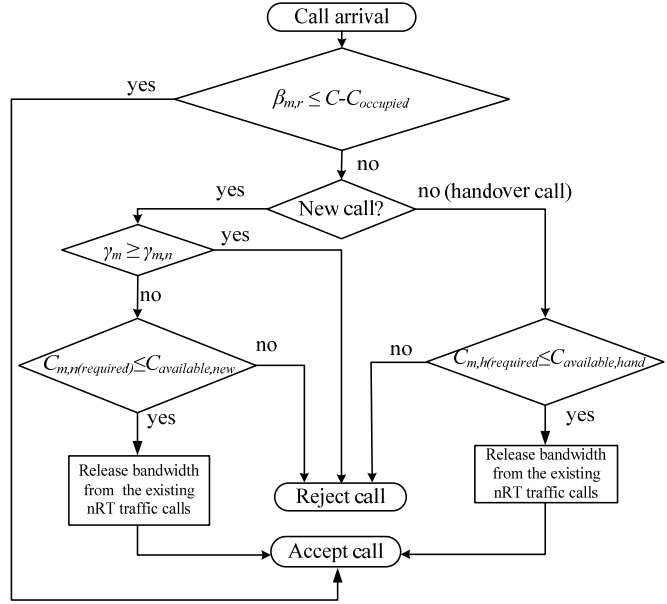

**Fig. 6.3:** The flow diagram of the proposed bandwidth-adaptive CAC.

## 6.4 Queuing Analysis

The proposed scheme can be modeled as an *M/M/K/K* queuing system (the value of *K* will be defined in the sequel). Suppose that the ratios of the calls arriving to the system for the *M* traffic classes are: $a_1 : a_2 : \ldots : a_M$, where:

$$\sum_{m=1}^{M} a_m = 1. \tag{19}$$

The Markov Chain for the queuing analysis of the traditional hard-QoS scheme with *G* guard channels is shown in Fig. 6.4, where the states of the system represent the number of calls in the system. The maximum number of calls that can be accommodated using the hard-QoS scheme is:

$$N = \left\lfloor \frac{C}{\sum_{m=1}^{M} \{a_m\, \beta_{m,r}\}} \right\rfloor. \tag{20}$$

The Markov Chain for the proposed scheme is shown in Fig. 6.5, where the states of the system represent the number of calls in the system. We define $\mu_i$ as the channel release rate when the system is in state $i$. The maximum number of additional calls that can be supported by the proposed adaptive bandwidth-allocation scheme is:

$$S = \left\lfloor \frac{C \sum_{m=1}^{M} a_m \gamma_{m,h} \beta_{m,r}}{\sum_{m=1}^{M} \{a_m(1 - \gamma_{m,h})\beta_{m,r}\} \sum_{m=1}^{M}\{a_m \beta_{m,r}\}} \right\rfloor. \quad (21)$$

The maximum number of calls that can be accommodated using the proposed adaptive bandwidth-allocation scheme is $K=(N+S)$. The maximal number of additional states of the Markov Chain in which the system accepts new call is:

$$L = \left\lfloor \frac{C \sum_{m=1}^{M} a_m \gamma_{m,n} \beta_{m,r}}{\sum_{m=1}^{M}\{a_m(1 - \gamma_{m,n})\beta_{m,r}\} \sum_{m=1}^{M}\{a_m \beta_{m,r}\}} \right\rfloor. \quad (22)$$

The average channel release rate ($\mu_c$) is given by ([52], [53]):

$$\mu_c = \mu + \eta. \quad (23)$$

However, as mentioned before, the average channel release rate of the proposed system is not the same as the channel release rate of the hard-QoS scheme. Due to the applied bandwidth degradation, the call duration of some of the non-real-time calls is increased, which results in a longer average channel holding time. Furthermore, with more calls in the system, the bandwidth allocated to the non-real-time calls decreases, which further prolongs the average call duration. If we label $\vec{\beta}_a = (\beta_{1,a}, \beta_{2,a}, \dots, \beta_{M,a})$ as the bandwidth allocation vector to the $M$ traffic classes, then the average call duration time, $1/\mu$, which we label as $T(\vec{\beta}_a)$ to indicate its dependence on the actual bandwidth allocation, is:

$$\left(\frac{1}{\mu}\right) = T(\vec{\beta}_a) = \frac{\sum_{m=1}^{q} N_m \cdot T_m(\beta_{m,a} = \beta_{m,r}) + \sum_{m=q+1}^{M} N_m \cdot T_m(\beta_{m,a} \leq \beta_{m,r})}{\sum_{m=1}^{M} N_m}. \quad (24)$$

We note that when all the $M$ traffic classes are allocated their requested bandwidth, $\vec{\beta}_a = (\beta_{1,r}, \beta_{2,r}, \dots, \beta_{M,r}) \triangleq \vec{\beta}_r$, equation (24) reduces to:

$$\left(\frac{1}{\mu}\right) = T(\vec{\beta}_r) = \frac{\sum_{m=1}^{M} N_m \cdot T_m(\beta_{m,r})}{\sum_{m=1}^{M} N_m}. \quad (25)$$

For the system states $0 < i \leq N$, when there is enough bandwidth in the system, all the $M$ traffic classes are allocated the requested bandwidth $\beta_{m,r}$. Thus, in these states, the average call duration time, $1/\mu$, equals $T(\vec{\beta}_r)$. Therefore, for the states $0 < i \leq N$, the average channel release rates ($\mu_c$) for the hard-QoS and for the proposed schemes are the same and are independent of the state $i$:

$$\mu_c = \eta + \left(\frac{1}{T(\vec{\beta}_r)}\right) \triangleq \mu_1 \qquad 0 < i \leq N .$$

However, when the proposed system is in a state $N < i \leq N + S$, some non-real-time calls are allocated less than the requested bandwidth, $\vec{\beta}_r$. But since the average call duration depends on the bandwidth allocation, this means that the average call duration now depends on the state that the system is in. In other words, the average call duration increases with the state. Consequently, the average channel release rate ($\mu_c$) is now state-dependent through the value of $\vec{\beta}_r$:

$$\mu_c = \eta + \left(\frac{1}{T(\vec{\beta}_a)}\right) \triangleq \mu_i(\vec{\beta}_a) \qquad N < i \leq N + S .$$

In Fig. 6.5, we refer to $\mu_i(\vec{\beta}_a)$ as simply $\mu_i$.

Using the *M/M/K/K* queuing analysis, where $K = N+S$, the probability that the system is in state $i$, is given by equation (26) below. In the proposed scheme a new call is blocked if the system is in the state $(N+L)$ or larger. However, a handover call is dropped if the system is in the state $(N+S)$. Thus, from equations (19) − (26), the call blocking probability of an originating new call ($P_B$) and the call dropping probability of a handover call ($P_D$) can be computed using equations (27) and (28), respectively.

For the non-prioritized bandwidth-adaptive scheme, where there is no priority of handover calls over new calls, $L=S$ and $\gamma_{m,n} = \gamma_{m,h}$. For the AQoS handover priority scheme, there are no additional states of the Markov Chain to accept new calls, thus $L=0$ and $\gamma_{m,n} = 0$.

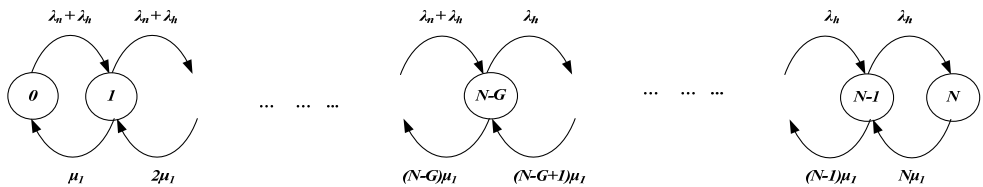

**Fig. 6.4:** The Markov Chain of the existing hard-QoS scheme with G guard channel

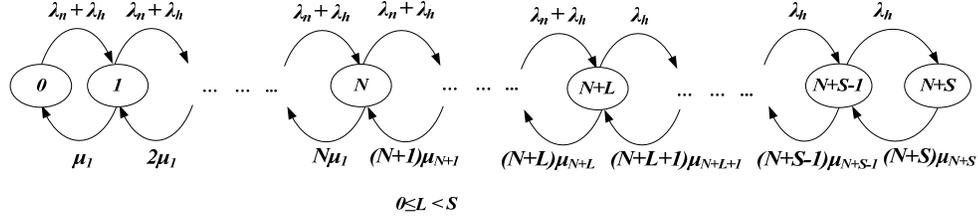

**Fig. 6.5:** The Markov Chain of the proposed bandwidth-adaptive CAC

$$P(i) = \begin{cases} \dfrac{(\lambda_n + \lambda_h)^i}{i!\,\mu_1^{\,i}} P(0), & 0 < i \leq N \\[2mm] \dfrac{(\lambda_n + \lambda_h)^i}{i!\,(\mu_1^{\,N})\prod_{p=N+1}^{i}\mu_p} P(0), & N \leq i \leq N+L \\[2mm] \dfrac{(\lambda_n + \lambda_h)^{\left\lfloor \frac{C}{\sum_{m=1}^{M}\{a_m(1-\gamma_{m,n})\beta_{m,r}\}}\right\rfloor}(\lambda_h)^{i-\left\lfloor \frac{C}{\sum_{m=1}^{M}\{a_m(1-\gamma_{m,n})\beta_{m,r}\}}\right\rfloor}}{i!\,(\mu_1^{\,N})\prod_{p=N+1}^{i}\mu_p} P(0), & N+L \leq i \leq N+S \end{cases} \quad (26)$$

where $P(0)$

$$= \left[ \sum_{i=0}^{N} \dfrac{(\lambda_n + \lambda_h)^i}{i!\,\mu_1^i} + \sum_{i=N+1}^{N+L} \dfrac{(\lambda_n + \lambda_h)^i}{i!\,(\mu_1^{\,N})\prod_{p=N+1}^{i}\mu_p} \right.$$

$$\left. + \sum_{i=N+L+1}^{N+S} \dfrac{(\lambda_n + \lambda_h)^{\left\lfloor \frac{C}{\sum_{m=1}^{M}\{a_m(1-\gamma_{m,n})\beta_{m,r}\}}\right\rfloor}(\lambda_h)^{i-\left\lfloor \frac{C}{\sum_{m=1}^{M}\{a_m(1-\gamma_{m,n})\beta_{m,r}\}}\right\rfloor}}{i!\,(\mu_1^{\,N})\prod_{p=N+L+1}^{i}\mu_p} \right]^{-1}$$

$$P_B = \sum_{i=N+L}^{N+S} P(i) = (\lambda_n + \lambda_h)^{\left\lfloor \frac{C}{\sum_{m=1}^{M}\{a_m(1-\gamma_{m,n})\beta_{m,r}\}}\right\rfloor} \sum_{i=N+L}^{N+S} \dfrac{(\lambda_h)^{i-\left\lfloor \frac{C}{\sum_{m=1}^{M}\{a_m(1-\gamma_{m,n})\beta_{m,r}\}}\right\rfloor}}{i!\,(\mu_1^{\,N})\prod_{p=N+1}^{i}\mu_p} P(0) \quad (27)$$

$$P_D = P(N+S) = \dfrac{(\lambda_n + \lambda_h)^{\left\lfloor \frac{C}{\sum_{m=1}^{M}\{a_m(1-\gamma_{m,n})\beta_{m,r}\}}\right\rfloor}\lambda_h^{S-L}}{(N+S)!\,(\mu_1^{\,N})\prod_{p=N+1}^{N+S}\mu_p} P(0) \quad (28)$$

## 6.5 Numerical Results

In this section, we present the numerical results of the analysis of the proposed scheme. We compared the performance of our proposed prioritized bandwidth-adaptive allocation scheme with the performance of the "Hard-QoS scheme", the "Non-prioritized bandwidth-adaptive scheme", the "Hard-QoS with 5% guard band scheme", and the "AQoS handover priority scheme". Table 6.1 shows the assumptions of the numerical evaluation. The call arriving process and the cell dwell times are assumed to be Poisson. The average cell dwell time is assumed to be 240 sec ([52]).

**Table 6.1:** The basic assumptions for the numerical analysis

| Assumptions for the traffic classes | | | | |
|---|---|---|---|---|
| Service type | Traffic class (m) | Requested bandwidth by each call $\beta_{m,r}$ | $\gamma_{m,n}$ | $\gamma_{m,h}$ |
| Real-time services | Conversational voice (m=1) | 25 kbps | 0 | 0 |
|  | Conversational video (m=2) (Live streaming) | 128 kbps | 0 | 0 |
|  | Real-time game gaming (m=3) | 56 kbps | 0 | 0 |
| Non-real-time services | Buffered streaming video (m=4) | 128 kbps | 0.4 | 0.6 |
|  | Voice messaging (m=5) | 13 kbps | 0.2 | 0.3 |
|  | Web-browsing (m=6) | 56 kbps | 0.2 | 0.5 |
|  | Background (m=7) | 56 kbps | 0.5 | 0.8 |
| **Assumptions for the traffic environment** | | | | |
| Average call duration at requested bandwidth ($T(\vec{\beta}_r)$) | | 120 sec | | |
| The average user's speed | | 7.5 km/hr | | |
| The cell radius | | 1 km | | |
| The average file size of background traffic | | 6 Mbit | | |
| $a_1: a_2: a_3: a_4: a_5: a_6: a_7$ | | 0.35:0.1:0.05:0.15:0.1:0.15:0.1 | | |

Fig. 6.6 shows that the proposed prioritized bandwidth-adaptive scheme can reduce the handover call dropping probability (HCDP) to less than 0.0005, even for very large traffic load. This HCDP is also smaller than the corresponding value of the "Hard-QoS with 5% guard band scheme" and almost equal to the corresponding value of the "AQoS handover priority scheme". Moreover, in the same scenario, the "Hard-QoS scheme", which operates without any guard band, causes significantly larger call dropping probability. Fig. 6.7 shows that the proposed scheme mildly increases the call blocking probability, but this call blocking probability is still smaller than that of the "Hard-QoS with 5% guard band scheme" and the "AQoS handover priority scheme". Indeed, the proposed scheme significantly decreases the call dropping probability at the expense of mildly increasing call blocking probability. Nevertheless, Fig. 6.8 shows that the bandwidth utilization of the proposed scheme is maximized. The bandwidth utilization for the "Hard-QoS with 5% guard band scheme" is very poor. Also the "AQoS handover priority scheme" cannot maximize the bandwidth utilization especially for the low and medium traffic condition.

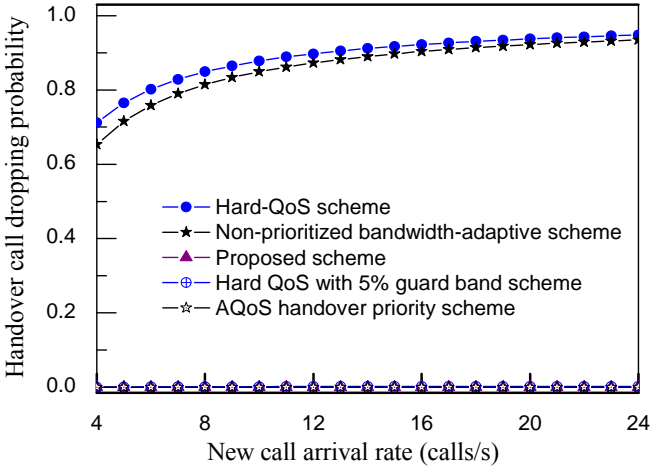

**Fig. 6.6:** Comparison of handover call dropping probability in heavy traffic conditions.

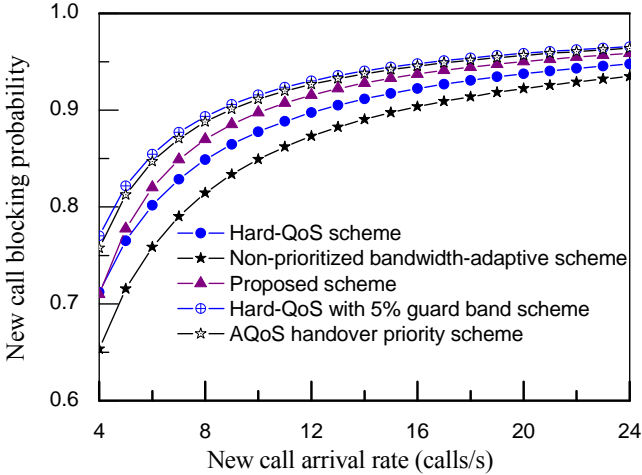

**Fig. 6.7:** Comparison of new call blocking probability in heavy traffic conditions.

The average number of handovers is also an important performance evaluation metric. The number of handovers is mainly related to the call blocking probability and the average call duration. As we have pointed out previously, it is commonly accepted that it is preferable to admit less calls, but to reduce the number of calls that are prematurely terminated (i.e., the dropping probability should be less than the blocking probability.) Fig. 6.9 shows that the proposed scheme results in somewhat additional handovers than the "Hard-QoS scheme", the "Hard-QoS with 5% guard band scheme," and the "AQoS handover priority scheme". But, at the same time, the proposed scheme also results in significantly less handovers compared to the "Non-prioritized bandwidth-adaptive scheme". The "Non-prioritized bandwidth-adaptive scheme" unnecessarily accepts too many new calls, causing longer call duration of some non-real-time traffic (e.g., background download traffic). The overall forced call termination probability is another key performance parameter. Fig. 6.10 shows that the "Non-prioritized bandwidth-adaptive scheme" can provide the lowest overall forced call termination probability. However, the proposed scheme also provides nearly equal overall forced call termination probability. The other schemes provide significantly higher overall forced call termination probability.

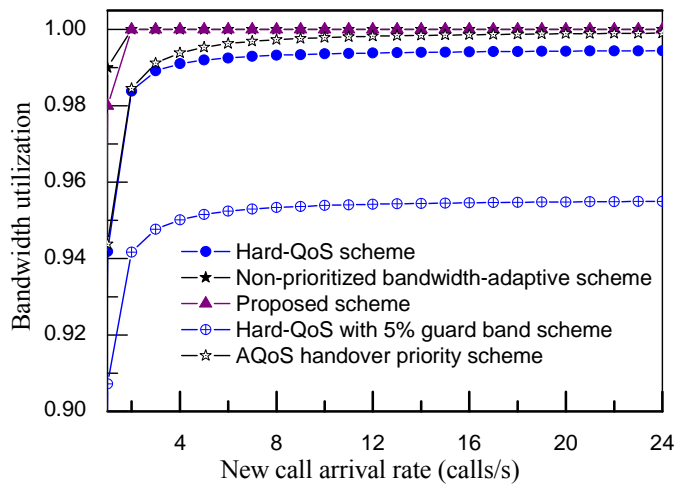

**Fig. 6.8:** Comparison of bandwidth utilization

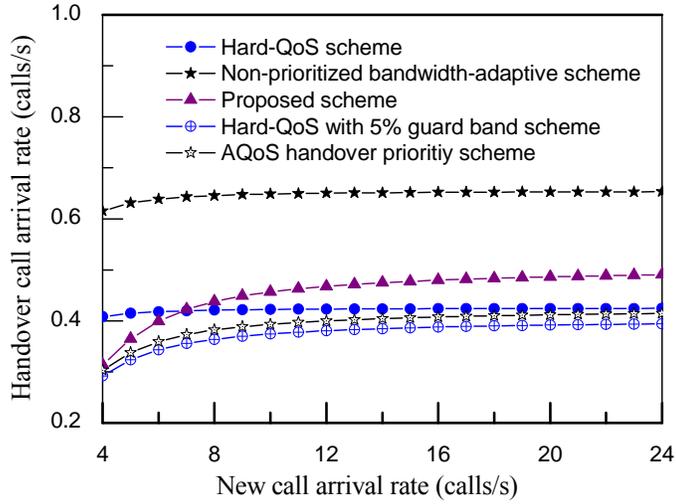

**Fig. 6.9:** Comparison of handover rates

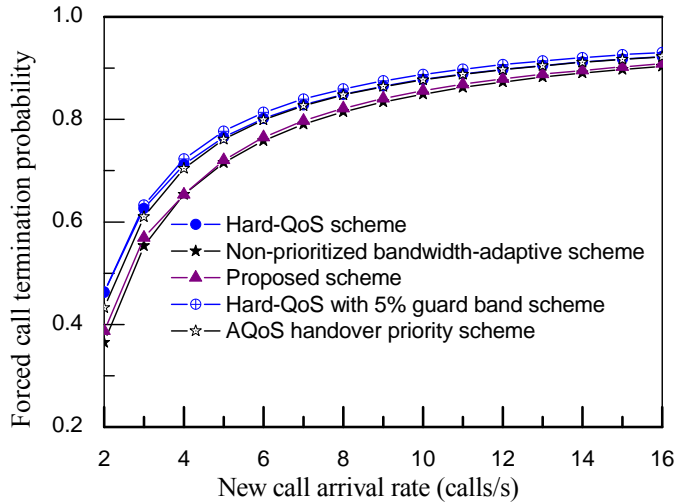

**Fig. 6.10:** Comparison of overall forced call termination probability

The numerical results from Fig. 6.6 − Fig. 6.10 demonstrate that, compared to the "Non-prioritized bandwidth-adaptive scheme" in which $\gamma_{m,n} = \gamma_{m,h}$, the proposed scheme supports negligible HCDP, about the same bandwidth utilization, and nearly equal overall forced call termination probability, even though the proposed scheme blocks a few more new calls. Although the "Hard-QoS with 5% guard band scheme" offers very small HCDP as well (alas, not less than our proposed scheme), however this scheme also causes very high call blocking probability. Our scheme offers about 4% more bandwidth utilization compared to the "Hard-QoS with 5% guard band scheme". Compared to the "AQoS handover priority scheme" in which $\gamma_{m,n} = 0$, the proposed scheme provides nearly equal HCDP, less new call blocking probability, better bandwidth utilization, and less overall forced call termination probability. In summary, the proposed scheme outperforms all the other schemes discussed in this paper.

# Chapter 7
# QoS Adaptive Radio Resource Allocation for Scalable Videos over Wireless Cellular Networks

Video over wireless networks is one of the fastest-growing data applications and mobile TV has become popular as it promises to deliver video contents to users whenever they want and wherever they are. Mobile TV has already proved to be a very promising ARPU (average revenue per user) generator for cellular operators [54]. In case of mobile TV, users can enjoy video services anywhere even with full mobility support through the several access networks. Worldwide Interoperability for Microwave Access (WiMAX) is a typical example of an emerging wireless network system to support high data rate services. The Mobile WiMAX (802.16e) is capable of providing high data rate with QoS mechanisms, making the support of mobile TV very attractive [55]. Therefore, with the rapid improvement of various wireless network technologies, now it is possible to provide high quality video transmission along with the voice, internet, and other background traffic over the wireless networks. Most of the existing wireless network technologies such as femtocell [6], Wi-Fi, Mobile WiMAX, 3G, and 4G support multicast/broadcast mechanisms [56]-[60]. However, the wireless bandwidth is still insufficient to support huge voice, data, and video services with full QoS especially for the congested traffic condition.

Good quality video services always require higher bandwidth. Hence, to provide the video services e.g., MBS and unicast services along with the existing voice, internet, and other background traffic services over the wireless cellular networks, it is essential to efficiently handle the wireless bandwidth in order to ensure the admission of more number of calls in the system during the congested traffic condition, to maximize the overall service quality, and to maximize the review. Fixed bandwidth allocation for the MBS sessions has several disadvantages. Whenever the MBS sessions are provided with the fixed bandwidth and minimum qualities, then the qualities of the MBS videos are lowest even if the traffic condition is very low. Therefore, for this condition, bandwidth utilization is also low. Similarly when the MBS sessions are provided with the fixed bandwidth and maximum qualities, then the MBS videos are provided with the best qualities but due to the shortage of the wireless bandwidth for the non-MBS traffic calls (e.g., voice, unicast video, internet, and other background traffic), the overall forced call termination probability and the handover call dropping probability are increased. Also the revenue is decreased for the operators.

SVC is an excellent solution to the problems raised by the diverse characteristics of high data rate video transmission through the wireless link. The SVC allows the elimination of some parts of the video bit stream in order to adapt it to the various needs or preferences of end users as well as to varying terminal capabilities or network conditions [61]. Therefore, scalable video technique [56], [61]-[63] allows the variable bit rate video broadcast/multicast/unicast over wireless networks. This technique utilizes multiple layering. Each of the layers improves spatial, temporal, or visual quality of the rendered video to the user [56]. Base layer or the highest priority layer guarantees the minimum quality of a video stream. Whereas the addition of enhanced layers or low priority layers improves the video quality. The number of layers for a video session (program) and the bandwidth per layer can be manipulated dynamically. Thus, to broadcast/multicast/unicast videos through a wireless environment, layered transmission is an effective approach for supporting heterogeneous receivers with varying bandwidth requirements [63]. Hence, if the system bandwidth is not sufficient to allocate the demanded bandwidth for all of the broadcasting/multicasting/unicasting video sessions, it is possible to reduce the bandwidth allocation for each of the video sessions. The QoS adaptability [11], [41], [42], [49], [50] of some multimedia traffic types is also an important technique for wireless communication to increase the admission of more number of calls in the system. This technique can be applied to support more number of calls during the congested traffic condition for the MBS supported wireless cellular networks.

## 7.1 System Model

The proposed scheme is based on the dynamic bandwidth allocation for the MBS sessions and non-MBS traffic calls. The allocated bandwidth for the MBS sessions is based on the system capacity and the traffic congestion. For the lower traffic condition, the MBS sessions are provided with the maximum demanded bandwidth for each of the sessions. With the increasing of the traffic arrival rates, the total allocated bandwidth for the MBS sessions is decreased. However, the minimum bandwidth is allocated for the MBS sessions to guarantee the minimum video qualities. The bandwidth allocation for the background traffic is also reduced with the increase of the traffic congestion. The bandwidth allocation for the unicast video calls is decreased only for highly congested traffic conditions. We consider non-MBS traffic (e.g., voice, unicast, and background) and MBS sessions. The MBS sessions are always active and they are provided with at least minimum amount of bandwidth. We give priority for each of the traffic types. The handover calls of any types of calls are considered as highest priority calls. The next priority is given to voice and the unicast video calls. The background traffic and the MBS sessions are given lowest priority. Fig. 7.1 shows the basic concept of bandwidth allocations for the MBS sessions and the non-MBS traffic calls. For the low traffic condition, all calls are provided with the maximum qualities. However, for the congested traffic condition, the bandwidth allocation for the MBS sessions is decreased. Based on the bandwidth allocation policy for different traffic calls, the system allocates bandwidth for the non-MBS traffic calls. Then, the remaining bandwidth is allocated for the MBS sessions. Suppose $C_{max,B}$ and $C_{min,B}$ are, respectively, the maximum allowable bandwidth and the minimum allocated bandwidth for the active MBS video sessions. $C_{max,nB}$ and $C_{min,nB}$ are, respectively, the maximum allowable bandwidth and the minimum allowable bandwidth for the non-MBS traffic calls. The bandwidth $C_{max,B}$ is provided to MBS sessions only if the allocated bandwidth for the non-MBS traffic calls is less than or equal to $C_{min,nB}$. Fig. 7.2 shows the variation of bandwidth allocations for the MBS sessions and the non-MBS traffic calls with the increasing of demanded bandwidth by the non-MBS traffic calls. During the lower traffic condition, all the the MBS sessions and non-MBS traffic calls are provided with the maximum demanded bandwidth for each of them. When the total demanded bandwidth exceeds the system capacity, the system reduces the bandwidth allocation both for the MBS sessions and non-MBS traffic calls. The reduction of bandwidth for the non-MBS traffic calls depends on the number of running QoS adaptive calls.

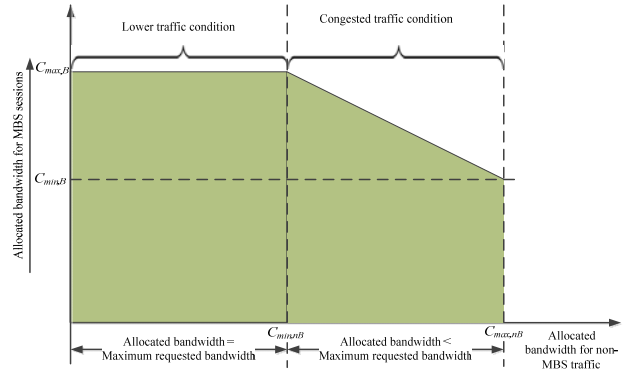

**Fig. 7.1:** Basic concept of bandwidth allocations among the MBS sessions and the non-MBS traffic calls.

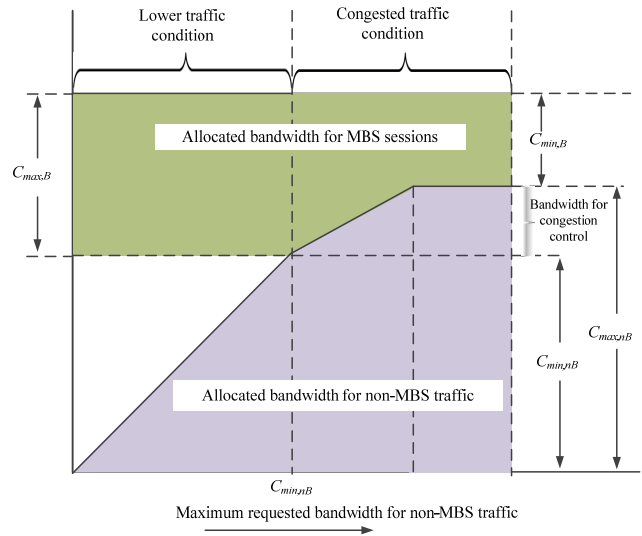

**Fig. 7.2:** Variation of bandwidth allocations for the MBS sessions and the non-MBS traffic calls.

## 7.2 Bandwidth Allocation and Adaptation

Efficient bandwidth allocation for the video sessions in broadcasting/multicasting/unicasting and for the other voice or data traffic is needed to make the best usage of the scare resources of wireless networks. An easy and straightforward approach is that all of the active broadcasting/multicasting video sessions and non-MBS traffic calls are provided by the requested bandwidth. However, such approach is not effective to serve huge traffic calls. Therefore, we propose an efficient bandwidth allocation scheme that makes the best utilization of the wireless bandwidth. The proposed scheme allows reclaiming some of the allocated bandwidth from already admitted QoS degradable traffic calls (e.g., background traffic and unicast video) and MBS sessions, as to accept more handover and new calls, when the system's resources are running low. Consequently, the scheme can accommodate more calls.

Fig. 7.3 shows the procedure of bandwidth degradation for the proposed scheme. The proposed scheme gives highest priority for any kinds of handover calls. Suppose $C_{req,max}$ and $C_{req,min}$ are, respectively, the maximum and the minimum required bandwidths for a requested call. The system accommodates a handover call if it can manage $C_{req,min}$ amount of bandwidth only. However, for a new arrival calls, it is equal to $C_{req,max}$. The qualities of the unicast video calls are degraded only to accept handover calls in the system. The overall resource allocation scheme is divided into four categories based on the traffic characteristics. The resource allocation and QoS adaptation for each of the traffic types is different.

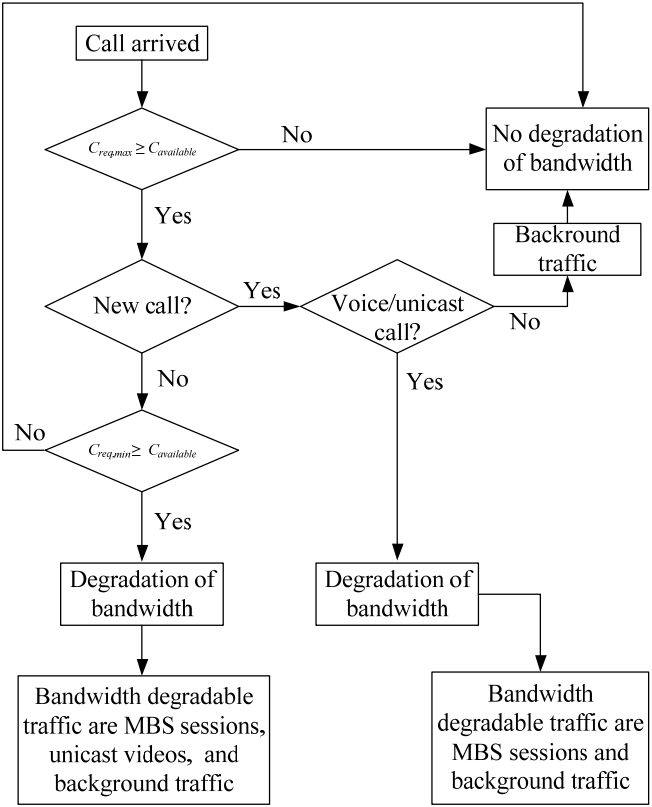

**Fig. 7.3:** Procedure of bandwidth degradation of the system traffic.

### 7.2.1 Voice Traffic

The proposed scheme gives highest priority for the voice calls followed by the handover calls. The QoS of this traffic is non-adaptive. Hence, the bandwidth allocation for the voice traffic is strict. However, the QoS levels of other classes of traffic are degraded to admit a voice call. Suppose $\beta_v$, $\beta_{min,v}$, and $\beta_{max,v}$ are, respectively, the currently allocated, minimum allocated, and maximum allocated bandwidths for a voice call. The bandwidth relation for the voice calls is found as:

$$\beta_v = \beta_{min,v} = \beta_{max,v} \tag{1}$$

### 7.2.2 Unicast Video (Scalable)

The unicast calls are also given same priority as the voice calls. The QoS of this traffic is adaptive. The number of enhanced layers is reduced to accommodate more handover calls in the system. The QoS levels of other classes of traffic are also degraded to admit a unicast video call. The unicast traffic calls shows the following characteristics for the proposed scheme.

- Same priority as the voice calls
- Multi-level of bandwidth allocations
    - ✓ Maximum allocated bandwidth for best quality (Maximum number of enhanced layers)
    - ✓ Minimum allocated bandwidth for guaranteed quality (Minimum number of enhanced layers)
- Number of enhanced layer can be controlled to accommodate more handover calls in the system
- QoS level of MBS sessions and the background traffic calls are degraded to admit a unicast call

The bandwidth relations for the unicast video calls are expressed as follows:

$$\beta_{min,uni} = \beta_{0,uni} + \beta_{1,uni} + \cdots + \beta_{K_{min},uni} \tag{2}$$

$$\beta_{max,uni} = \beta_{0,uni} + \beta_{1,uni} + \cdots + \beta_{K_{min},uni} + \cdots + \beta_{K_{max},uni} \tag{3}$$

$$\beta_{uni} = \beta_{0,uni} + \beta_{1,uni} + \cdots + \beta_{K_{min},uni} + \cdots + \beta_{k,uni} \tag{4}$$

where $β_{uni}$, $β_{min,uni}$, and $β_{max,uni}$ are, respectively, the currently allocated, minimum allocated, and maximum allocated bandwidths for a unicast video calls. $β_{0,uni}$ is the allocated bandwidth for the base layer of a unicast video call. $K_{max}$ and $K_{min}$ are, respectively, the maximum and the minimum numbers of supported enhanced layers for each of the unicast video calls. $β_{k,uni}$ is the required bandwidth for the *k-th* layer of a unicast call.

### *7.2.3 Multicast/Broadcast Video (Scalable)*

The QoS of this traffic is adaptive. The numbers of enhanced layers are reduced to accommodate more handover calls as well as new voice and unicast video calls. The MBS sessions show the following characteristics for the proposed scheme.

- Multiple number of videos are continuously broadcasted/ multicasted
- Multi-level of bandwidth allocations
    - ✓ Maximum allocated bandwidth for best quality (Maximum number of enhanced layers)
    - ✓ Minimum allocated bandwidth for guaranteed quality (Minimum number of enhanced layers)
- Number of enhanced layers is controllable
    - ✓ To accommodate more unicast video calls (new originating and handover calls)
    - ✓ To accommodate more voice calls (new originating and handover calls)
    - ✓ To accommodate handover calls of background traffic
- Different level of bandwidth allocation for different multicasting/broadcasting video sessions
    - ✓ Based on the priority of video sessions, bandwidth can be allocated for each of them

The bandwidth relations for the MBS video sessions are expressed as follows:

$$β_{min,m} = β_{0,m} + β_{1,m} + \cdots + β_{N_{min,m},m} \tag{5}$$

$$β_{max,m} = β_{0,m} + β_{1,m} + \cdots + β_{N_{min,m},m} + \cdots + β_{N_{max,m},m} \tag{6}$$

$$β_{B,m} = β_{0,m} + β_{1,m} + \cdots + β_{N_{min,m},m} + \cdots + β_{N_m,m} \tag{7}$$

$$C_{min,B} = \beta_{min,1} + \beta_{min,2} + \cdots + \beta_{min,m} + \cdots + \beta_{min,M}$$
$$= \sum_{m=1}^{M} \beta_{0,m} + \sum_{m=1}^{M} \sum_{n=1}^{N_{min,m}} \beta_{n,m} \quad (8)$$

$$C_{max,B} = \beta_{max,1} + \beta_{max,2} + \cdots + \beta_{max,m} + \cdots + \beta_{max,M}$$
$$= \sum_{m=1}^{M} \beta_{0,m} + \sum_{m=1}^{M} \sum_{n=1}^{N_{min,m}} \beta_{n,m} + \sum_{m=1}^{M} \sum_{n=N_{min,m}+1}^{N_{max,m}} \beta_{n,m} \quad (9)$$

where $\beta_{B,m}$, $\beta_{min,m}$, and $\beta_{max,m}$ are, respectively, the currently allocated, minimum allocated, and maximum allocated bandwidths for *m-th* MBS session. $\beta_{0,m}$ is the allocated bandwidth for the base layer of the *m-th* MBS session. $N_{max,m}$ and $N_{min,min}$ are, respectively, the maximum and the minimum numbers of supported enhanced layers for the *m-th* MBS session. $\beta_{n,m}$ is the required bandwidth for the *n-th* layer of the *m-th* MBS session. *M* is the number of active MBS sessions.

If $C_B$ and $C_{nB}$ are, respectively the allocated bandwidths for the MBS sessions and the non-MBS traffic calls, then the lower traffic condition is defined as $C - C_{nB} \geq \sum_{m=1}^{M} \sum_{n=0}^{N_{max,m}} \beta_{n,m}$. For this condition, the allocated bandwidth for the non-MBS traffic calls is less than or equal to the $C_{min,nB}$. Therefore, all the MBS sessions are provided with the maximum allowable bandwidth. The allocated bandwidth for the MBS sessions during this traffic condition is calculated as,

$$C_B = C_{max,B} = \sum_{m=1}^{M} \sum_{n=0}^{N_{max,m}} \beta_{n,m} \quad (10)$$

$$\beta_{B,m} = \beta_{max,m} = \beta_{0,m} + \beta_{1,m} + \cdots + \beta_{N_{min,m},m} + \cdots + \beta_{N_{max,m},m} \quad (11)$$

Congested traffic condition is defined as, $C - C_{nB} < \sum_{m=1}^{M} \sum_{n=0}^{N_{max,m}} \beta_{n,m}$. For this condition, the allocated bandwidth for the non-MBS traffic call is greater than $C_{min,nB}$. Therefore, all the MBS sessions are not provided with the maximum allowable bandwidth. For the congested traffic condition, the allocated bandwidth for the MBS sessions is calculated using two separate proposed techniques. For each of the techniques, the total allocated bandwidths for the MBS sessions are equal. However, the allocated bandwidths for different active sessions are different.

***Techniques 1 (Two level bandwidth allocation):*** This technique is applicable when the video qualities of all the MBS sessions need to be equally degraded. The proposed scheme provides almost equal degradation of MBS video qualities. The maximum difference between the reduced numbers of enhanced layers for two MBS sessions is one. Therefore, if the reduced number of enhanced layers for the most popular video session is $P$, the reduced number of enhanced layers for the lowest popular video session is either $P$ or $(P+1)$. The bandwidth for each of the MBS sessions is calculated as:

$$C_B = \sum_{m=1}^{M_1} \sum_{n=0}^{N_{max,m}-P} \beta_{n,m} + \sum_{m=M_1+1}^{M} \sum_{n=0}^{N_{max,m}-P-1} \beta_{n,m} \qquad (12)$$

$$\beta_{B,m} = \begin{cases} \beta_{0,m} + \beta_{1,m} + \cdots + \beta_{N_{min,m},m} + \cdots + \beta_{N_{max,m}-P,m}, & 1 \leq m \leq M_1 \\ \beta_{0,m} + \beta_{1,m} + \cdots + \beta_{N_{min,m},m} + \cdots + \beta_{N_{max,m}-P-1,m}, & M_1 < m \leq M \end{cases} \qquad (13)$$

where $P$ is the minimum number of enhanced layers that must be removed from every active MBS sessions due to the congested traffic. $M_1$ is the minimum number of MBS sessions for which $P$ number of enhanced layers are removed and for the remaining $(M-M_1)$ number of MBS sessions $(P+1)$ number of enhanced layers are removed.

$P$ is the only value that satisfies the following equation:

$$\frac{C - C_{nB}}{\sum_{m=1}^{M} \sum_{n=0}^{N_{max,m}-P-1} \beta_{n,m}} \geq 1 \text{ and } \frac{C - C_{nB}}{\sum_{m=1}^{M} \sum_{n=0}^{N_{max,m}-P} \beta_{n,m}} < 1 \qquad (14)$$

$M_1$ is the only value that satisfies the following equation:

$$\frac{C - C_{nB} - \sum_{m=1}^{M} \sum_{n=0}^{N_{max,m}-P-1} \beta_{n,m}}{\sum_{m=1}^{M_1} \beta_{(N_{max,m}-P),m}} \geq 1 \text{ and } \frac{C - C_{nB} - \sum_{m=1}^{M} \sum_{n=0}^{N_{max,m}-P-1} \beta_{n,m}}{\sum_{m=1}^{M_1+1} \beta_{(N_{max,m}-P),m}} < 1 \qquad (15)$$

***Techniques 2 (Multi-level bandwidth allocation):*** This technique is applicable when the MBS video qualities are provided according to the priority of each of the sessions. In this scheme, all of the enhanced layers for the lowest priority sessions are removed first. The bandwidth for the MBS sessions is calculated as:

$$C_B = \sum_{m=1}^{M_2} \sum_{n=0}^{N_{max,m}} \beta_{n,m} + \sum_{m=M_2+1}^{M} \sum_{n=0}^{N_{min,m}} \beta_{n,m} \qquad (16)$$

$$\beta_{B,m} = \begin{cases} \beta_{0,m} + \beta_{1,m} + \cdots + \beta_{N_{min,m},m} + \cdots + \beta_{N_{max,m},m}, & 1 \leq m \leq M_2 \\ \beta_{0,m} + \beta_{1,m} + \cdots + \beta_{N_{min,m},m} + \cdots + \beta_{N_{min,m},m}, & M_2 < m \leq M \end{cases} \qquad (17)$$

where $M_2$ is the minimum number of MBS sessions for which the system provides best quality and for the remaining ($M$-$M_2$) number of MBS sessions, the system provides minimum quality.

$M_2$ is the only value that satisfies the following equation:

$$\frac{C-C_{nB}}{\sum_{m=1}^{M}\sum_{n=0}^{N_{min,m}}\beta_{n,m}+\sum_{m=1}^{M_2}\sum_{N_{min,m}+1}^{N_{max,m}}\beta_{n,m}} \geq 1 \text{ and } \frac{C-C_{nB}}{\sum_{m=1}^{M}\sum_{n=0}^{N_{min,m}}\beta_{n,m}+\sum_{m=1}^{M_2+1}\sum_{N_{min,m}+1}^{N_{max,m}}\beta_{n,m}} < 1 \quad (18)$$

### *7.2.4 Background Traffic (e.g., file transfer)*

The QoS of background traffic is adaptive. The QoS adaptability [11], [41], [42], [49], [50] of these traffic allows the reclaiming of system resources to support more number of higher priority calls without reducing the bandwidth utilization. Two level of bandwidth adaptation for the background traffic calls is proposed. The first level (higher) is used to accommodate handover calls and the second one (lower) is used to accommodate more new calls in the system. The background traffic shows the following characteristics for the proposed scheme.

- Lowest priority
- QoS adaptive
- QoS level is degradable
    - ✓ To accommodate more unicast video calls (new originating and handover calls)
    - ✓ To accommodate more voice calls (new originating and handover calls)
    - ✓ To accommodate handover calls of background traffic

The bandwidth relations for the background traffic are expressed as follows:

$$\beta_{min,back(i)} = \beta_{hand,back(i)} = (1-\xi_i)\beta_{r,back(i)} \quad (19)$$

$$\beta_{max,back(i)} = \beta_{r,back(i)} \quad (20)$$

$$\beta_{new,back(i)} = (1-\xi'_i)\beta_{r,back(i)} \quad (21)$$

where $\beta_{min,back(i)}$, and $\beta_{max,back(i)}$ are, respectively, the minimum allocated and maximum allocated bandwidths for a background traffic call of *i-th* class. $\beta_{hand,back(i)}$ ($\beta_{new,back(i)}$) is the minimum required bandwidth to accept a handover (new) background traffic call of *i-th* class or minimum allocated bandwidth for each of the background traffic calls of class *i* after accepting any handover (new) calls. $\xi_i$ and $\xi_i'$ are, respectively, the maximum levels of bandwidths that can be degraded for a background traffic call of *i-th* class to accept a handover call and new call.

We may compare our proposed scheme with few other schemes whrere the allocated bandwidths for the MBS sessions are fixed. Few possible bandwidth allocation schemes are summarized as follows:

### Scheme #1 (proposed scheme):
- Proposed dynamic bandwidth allocation for the MBS sessions
- Priority based proposed dynamic bandwidth allocation for the non-MBS traffic calls

### Scheme #2:
- Fixed $C_{max,B}$ amount of bandwidth allocation for the MBS sessions
- Priority based proposed dynamic bandwidth allocation for the non-MBS traffic calls

### Scheme #3:
- Fixed $C_{max,B}$ amount of bandwidth allocation for the MBS sessions
- QoS degradable but non-prioritized bandwidth allocation for the non-MBS traffic calls

### Scheme #4:
- Fixed $C_{max,B}$ amount of bandwidth allocation for the MBS sessions
- Non-QoS degradable as well as non-prioritized bandwidth allocation for the non-MBS traffic calls

### Scheme #5:
- Fixed $C_{min,B}$ amount of bandwidth allocation for the MBS sessions
- Priority based proposed dynamic bandwidth allocation for the non-MBS traffic calls

### Scheme #6:
- Fixed $C_{min,B}$ amount of bandwidth allocation for the MBS sessions
- QoS degradable but non-prioritized bandwidth allocation for the non-MBS traffic

calls

### *Scheme #7:*

- Fixed $C_{min,B}$ amount of bandwidth allocation for the MBS sessions
- Non-QoS degradable as well as non-prioritized bandwidth allocation for the non-MBS traffic calls

Scheme #2 - scheme #4 provide always best quality for all the MBS sessions even for the very congested traffic condition. For these schemes $C_{max,B}$ amount of bandwidth is reserved for the MBS sessions. This $C_{max,B}$ amount of bandwidth is able to provide best quality videos for all the MBS sessions. These schemes can provide better bandwidth utilization but cannot improve the system performances in terms of overall forced call termination probability due to the reduced bandwidth allocation for the non-MBS sessions. Among these three schemes, only the scheme #2 can moderately improve the handover call dropping probability performance due to the presence of the proposed priority scheme for the non-MBS traffic calls. The scheme #2 can also reduce the new call blocking probabilities for the voice and the unicast traffic calls because of the priority based admission control but these reduced new call blocking probabilities are not significant. For the scheme #3 and scheme #4, the handover call dropping probability is very high because of the non-prioritized call admission control. Scheme #4 provides lowest number of call admission in the system.

Scheme #5 - scheme #7 provide always lowest quality for all the MBS sessions even for the very low traffic condition. For these schemes only $C_{min,B}$ amount of bandwidth is reserved for the MBS sessions. This $C_{min,B}$ amount of bandwidth is able to provide only the lowest quality videos for all the MBS sessions. These schemes cannot perform well in terms of bandwidth utilization for the lower traffic condition. For the lower traffic condition, even the bandwidth is empty but the MBS sessions are provided with the lowest qualities. Scheme #5 and scheme #6 can maximize the number of call admission because of the increased bandwidth for the non-MBS sessions and the presence of the QoS degradation policy for the non-MBS traffic calls. Among these three schemes, only the scheme #5 can significantly improve the handover call dropping probability performance due the increased bandwidth for the non-MBS sessions and the presence of the proposed priority based QoS degradation policy for the non-MBS traffic calls.

The scheme #5 can also significantly reduce the new call blocking probabilities for the voice and the unicast traffic calls. For the scheme #6 and scheme #7, the handover call dropping probability is very high because of the non-prioritized call admission control.

Each of the schemes from scheme #2 – scheme #7 has some limitations. Therefore, we propose a scheme (scheme #1) that optimizes all the limitations and provides efficient utilization of the wireless bandwidth. Our scheme can provide maximum number of call admission, significantly reduced handover call dropping probability, significantly reduced new call blocking probabilities for the voice and the unicast traffic calls, and maximized bandwidth utilization.

## 7.3 Queuing Model

Our proposed scheme can be modeled using *M/M/K/K* queuing analysis. The resource allocation policy is varied with the increase of system states. Fig. 7.4. shows the state transition rate diagram for the proposed resource allocation scheme. We define $1/\mu$ as the average channel holding time (exponentially distributed). In this figure, $\lambda_n$ and $\lambda_h$ are, respectively, the total new call arrival rate and the handover call arrival rate. $\lambda_{n,x}$ is the new call arrival rate of *x* ( *v* or *uni* or *b*) type of traffic calls (*v* for voice, *uni* for unicast, and *b* for the background traffic calls). The *M* number of MBS sessions are continuously provided. Therefore, the minimum number of states in the system is *M*. The QoS degradation status for a traffic type is also obtainable form the state of the system.

The probability that the system is in state *i,* is given by $P_i$. From the Fig. 7.4, the state balance equations are expressed as (22). In the proposed scheme, a new voice call or a unicast video call is blocked if the system is in the state (*N+L*) or larger and a new background traffic call is blocked if the system is in the state *N* or larger. However, a handover call is dropped only if the system is in the state (*N+S*). Thus, the handover call dropping probability ($P_D$) is calculated using (23). The call blocking probability of an originating new voice call ($P_{B,v}$) or unicast call ($P_{B,uni}$) can be computed using (24). Finally, the call blocking probability of an originating new background traffic call ($P_{B,back}$) can be computed using (25).

$$\begin{aligned}
(i-M)\mu P_i &= \lambda_T P_{i-1}, & M \leq i \leq N \\
(i-M)\mu P_i &= (\lambda_{n,v} + \lambda_{n,uni} + \lambda_h) P_{i-1}, & N \leq i \leq N+L \\
(i-M)\mu P_i &= \lambda_h P_{i-1}, & N+L \leq i \leq N+S
\end{aligned} \quad (22)$$

$$P_D = P_{N+S} = \frac{\lambda_T^{N-M} (\lambda_{n,v} + \lambda_{n,uni} + \lambda_h)^L (\lambda_h)^{S-L}}{\mu^{N+S-M}(N+S-M)!} P_M \quad (23)$$

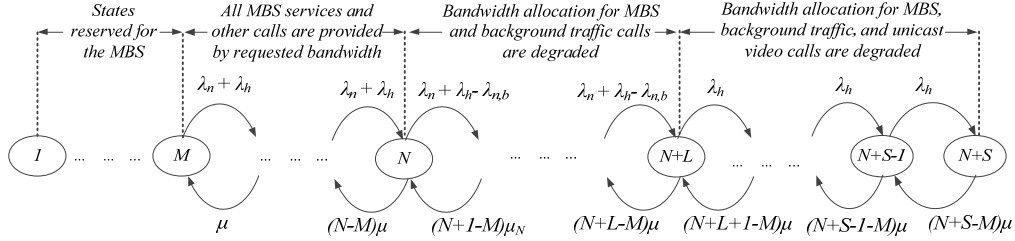

**Fig. 7.4:** State transition rate diagram for the proposed resource allocation scheme.

$$P_{B,v} = P_{B,uni} = \sum_{i=N+L}^{N+S} P_i = \sum_{i=N+L}^{N+S} \frac{(\lambda_T)^{N-M}(\lambda_{n,v}+\lambda_{n,uni}+\lambda_h)^L (\lambda_h)^{i-N-L}}{\mu^{i-M}(i-M)!} P_M \quad (24)$$

$$P_{B,back} = \sum_{i=N}^{N+S} P_i = \sum_{i=N+L}^{N+S} P_i + \sum_{i=N}^{N+L-1} P(i) = P_{B,v} + \sum_{i=N}^{N+L-1} \frac{(\lambda_T)^{N-M}(\lambda_{n,v}+\lambda_{n,uni}+\lambda_h)^{i-N}}{\mu^{i-M}(i-M)!} P_M \quad (25)$$

$$P_M = \left[ \sum_{i=M}^{N} \frac{(\lambda_T)^{i-M}}{\mu^{i-M}(i-M)!} + \sum_{i=N+1}^{N+L} \frac{(\lambda_T)^{N-M}(\lambda_{n,v}+\lambda_{n,uni}+\lambda_h)^{i-N}}{\mu^{i-M}(i-M)!} + \sum_{i=N+L+1}^{N+S} \frac{(\lambda_T)^{N-M}(\lambda_{n,v}+\lambda_{n,uni}+\lambda_h)^L \lambda_h^{i-N-L}}{\mu^{i-M}(i-M)!} \right]^{-1} \quad (26)$$

## 7.4 Performance Analysis

In this section, we present the results of the numerical analysis of the proposed scheme. We compare the performance of our proposed scheme with the performance of the "fixed bandwidth allocation for MBS sessions" schemes. Table 7.1 shows the assumptions of the summary of the parameter values used in analysis. The call arriving process and the cell dwell times are assumed to be Poisson. The average cell dwell time is assumed to be 540 sec (exponentially distributed) [52].

**Table 7.1:** Summary of the parameter values used in analysis

| Parameter | Value |
|---|---|
| Bandwidth capacity ($C$) | 20 Mbps |
| Required bandwidth for each of the voice calls ($\beta_v$) | 64 kbps |
| Maximum allocated bandwidth for each of the unicast video calls ($\beta_{max,uni}$) | 0.5 Mbps |
| Maximum number of enhanced layers for each of the unicast calls ($K_{max,uni}$) | 10 |
| Minimum number of enhanced layers for each of the unicast calls ($K_{min,uni}$) | 0 |
| Allocated bandwidth for each of the enhanced layers of the unicast calls ($\beta_{m,uni}$) | 20 kbps |
| Maximum allocated bandwidth for each of the MBS sessions ($\beta_{max,m}$) | 1 Mbps |
| Minimum allocated bandwidth for each of the MBS video sessions ($\beta_{min,m}$) | 0.5 Mbps |
| Maximum number of enhanced layers for each of the MBS sessions ($N_{max,m}$) | 10 |
| Minimum number of enhanced layers for each of the MBS sessions ($N_{min,m}$) | 0 |
| Allocated bandwidth for each of the enhanced layers of each of the MBS video sessions ($\beta_{m,uni}$) | 50 kbps |
| Number of MBS sessions ($M$) | 12 |
| Maximum required/allocated bandwidth for each of the background traffic calls ($\beta_{max,back}$) | 120 kbps |
| Minimum required/allocated bandwidth for each of the background traffic calls ($\beta_{min,back}$) | 60 kbps |
| Maximum portion of bandwidth that can be degraded for a background traffic call of $i$-th class ($\xi_i$) | 0.5 |
| Maximum portion of bandwidth that can be degraded for a background traffic call of $i$-th class to accept a new originating call ($\xi_i'$) | 0.3 |
| Average call duration time considering all calls (exponentially distributed) | 120 sec |
| Ratio of call arrival rates (voice: unicast call: background traffic ) | 5:1:4 |

Fig. 7.5 shows the variation of bandwidth allocations for the MBS sessions and the non-MBS traffic calls with the increase of call arrival rate for our proposed scheme. It shows that the bandwidth allocation for the MBS sessions is decreased with the increase of demanded bandwidth by the non-MBS traffic calls. However, the minimum allocated bandwidth for the MBS sessions is 6 Mbps. Therefore, the maximum allocated bandwidth for the the non-MBS traffic calls is 14 Mbps. Fig. 7.6 shows the number of provided enhanced layers for a MBS session and a unicast video call with the increase of call arrival rate. Its shows that the proposed system provides less quality degradation of unicast video calls compared to the MBS video sessions. The bandwidth allocation for the MBS sessions using "two level technique" causes maximum difference of one enhanced layer between the highest priority and the lowest priority sessions. The bandwidth allocation for the MBS sessions using "multi-level technique" causes difference of zero to ten enhanced layers between the highest priority and the lowest priority sessions.

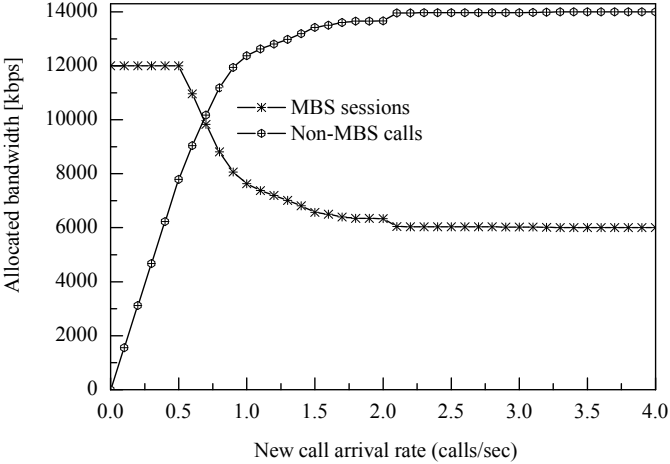

**Fig. 7.5:** The variation of bandwidths allocations for the MBS sessions and the non-MBS traffic calls with the increase of call arrival rate.

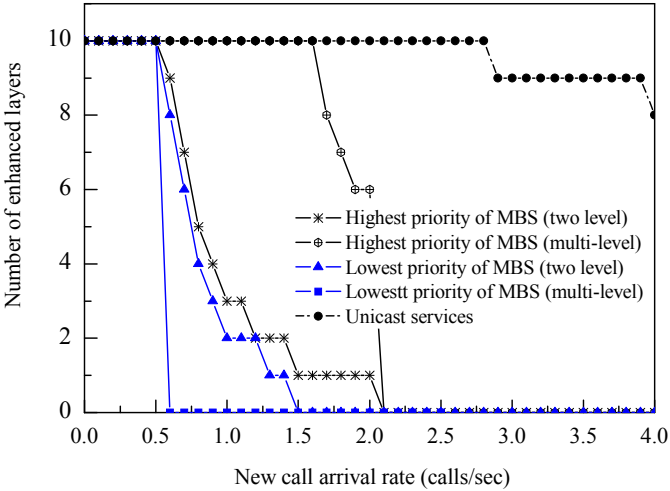

**Fig. 7.6:** Number of provided enhanced layers for a MBS session and a unicast video call with the increase of call arrival rate.

Fig. 7.7 shows that the proposed scheme provides negligible handover call dropping probability even for very high traffic condition. The scheme #2 is also based on the QoS adaptive and priority of traffic classes. However, the reduced maximum bandwidth allocation for the non-MBS traffic calls causes the higher handover call dropping probability. Even the QoS adaptation policy is applicable for the scheme #6 but it causes very high handover call dropping probability due the non-priority of traffic classes. Among the scheme #2 – scheme #7, only the scheme #5 can provide negligible handover call dropping probability but the MBS sessions are always provided with the minimum qualities for this scheme. The scheme #3 and scheme 6 causes very high handover call dropping probability and new call blocking probabilities for voice and unicast calls due the the non-priority of traffic classes. The performance of scheme #4 is poorer than scheme #3 and performance of scheme #7 is poorer than scheme #6 in terms of handover call dropping probability for all traffic types and new call blocking probabilities for voice and unicast calls because these two schemes do not support QoS adaptability and the priority of traffic classes. Fig. 7.8 shows that our proposed scheme provides comparatively lower new call blocking probabilities for the voice and unicast traffic calls. Our scheme provides only higher new call blocking probability for the background traffic calls but that is still lower than the scheme #2. The scheme #2 cannot significantly reduce the new call blocking probabilities for the voice and unicast traffic calls due to the reduced maximum bandwidth allocation for the non-MBS traffic calls. Fig. 7.9 shows the overall forced call termination probability performance comparison. Our proposed scheme provides best performance due to the dynamic nature of bandwidth allocation both for the MBS sessions and the non-MBS traffic calls. The scheme #2, scheme #3, and scheme #4 cannot improve the overall forced call termination performance due to the reduced maximum bandwidth allocation for the non-MBS traffic calls. Among the scheme #2 – scheme #7, only the scheme #5 and scheme #6 can maximize the number of call admission. However, the handover call dropping performance of the scheme #6 is poor. Scheme #7 performs poorer than scheme 6 and scheme #5 in terms of overall forced call termination probability because this scheme does not support QoS adaptability. During the lower traffic condition, the proposed scheme allocates higher bandwidth for the MBS sessions. The proposed scheme reduces the bandwidth allocation for the MBS sessions for the congested traffic condition only. Hence, our scheme effectively uses the system bandwidth. Fig. 7.10 shows the bandwidth utilization comparison. Even though the scheme #5 can maximize the number of call admission but bandwidth utilization is poor for this scheme especially for the lower traffic condition. Therefore, the proposed scheme maximizes bandwidth utilization also.

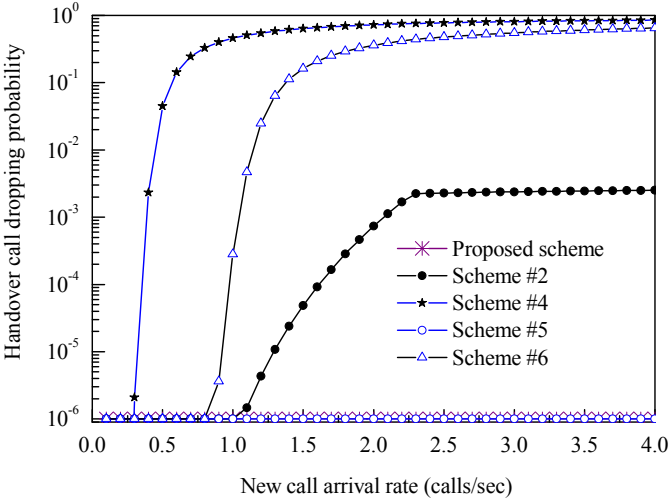

**Fig. 7.7:** Comparison of handover call dropping probability.

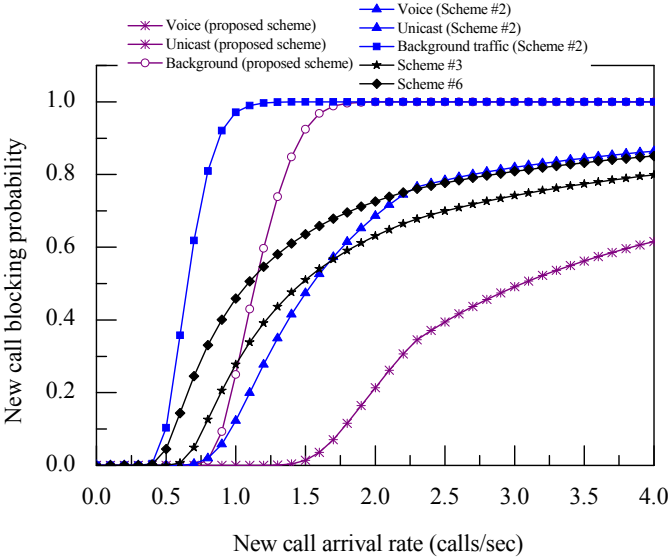

**Fig. 7.8:** Comparison of new originating call blocking probability.

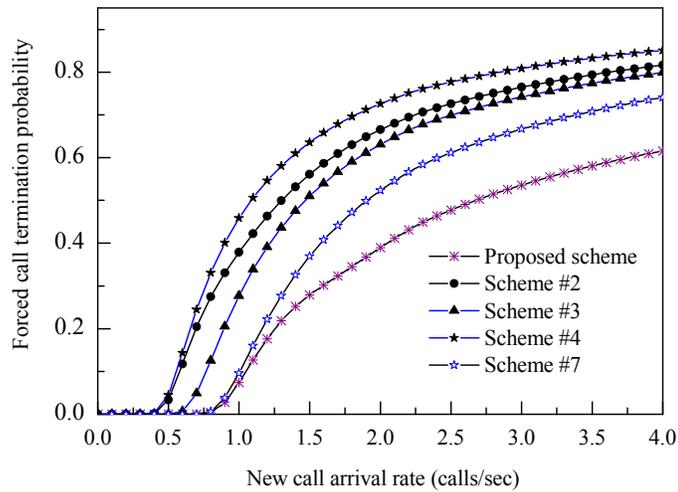

**Fig. 7.9:** Comparison of overall forced call termination probability.

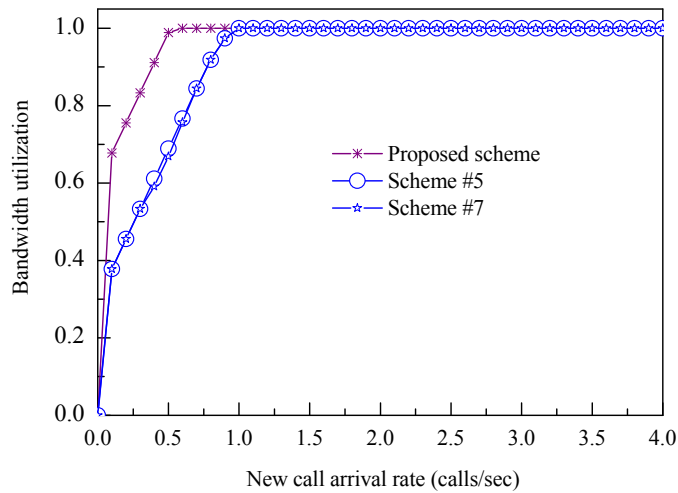

**Fig. 7.10:** Comparison of bandwidth utilization.

The results in Figs. 7.7 – 7.10 show that the proposed scheme quite effective for the MBS services over the wireless networks. The only disadvantage of the proposed scheme is that the video quality of the MBS sessions and QoS level of background traffic calls are degraded during the congested traffic condition.

# Chapter 8

# Popularity based Bandwidth Allocation for Scalable Video over Wireless Networks

Due to the limited data rates of wireless networks, it is not possible to provide the best quality for the entire active broadcasting/multicasting video sessions. Hence, equal bandwidth allocation for all of the broadcasting/multicasting video sessions is an easy and simple method. The service qualities of all broadcasting/multicasting video sessions are equally degraded when the total wireless bandwidth is not sufficient to provide the maximum demanded bandwidth to all. Instead of allocating equal bandwidths to all of the sessions during an insufficient bandwidth condition, our proposed scheme efficiently allocates the total system bandwidth among them in such a way that larger bandwidth is allocated to the video session of higher popularity. Thus, the average user satisfaction level is increased significantly. However, a minimum quality for the lowest popular video session is guaranteed by assigning a minimum amount of bandwidth.

## 8.1 Proposed Bandwidth Allocation Scheme

Even though the effective bandwidth of wireless links is growing very rapidly, fully deployed 4G wireless network will not even enough to accommodate many best quality video services simultaneously. The wireless link will always have less bandwidth than the wired link and it will continue in further. Efficient bandwidth allocation for the video sessions in broadcasting/multicasting is needed to make the best usage of the scare resources of wireless networks. An easy and straightforward approach is that all of the active broadcasting/multicasting video sessions share the total system bandwidth equally. However, such approach is not sensible. Because a popular video program attracting a large number of subscribers should be allocated with more bandwidth compared to the less popular one, if allocation of total demanded bandwidth is not possible. Therefore, we propose an efficient bandwidth allocation scheme that makes the best utilization of the bandwidth.

Let the total system bandwidth capacity and the total number of active broadcasting/multicasting video sessions are $C$ and $M$, respectively. $\beta_{max}$ and $\beta_{min}$ are, respectively, the maximum allocated bandwidth and the minimum allocated bandwidth for each of the active broadcasting/multicasting video sessions. Then the allocated bandwidth for each of the active sessions in the equally shared bandwidth allocation scheme is:

$$\beta = \begin{cases} \beta_{max}, & \beta_{max}M \leq C \\ \dfrac{C}{M}, & \beta_{max}M > C \end{cases} \quad (1)$$

As we mentioned before, larger amount of allocated bandwidth for a video session makes the chance of increasing the number of enhanced layers and thus improving the video quality for that session. User satisfaction level depends on received video quality. Therefore, we assume that user satisfaction level is directly proportional to allocated bandwidth for a video session. User satisfaction level becomes maximum (equal to 1) when the demanded bandwidth ($\beta_{max}$) is allocated for a broadcasting/multicasting video session. The satisfaction level of a user in the equally shared bandwidth allocation scheme can be written as:

$$S_L = \dfrac{\beta}{\beta_{max}} = \begin{cases} 1, & \beta_{max}M \leq C \\ \dfrac{C}{\beta_{max}M}, & \beta_{max}M > C \end{cases} \quad (2)$$

Our proposed scheme allocates different amount of bandwidths for different broadcasting/multicasting video sessions based on popularity of video programs. However, the maximum allocated bandwidth to a broadcasting/multicasting video session is $\beta_{max}$ and the minimum allocated bandwidth to a broadcasting/multicasting video session is $\beta_{min}$. Where $\beta_{min}$ ensures the minimum quality of a video session. An active broadcasting/multicasting video session is ranked based on the number of users currently watching the program on that session. The most popular broadcasting/multicasting video session (program) is ranked as 1. Where the lowest popular one is ranked as $M$. The numbers of active users for different broadcasting/multicasting video sessions are related as:

$$K_1 \geq K_2 \geq \cdots \geq K_m \geq \cdots \geq K_M \quad (3)$$

where $K_m$ is the number of users watching the *m-th* video program. *m=1* indicates that program which is being watched by the maximum number of users. Whereas *m=M* indicates that with the minimum users. *K* is the total number of active users in the system.

Total number of active users, *K,* in the system is:

$$K = K_1 + K_2 + \cdots + K_m + \cdots + K_M \qquad (4)$$

Equation (3) can be rewritten as:

$$K_1 \geq \frac{K}{M} \text{ and } K_M \leq \frac{K}{M} \qquad (5)$$

Minimum number of broadcasting/multicasting video sessions, $N_{HQ}$, that can be provided simultaneously with the allocated bandwidth $\beta_{max}$ (best quality) for each of the broadcasting/multicasting video sessions is:

$$N_{HQ} = \left\lfloor \frac{C}{\beta_{max}} \right\rfloor \qquad (6)$$

Maximum number of broadcasting/multicasting video sessions, $N_{LQ}$, that can be provided simultaneously with the allocated bandwidth $\beta_{min}$ (lowest quality of video) for each of the broadcasting/multicasting video sessions is:

$$N_{LQ} = \left\lfloor \frac{C}{\beta_{min}} \right\rfloor \qquad (7)$$

Fig. 8.1 shows the basic concepts of bandwidth allocation per broadcasting/multicasting video session by the equally shared and the proposed popularity based bandwidth allocation schemes when the system bandwidth is not sufficient to allocate $\beta_{max}$ for each of the active broadcasting/multicasting video sessions. Fig. 8.1(a) shows that an equal bandwidth $\beta$ is allocated to each of the broadcasting/multicasting video sessions by the equally shared bandwidth allocation scheme. On the other hand, Fig. 8.1(b) shows that the same bandwidth is not allocated to each of the active video sessions by the proposed popularity based bandwidth allocation scheme.

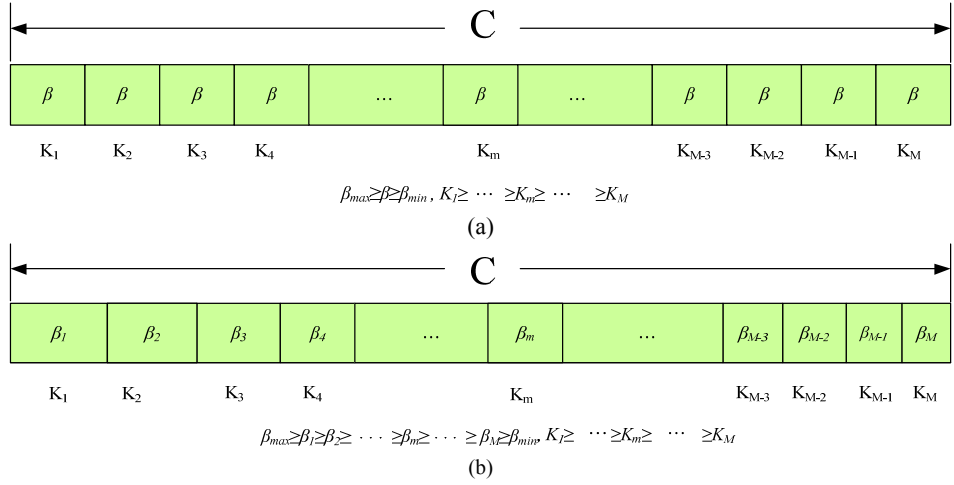

**Fig. 8.1:** An example of bandwidth allocation when the system bandwidth is not sufficient to allocate $\beta_{max}$ for each of the broadcasting/multicasting video sessions (a) equal allocated bandwidths to all of the broadcasting/multicasting video sessions by the equally shared bandwidth allocation scheme, (b) different allocated bandwidths to different broadcasting/multicasting video sessions by the proposed popularity based bandwidth allocation scheme.

Maximum bandwidth $\beta_1$ is allocated to the broadcasting/multicasting video session #1 which is enjoyed by the maximum number of subscribers. On the other hand, minimum bandwidth $\beta_M$ is allocated to the broadcasting/multicasting video session #M which is received by the minimum number of subscribers.

Bandwidth $\beta_{max}$ is allocated for each of the broadcasting/multicasting video sessions whenever $\beta_{max} M \leq C$. However, if $\beta_{max} M > C$, then the allocated bandwidth $\beta_m$ for *m-th* broadcasting/multicasting video session in the proposed scheme is calculated by the following procedures,

$$X_m = \begin{cases} 0, & \left(aK_m + \sum_{j=1}^{m-1} X_j\right) \leq \beta_{diff} \\ \dfrac{aK_m + \sum_{j=1}^{m-1} X_j - \beta_{diff}}{M - m}, & \left(aK_m + \sum_{j=1}^{m-1} X_j\right) > \beta_{diff} \end{cases} \quad (8)$$

where $a = \dfrac{M}{K}\left(\dfrac{C}{M} - \beta_{min}\right)$ and $\beta_{diff} = \beta_{max} - \beta_{min}$

$$\beta_m = \begin{cases} \beta_{max}, & \left(aK_m + \sum_{j=1}^{m-1} X_j\right) \geq \beta_{diff} \\ \beta_{min} + aK_m + \sum_{j=1}^{m-1} X_j, & \left(aK_m + \sum_{j=1}^{m-1} X_j\right) < \beta_{diff} \end{cases} \quad (9)$$

Hence, the dedicated bandwidth for the *m-th* broadcasting/multicasting video session is $\beta_m$. However, for instance, a receiving device with limited resources e.g., restricted display resolution, processing capacity, and battery power decodes only a part of the broadcasted/multicasted bit stream. Therefore, in a broadcast/multicast cases, terminals with different capabilities can be served through a single scalable bit stream.

The difference between the allocated bandwidth for the *m-th* and the *(m+1)-th* broadcasting/multicasting video session is:

$$\beta_m - \beta_{m+1} = \begin{cases} 0, & \left(aK_m + \sum_{j=1}^{m+1} X_j\right) \geq \beta_{diff} \text{ or } (K_m = K_{m+1}) \\ a(K_m - K_{m+1}) - X_m, & \left(aK_m + \sum_{j=1}^{m+1} X_j\right) < \beta_{diff} \end{cases} \quad (10)$$

Hence, from (8)-(10), the allocated bandwidths of the active broadcasting/multicasting video sessions for the proposed scheme are related as follows,

Case 1: when $\beta_{max} M \leq C$,

$$\beta_1 = \beta_2 = \cdots = \beta_m = \cdots = \beta_M = \beta_{max} \quad (11)$$

Case 2: when $\beta_{max} M > C$,

$$\left. \begin{aligned} \beta_1 \geq \beta_2 \geq \cdots \geq \beta_m \geq \cdots \geq \beta_M \\ \beta_{max} \geq \beta_1 \geq \frac{C}{M} \\ \beta_{min} \leq \beta_M \leq \frac{C}{M} \end{aligned} \right\} \quad (12)$$

Whenever the number of users for an active video session or the total number of active video sessions is changed, the allocated bandwidth for each of the active broadcasting/multicasting video sessions is also dynamically changed. As a consequence, the number of enhanced layers per session and the allocated bandwidth per enhanced layer may also be changed. It can be mentioned that a receiver cannot subscribe to a fraction of a layer.

In our proposed scheme, the satisfaction level of the users who are connected with the *m-th* broadcasting/multicasting video session is:

$$S_{L(m)} = \begin{cases} 1, & \beta_{max}M \leq C \\ \dfrac{\beta_m}{\beta_{max}}, & \beta_{max}M > C \end{cases} \quad (13)$$

From (11)-(13), the relation between the satisfaction levels of different users can be written as:

$$1 \geq S_{L(1)} \geq S_{L(2)} \geq \cdots \geq S_{L(m)} \geq \cdots \geq S_{L(M)} \geq \dfrac{\beta_{min}}{\beta_{max}} \quad (14)$$

The average user satisfaction level for the proposed scheme is calculated as:

$$S_{L(av)} = \begin{cases} 1, & \beta_{max}M \leq C \\ \dfrac{\sum_{m=1}^{M} S_{L(m)}K_m}{K}, & \beta_{max}M > C \end{cases} \quad (15)$$

where $SL_{(av)}$ is the average user satisfaction level for the proposed scheme considering all the active users in the system.

The relation between the average user satisfaction levels for the proposed popularity based bandwidth allocation scheme and the equally shared bandwidth allocation scheme can be written as:

$$\left.\begin{array}{ll} S_{L(av)} = S_L = 1, & \beta_{max}M \leq C \\ S_{L(av)} = S_L, & K_1 = K_M \\ S_{L(av)} > S_L, & K_1 \neq K_M \text{ and } \beta_{max}M > C \end{array}\right\} \quad (16)$$

It seems that, if larger number of users watch the program of a broadcasting/multicasting video session, then higher bandwidth is allocated for that video session to provide better quality of service for those. Thus, the average user satisfaction level is increased significantly.

## 8.2 Performance Evaluation

In this section, we verified performance of the proposed scheme using simulation results. The basic assumptions for the performance analysis are shown in Table 8.1. We performed the analysis for two scenarios of traffic environment while the total number of users in the system is fixed. Scenario 1 considers random number of active users for per video session. Scenario 2 also considers random manner. However, in second scenario, 50% users watch one video program and the remaining 50% users watch other video programs.

**Table 8.1:** Basic assumptions

| Parameter | Value |
|---|---|
| The total system bandwidth capacity $(C)$ | 30 Mbps |
| Maximum allocated bandwidth for each of the broadcasting/multicasting video sessions $(\beta_{max})$ | 2 Mbps |
| Minimum allocated bandwidth for each of the broadcasting/multicasting video sessions $(\beta_{min})$ | 0.6 Mbps |
| Number of users with per video session | Random |
| Total number of active users in the system $(K)$ | 200 |

Firstly, we verify the improvement of average user satisfaction level for our proposed scheme compared to the equally shared bandwidth allocation scheme. Fig. 8.2 shows that the proposed scheme provides much better average user satisfaction level compared to the equally shared bandwidth allocation scheme. The user satisfaction level decreases with the increase of active video sessions due to the limited bandwidth capacity of the network. Fig. 8.2(b) shows that our proposed scheme is even more effective when large number of users watch the program of a common broadcasting/multicasting video session.

Fig. 8.3 shows the allocated bandwidth for each of the video sessions when 30 video sessions are active. It shows that the allocated bandwidth to a video session is gradually decreased with the decrease of popularity in the proposed scheme. The maximum allowable bandwidth $\beta_{max}$ can be allocated for more than one video sessions depending on the network bandwidth and the traffic conditions. However, the allocated bandwidth for any of the active broadcasting/multicasting video sessions does not go below a threshold level to guarantee the minimum video quality for all the active video sessions. Hence, the allocated bandwidths for some broadcasting/multicasting video sessions are increased and for some broadcasting/multicasting video sessions are decreased compared to the equally shared bandwidth scheme. In Fig. 8.3(a) case, 168 users enjoy improved quality videos whereas 32 users receive degraded quality of videos. In Fig. 8.3(b) case, 177 users enjoy improved quality videos whereas 23 users receive degraded quality of videos. However, for both the cases minimum quality of video is assured.

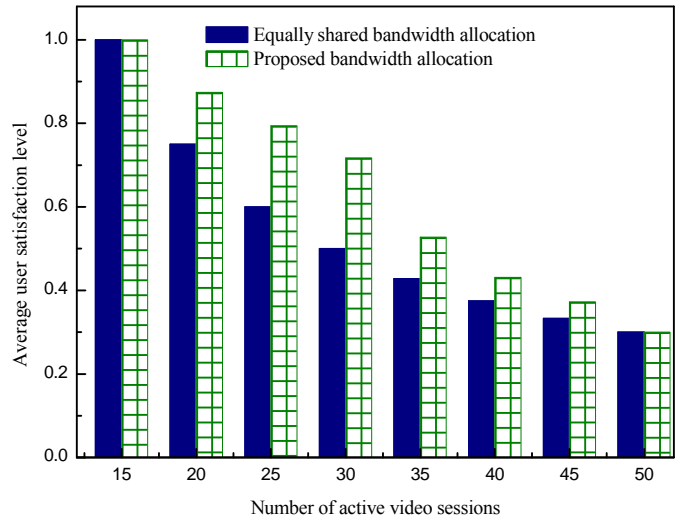

(a)

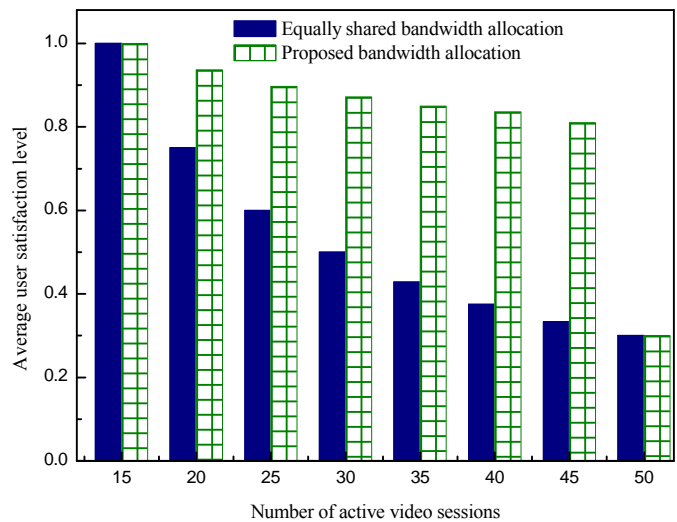

(b)

**Fig. 8.2:** A comparison of the average user satisfaction levels for various numbers of active video sessions (a) scenario 1 traffic environment, (b) scenario 2 traffic environment.

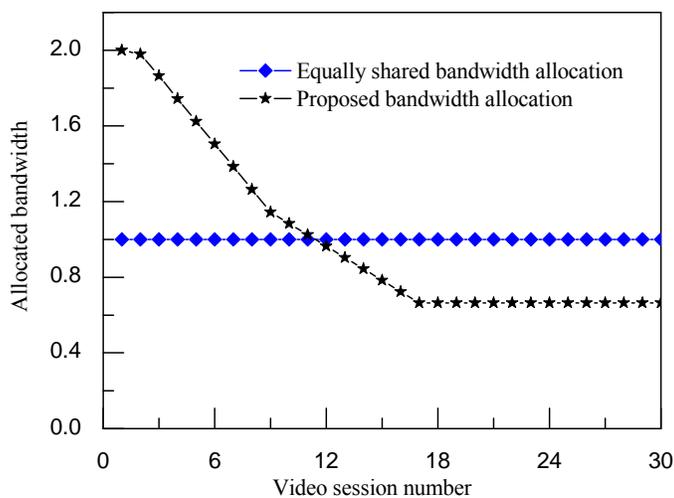

(a)

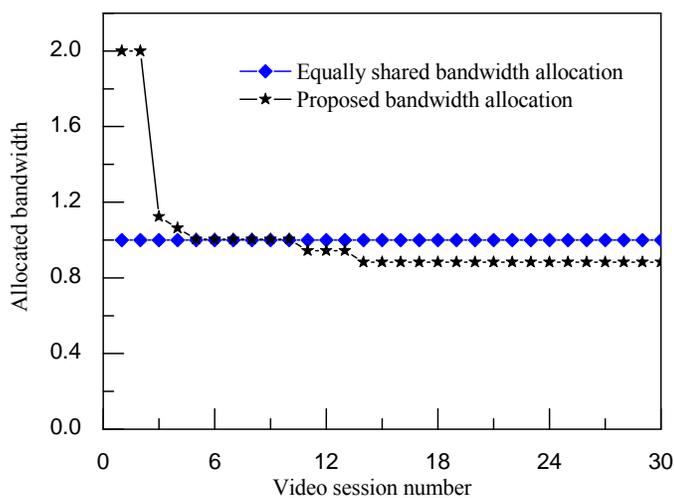

(b)

**Fig. 8.3:** Allocated bandwidths for various video sessions when 30 video sessions are active (a) scenario 1 traffic environment, (b) scenario 2 traffic environment.

Fig. 8.4 shows a comparison between the numbers of users to whom the video quality is improved and the users to whom it is degraded in the proposed scheme compared to the equally shared bandwidth allocation scheme. Both Figs. in 8.4(a) and 8.4(b) indicate that huge number of users can enjoy improved video quality. To improve the video quality for these large number of users, the only adjustment is that a very few users receive slightly degraded video quality. Hence, a large number of users enjoy the significantly improved video quality.

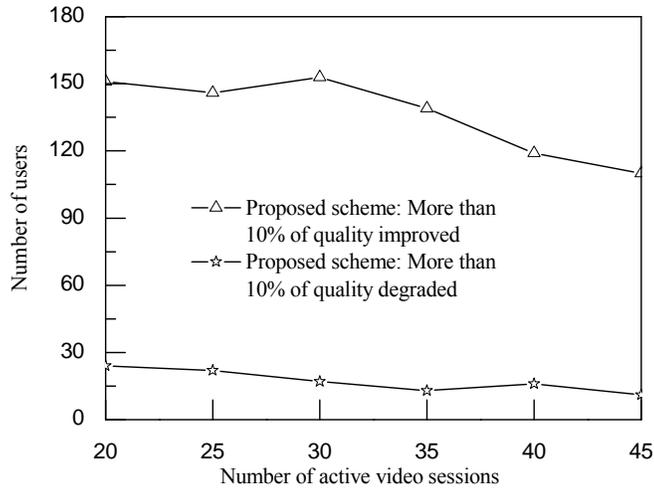

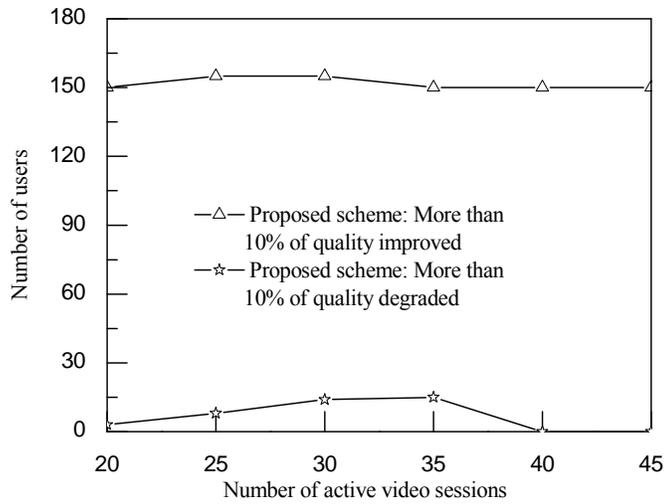

**Fig. 8.4:** A comparison to show the number of users to whom video quality is improved or degraded with respect to the equally shared bandwidth allocation scheme (a) scenario 1 traffic environment, (b) scenario 2 traffic environment.

Fig. 8.5 demonstrates the bandwidth allocation for the most popular broadcasting/multicasting video session and lowest popular broadcasting/multicasting video session in the proposed scheme. Both Figs. 8.5(a) and 8.5(b) designate that the most popular video session is guaranteed with the sufficient bandwidth except the case where entire system bandwidth is required to provide the minimum bandwidth for each of the video sessions. For the Fig. 8.5(b), even the bandwidth allocation for the lowest popular video session is quite close to that of the equally shared bandwidth allocation scheme.

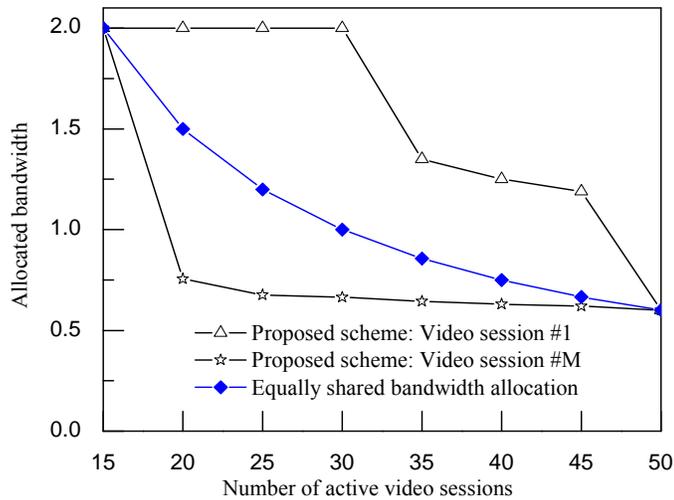

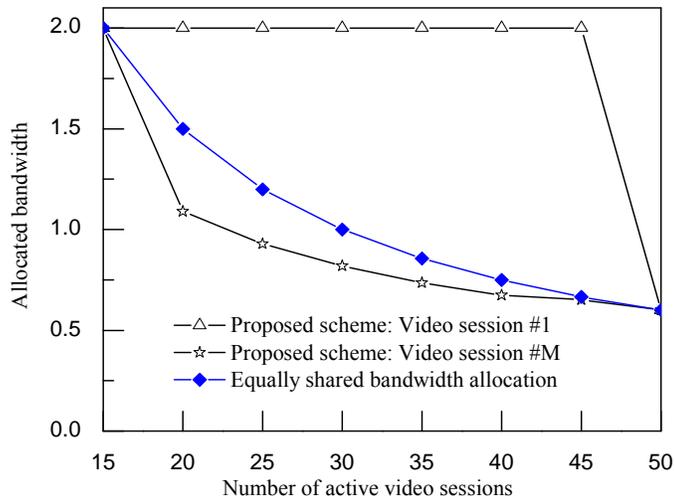

**Fig. 8.5:** A comparison of the allocated bandwidths for the most popular video session and the lowest popular video session (a) scenario 1 traffic environment, (b) scenario 2 traffic environment.

    The results of the performance analyses show that our proposed popularity based bandwidth allocation scheme is able to improve average user satisfaction level. The proposed scheme is even more effective when large number of users watch the program of a common broadcasting/multicasting video session.

.

# Chapter 9
# Conclusions

The tremendously increasing of the use of wireless networks for various high data rate applications have been seen in the recent years and it will continue in the future. Different wireless technologies have been developed due to these huge demands, varieties of user types, and varieties of user's requirement. However, the capacity of wireless links are not increasing compared to the increasing rate of demand. Moreover, many wired services are converted to wireless services (e.g., mobile IPTV). Therefore, the limited wireless resources should be utilized properly.

The research was mainly divided into two parts:

a. Femtocellular network deployment and resource allocation
b. Resource management for macrocellular networks.

Chapter 1 discusses the motivation behind this research study. The femtocellular network deployment scenarios, advantages of femtocellular network deployment, and the comparisons of femtocell technology with other indoor technologies are presented in chapter 2. Chapter 3 provides the integrated femtocell/macrocell network architectures. The network architectures for different scale of femtocellular network deployments are also presented in this chapter. Cost-effective frequency planning for capacity enhancement of femtocellular networks is presented in Chapter 4. Firstly, the interference scenarios are presented and then, shared frequency band, sub-frequency band, frequency-reuse, static frequency re-use, and dynamic frequency re-use schemes are discussed in details in this chapter. The outage probability as well as capacity are analyzed and compared for different frequency allocation schemes and different scales of femtocellular network deployment. Chapter 5 discusses the mobility management issues for the dense femtocellular network deployment. We present a complete tutorial of mobility management scheme for the integrated macrocell/femtocell network. We suggest the SON network architecture to support the dense femtocellular networks. Then, we propose, an algorithm to make neighbor cell list with minimum number but appropriate cells for the handover, detail handover procedures, and a traffic model for the integrated macrocell/femtocell network in this chapter. We provide a CAC based on adaptive bandwidth allocation for the wireless network in Chapter 6. This chapter provides a CAC that relies on adaptive multi-level bandwidth-allocation scheme for non-real-time calls. The scheme allows reduction of the call dropping probability along with increase of the bandwidth utilization. Numerical results analyses are also presented to verify the performance of the proposed scheme. In Chapter 7, we provide a QoS adaptive radio resource allocation for scalable videos over wireless cellular networks. The proposed bandwidth allocation scheme efficiently allocates bandwidth among the MBS sessions and the non-MBS traffic calls. Popularity based bandwidth allocation for scalable video over wireless networks is proposed in Chapter 8.

The research results presented in this dissertation clearly imply the advantages of our proposed schemes. Dense deployment of femtocells will offload huge traffic from the macrocellular network to femtocellular network by the successful integration of macrocellular and femtocellular networks. The proposed three possible architectures for the evolution of the integrated network architectures are economical for the expansion of femtocellular network deployment. The SON-based coordination and cooperative communications among FAPs and macrocellular Base Stations can improve the spectral utilization and QoS performances. Our study shed some light onto the challenging problem of interference mitigation in an integrated femtocellular/macrocellular networks. The analyses of the mobility management for the dense femtocellular network deployment indicate the effect of femtocellular network deployment.

The proposed handover priority scheme provides an efficient CAC that relies on adaptive multi-level bandwidth-allocation scheme for non-real-time calls. The scheme allows reduction of the call dropping probability along with increase of the bandwidth utilization.

Video over wireless networks is proposed in this research. The proposed bandwidth allocation scheme efficiently allocates bandwidth among the MBS sessions and the non-MBS traffic calls. The proposed scheme reduces the bandwidth allocation for the MBS sessions during the congested traffic condition only to accommodate more non-MBS traffic calls in the system. For the proposed popularity based bandwidth allocation scheme for the MBS, instead of allocating equal bandwidths to all of the sessions during an insufficient bandwidth condition, our proposed scheme efficiently allocates the total system bandwidth among them in such a way that higher bandwidth is allocated to the video session of higher popularity. Thus, the average user satisfaction level is increased significantly. However, a minimum quality for the lowest popular video session is guaranteed by assigning a minimum amount of bandwidth.

Our proposed schemes are able to efficiently manage the valuable wireless resources as well as can provide high data rate various services for the mobile users in various environments.

# List of Publications (Sept. 2008~)

# List of Contribution for Standardization (Sept. 2008~)

[1] Young-Il Kim , Dae Geun Park, Chun Sun Sim, Won Ryu, Yeong Min Jang, **Mostafa Zaman Chowdhury**, and Ilyoung Chong, "Proposed modification of Draft Recommendation Y.miptv-req (Functional requirements for Mobile IPTV)," Presented at ITU-T Meeting, Geneva, Switzerland, September 2010.

[2] Yeong Min Jang and **Mostafa Zaman Chowdhury** "MRPM for Longer Battery Lifetime," *IEEE 802.21 Media Independent Handover,* Presented at IEEE 802.21 sessions #33 in SFO, USA, July 2009.

[3] Yeong Min Jang, Le Nam Tuan, **Mostafa Zaman Chowdhury,** and Muhammad Shahin Uddin, "Network Composition for Interconnecting Cellular Networks and Visible Light Communication (VLC)," Presented at IEEE 802.15 Hawaii, USA, September 2009.